\documentclass{jfm}
\usepackage{natbib}
\usepackage{upmath}
\usepackage[british]{babel}
\usepackage{amsmath,bm}
\usepackage{amssymb}
\usepackage{amsbsy}
\usepackage{amscd}
\usepackage{amstext}
\usepackage{tabularx}
\usepackage{float}
\usepackage{makeidx}
\usepackage{amsmath}
\usepackage{subfigure}
\usepackage{afterpage}
\usepackage[T1]{fontenc}
\usepackage[latin1]{inputenc}
\usepackage{multirow}
\usepackage{color}
\usepackage{url}

\usepackage{graphicx} 
\usepackage{graphics,epsfig}
\usepackage{amsfonts}
\usepackage{psfrag}

\ifCUPmtlplainloaded \else
  \checkfont{eurm10}
  \iffontfound
    \IfFileExists{upmath.sty}
      {\typeout{^^JFound AMS Euler Roman fonts on the system,
                   using the 'upmath' package.^^J}%
       \usepackage{upmath}}
      {\typeout{^^JFound AMS Euler Roman fonts on the system, but you
                   dont seem to have the}%
       \typeout{'upmath' package installed. JFM.cls can take advantage
                 of these fonts,^^Jif you use 'upmath' package.^^J}%
      }
  \else
  \fi
\fi

\ifCUPmtlplainloaded \else
  \checkfont{msam10}
  \iffontfound
    \IfFileExists{amssymb.sty}
      {\typeout{^^JFound AMS Symbol fonts on the system, using the
                'amssymb' package.^^J}%
       \usepackage{amssymb}%
         \let\leq=\leqslant
         \let\geq=\geqslant
      }{}
  \fi
\fi

\ifCUPmtlplainloaded \else
  \IfFileExists{amsbsy.sty}
    {\typeout{^^JFound the 'amsbsy' package on the system, using it.^^J}%
     \usepackage{amsbsy}}
    {\providecommand\boldsymbol[1]{\mbox{\boldmath $##1$}}}
\fi

\providecommand\bcdot{\boldsymbol{\cdot}}

\newsavebox{\astrutbox}
\sbox{\astrutbox}{\rule[-5pt]{0pt}{20pt}}

\def\Lh{\mathbf{L}}

\def\Ey{{\bm{E}^\ast}}

\def\Qy{Q^\ast}
\def\Uy{{\bm{U}^\ast}}

\def\Ky{K^\ast}

\def\Py{P^\ast}
\def\ty{t^\ast}
\def\phiy{\phi^\ast}

\def\Ub{\mathbf{U}}
\def\Eb{\mathbf{E}}
\def\ub{\mathbf{u}}
\def\eb{\mathbf{e}}

\def\Ub{\mathbf{U}}
\def\Eb{\mathbf{E}}
\def\ub{\mathbf{u}}

\def\eb{\mathbf{e}}

\title[Moffatt eddies in EHD flows: numerical simulations and analyses]
{Moffatt eddies in electrohydrodynamic flows: numerical simulations and analyses}

\author[X. He, Z. Sun, M. Zhang]%
{
Xuerao He$^1$, Zhihao Sun$^{1,2}$ \and Mengqi Zhang$^1$\thanks{Email address for correspondence: {mpezmq@nus.edu.sg}}
}

\affiliation{
  $^1$ Department of Mechanical Engineering, National University of Singapore, 9 Engineering Drive 1, 117575 Singapore 
  \\
  $^2$ School of Energy Science and Engineering, Harbin Institute of Technology, Harbin 150001, PR China\\ [\affilskip]
}

\date{\today}

\graphicspath{{figures/}}
\newlength\savewidth

\begin{document}
\maketitle

\begin{abstract}
We study numerically a sequence of eddies in two-dimensional electrohydrodynamic (EHD) flows of a dielectric liquid, driven by an electric potential difference between a hyperbolic blade electrode and a flat plate electrode (or the blade-plate configuration). The electrically-driven flow impinges on the plate to generate vortices, which resemble Moffatt eddies (Moffatt, \textit{J. Fluid Mech.} vol. 18, 1-18, 1964). Such a phenomenon in EHD was first reported in the experimental work of Perri \textit{et al., J. Fluid Mech.} vol. 900, A12, 2020. We conduct direct numerical simulations of the EHD flow with three Moffatt-type eddies in a large computational domain at moderate electric Rayleigh numbers ($ T $, quantifying the strength of the electric field). The ratios of size and intensity of the adjacent eddies are examined and they can be compared favourably to the theoretical prediction of Moffatt; interestingly, the quantitative comparison is remarkably accurate for the two eddies in the farfield. Our investigation also shows that a larger $T$ strengthens the vortex intensity, and a stronger charge diffusion effect enlarges the vortex size. A sufficiently large $T$ can further result in an oscillating flow, consistent with the experimental observation. In addition, a global stability analysis of the steady blade-plate EHD flow is conducted. The global mode is detailedly characterised at different values of $T$. When $T$ is large, the confinement effect of the geometry in the center region may lead to an increased oscillation frequency. This work contributes to the quantitative characterisation of the Moffatt-type eddies in electrohydrodynamic flows. 

\end{abstract}

\section{Introduction}

The formation and evolution of vortices in fluid flows have long been of great interest to fluid dynamicists. In a Stokes flow, \cite{moffatt} first theoretically determined and characterised a sequence of counter-rotating vortices in the corner flow between two planes (at least one of which is a solid boundary). Moffatt analytically calculated the size ratio and the intensity ratio of successive vortices. After this seminal work, Moffatt-like eddies have been observed and studied in many other flow settings, such as the flow in a sudden expansion \citep{ALLEBORN1997}, backward-facing step flows \citep{Biswas2004}, lid-driven cavity flows \citep{biswas2018moffatt}, etc. In the context of electrohydrodynamics (EHD) flows, the Moffatt-like eddies were only observed very recently by \cite{perri2020electrically,perri2021particle} in their experiments. Their experimental device consists of a grounded plate and a needle electrode placed vertically above it. Their results showed that the Moffatt-like vortices appeared when a sufficiently high voltage was applied and the experimental results agreed with the theoretical predictions of the Moffatt-type vortices between two concentric conical surfaces \citep{malhotra2005nested}. Inspired by their work, we are interested in numerically exploring and characterising the Moffatt-type eddies in the EHD flow between a hyperbolic blade electrode and a flat plate electrode. In the following, we will first review works on Moffatt eddies in general and then summarise studies on the EHD flow in a blade/needle-plate configuration. We will in the end explain the motivation of this work and define its position in the literature.

\subsection{Moffatt eddies}

Moffatt eddies refer to a sequence of eddies that develop in a corner between two planes as a result of an arbitrary disturbance imposed at a large distance or near the corner \citep{moffatt}. Moffatt showed that such eddies in a Stokes flow will form when the angle between the two planes is less than about $146^\circ$. {The formation of these eddies has also been mathematically explained from the perspective of flow singularities in the Navier-Stokes equations at the perfectly sharp corner \citep{moffatt1999corner,moffatt2019singularities}.} The existence of these vortices was verified in flow visualization experiments by \cite{taneda1979visualization}. Meanwhile, the Moffatt-type eddies in different geometries have been observed and studied.  A sequence of viscous eddies was found and studied between two spherical surfaces \citep{davis1976separation} and between a cylinder and a plane \citep{davis1977separation}. {The axisymmetric flow of a viscous fluid within rigid conical surfaces was investigated by \cite{wakiya1976axisymmetric}, and the largest semi-angle of the cone generating Moffatt-like eddies was found to be $80.9^\circ$. \cite{moffatt1980local} examined the pressure-driven flow along a duct with a sharp corner and found that when the corner angle is larger than $90^\circ$, the local similarity solution is valid.} \cite{weidman1999instantaneous} studied the Stokes flow in a cone bounded by stress-free surfaces and driven by gravity parallel to the conical axis. {More recently, Shankar conducted a series of work on the Moffatt-type eddies in a cylindrical container \citep{shankar1997three,shankar1998three}, a semi-infinite wedge \citep{shankar2000stokes} and a cone \citep{shankar2005moffatt}.} The Moffatt eddies in a circular cone were also examined by \cite{malyuga2005viscous}, where the flow is driven by a non-zero velocity applied to the boundary. \cite{malhotra2005nested} investigated the Moffatt vortices in an asymmetric double-cone geometry. Later, \cite{scott2013moffatt} explored the three-dimensional Moffatt eddies in a trihedral cone formed by three orthogonal planes.  \cite{kirkinis2014moffatt} predicted the presence of Moffatt vortices in a moving liquid wedge between a gas-liquid interface and a rigid boundary. It can be summarised that theoretical analyses of the Moffatt eddies in different geometries have been of interest for fluid dynamicists for a long time.

In addition to the theoretical work, numerical simulation techniques have also been adopted to study Moffatt eddies. The first numerical computation mentioning Moffatt vortices, to the best of our knowledge, was the work of \cite{burggraf1966analytical} on the lid-driven cavity flow at a moderate Reynolds number (quantifying the ratio of inertia to viscosity). Much later, \cite{magalhaes2013adaptive} performed the numerical simulations of lid-driven cavity flows and found that the small eddies in the corner of a creeping flow agreed quantitatively well with the theory of \cite{moffatt}. In the numerical simulations of the lid-driven cavity flow by \cite{biswas2016moffatt,biswas2018moffatt}, three eddies of Moffatt type were observed and their size and intensity ratios corroborated the theoretical values in \cite{moffatt}.  In addition, the Moffatt eddies were also observed and discussed in other numerically-simulated flows. \cite{Biswas2004} numerically investigated two- and three-dimensional laminar backward-facing step flows within a wide range of Reynolds numbers. For the two Moffatt eddies that the authors simulated at a finite Reynolds number, their size ratio agreed well with the theoretical value in \cite{moffatt}. Moreover, Moffatt eddies have also been observed in a sudden expansion by \cite{ALLEBORN1997} (in the limit of creeping flows). Their numerical results of the streamfunction field showed a high degree of resemblance with the theoretical result in \cite{moffatt}. Additionally, Moffatt vortices can also exist in some multi-physical flows. For example, the Moffatt-type vortices in thermocapillary convection were analysed by \cite{davis1989thermocapillary} and \cite{kuhlmann1999local}. As mentioned above, the Moffatt-like eddies in EHD flows have been reported by \cite{perri2020electrically,perri2021particle} in their experiments and simulations. In order to help the reader understand the EHD flows in general, we will introduce below a literature survey of the EHD flows and then discuss in detail the work of \cite{perri2020electrically,perri2021particle}.

\subsection{EHD flows in a blade/needle-plate configuration}
\begin{figure}
	\centering
	\subfigure{
		\begin{minipage}[h]{0.4\textwidth}
			\centering
			\includegraphics[height=3.7cm]{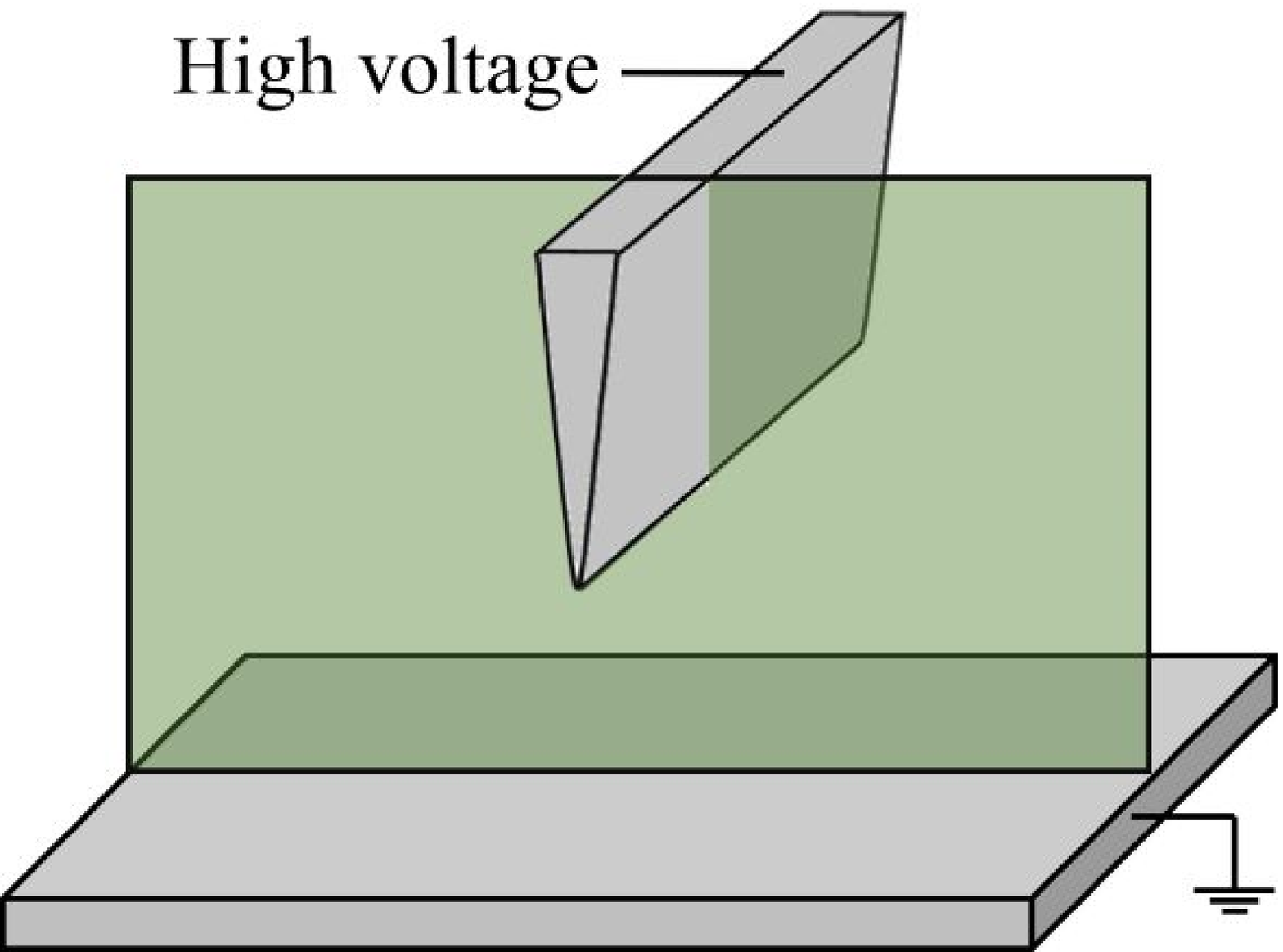}
			\put(-150,90){(a)}
			\label{fig.fig1a}
		\end{minipage}
	}
	\hspace{20pt}
	\subfigure{
		\begin{minipage}[h]{0.4\textwidth}
			\centering
			\includegraphics[height=3cm]{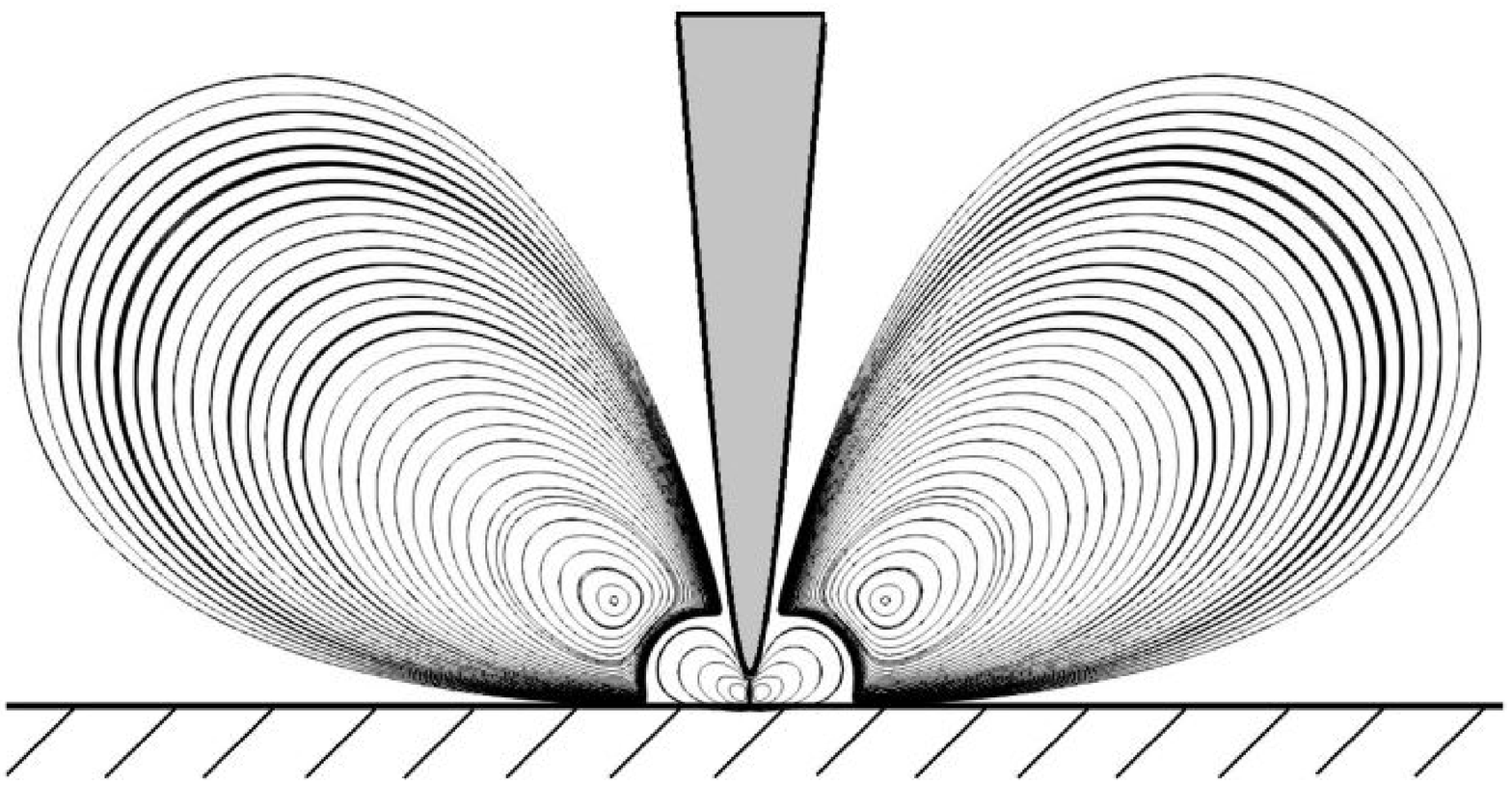}
			\put(-180,85){(b)}
			\label{fig.fig1b}
		\end{minipage}
	}
	\caption{(a) Sketch of the blade-plate electrodes in our EHD flow problem. (b) Streamlines in a two-dimensional cross section (green shade in panel (a)).}
	\label{fig.fig1}
\end{figure}

Electrohydrodynamics (EHD) studies the complex interaction between an electric field and a flow field; the flow field is driven by the electric force and at the same time influences the latter. This is an interdisciplinary subject of electromagnetism and hydrodynamics \citep{castellanos1998electrohydrodynamics}. It has broad prospects of applications in many scientific and industrial fields. In the study of EHD, the characteristics of the electric field play an important role in determining the dynamics of the electrified flow, such as the geometry of the electrodes and their arrangement. Accordingly, the research on EHD flow can be roughly divided into two categories: uniform electric fields (including parallel plates, concentric rings and balls) and non-uniform electric fields (including needle-plate, blade-plate, wire-plate, needle-ring, sphere-plane, etc.) \citep{suh2012modeling}. The configuration of a uniform electric field is beneficial for fundamental studies as it peels off unnecessary components that may complicate the theoretical treatment. This line of research has been followed by many researchers in the past decades, including, most notably, Castellanos and co-workers \citep{castellanos1998electrohydrodynamics,castellanos1991coulomb,chicon1997numerical,vazquez1996thermal,P1989Role} and Atten and co-workers \citep{atten1979non,lacroix1975electro,Malraison1982,mccluskey1988modifications}. However, although much useful flow information has been extracted in studying this (geometrically) simple EHD flow, when it comes to the practical applications of EHD, the non-uniform electric field electrode configuration is more relevant. For example, the needle-plate EHD configuration is widely used in electrospray techniques \citep{Fenn1989}. In fact, it is easier for the ions to overcome the potential energy barrier and be pushed
to the collector using a sharp electrode \citep{grassi2006heat}. In the electrode structure with a non-uniform electric field, the flow will be strongly non-parallel and non-stationary. We will in this work focus on the blade-plate configuration, as illustrated in figure \ref{fig.fig1}. We summarise in the following the research on the EHD flow in the blade-plate and needle-plate configurations.

The blade-plate EHD flow has been studied in some early experiments by \cite{tobazeon1984ion,haidara1985role}.
\cite{atten1997electrohydrodynamic} investigated the EHD flow in a needle-plate geometry both experimentally and theoretically. {The electrical current versus the tip-plate distance and the applied voltage were determined experimentally. In addition, their theoretical analysis estimated the axial velocity, although at variance with the experimental results.} The linear stability of a laminar EHD flow between a blade and a plate has been analysed using a small-disturbance method by  \cite{perez1995dynamics} and the results indicated that a smaller charged layer thickness renders the flow more unstable. Their experimental results were qualitatively consistent with their stability analysis, but quantitative differences existed. In the blade/needle-plate EHD flow, the so-called EHD plume structure will emerge, similar to the thermal plumes in the natural convection. \cite{vazquez1996thermal} studied and compared the dynamics of the thermal plumes and the EHD plumes. Their results showed that the equations describing the EHD plumes and the thermal plumes are equivalent under some assumptions at very large Prandtl number (ratio of momentum diffusivity to thermal diffusivity). 
Later, \cite{perez2009numerical} analysed the blade-plate EHD flow driven by the charge injection from a grid point. They found that with the increase of electric Rayleigh number ($ T $, quantifying the ratio between the Coulomb force and the viscous force), three different regimes could be observed: steady laminar, periodic and fully turbulent. Moreover, their results showed that the parameter $ M $ (which is the ratio between the hydrodynamic mobility to the ionic mobility) did not affect the dynamic of EHD plume. \cite{wu2013direct,traore2014electrohydrodynamic} numerically studied the EHD flow between a hyperbolic blade and a plate electrode. In their work, a non-autonomous injection law was considered and was compared with the classical autonomous injection law. It is noted that all the above numerical work ignored the effect of charge diffusion.

Experimental research on the blade/needle-plate EHD flow has been conducted in recent years. \cite{daaboul2017study} carried out experiments to investigate the EHD flow of blade-plate geometry based on a Particle Image Velocimetry system. The transition from conduction to injection with the increase of a DC (direct current) voltage was studied and complex flow patterns related to different charge generation mechanisms were observed. \cite{sankaran2018faradaic} examined the kinetic mechanism of the electrode Faradaic reaction in a vegetable oil between a needle and a plate electrode. Their experiment showed that redox reactions occurred on the needle electrode at a high voltage (about 4kV), leading to the emergence of Coulomb force acting on the charged oil and the EHD plume. In addition, the needle-plate EHD flow under DC or alternating-current (AC) electric field was investigated in detail by \cite{sun2020experimental}. The velocity field and current-voltage characteristics were discussed and analysed. 

In the aforementioned works on the blade-plate or needle-plate EHD configuration, the Moffatt eddies have not been mentioned or characterised. It was \cite{perri2020electrically,perri2021particle} who first discussed the Moffatt eddies in the needle-plate EHD flow. In their experiment, the flow was induced in a canola oil between a high-voltage needle electrode and a grounded plate. Three consecutive counter-rotating Moffatt vortices were observed once the voltage difference was above a threshold, determined to be between -8kV and -12kV. In addition, the position and structure of the vortices agreed well with the theoretical solutions of the Moffatt vortices between two concentric conical surfaces  \citep{malhotra2005nested}. They also observed transient flow phenomena, indicative of a flow bifurcation to another state. Finally, \cite{perri2020electrically} also performed numerical simulations of the EHD flow corresponding to their experimental setup. They presented numerical results on the electric field strength magnitude and charge density. This motivates the current work to further investigate the Moffatt-like eddies in EHD flows by numerical means.

\subsection{The current work}
From the literature review above, we realise that the Moffatt-like eddies in the blade-plate EHD flow has not been studied thoroughly from a  numerical perspective. We will in this work characterise the intrinsic flow characteristics and properties of the Moffatt-like eddies in the blade-plate EHD flow, supplementing the previous works of \cite{perri2020electrically,perri2021particle} in a needle-plate configuration. 
More specifically, we will conduct direct numerical simulations (DNS) and global stability analyses of 2-D EHD flows between a high-voltage hyperbolic blade electrode and a grounded plate electrode. 

In the first part of this work, the steady EHD flow, manifesting itself in the form of Moffatt-like eddies, will be numerically solved using DNS. Different from most previous numerical works reporting Moffatt-like eddies induced by a disturbance at a large distance from the corner, our simulations pertain to the case where the eddies are engendered by the disturbance near the corner. So the intensity of the eddies decreases with increasing distance from the corner. We will compare our numerical results with the theoretical predictions of \cite{moffatt}. The effect of charge diffusion in the EHD flow will be considered, which was omitted in the numerical simulations of \cite{perri2020electrically}. In the second part of this work, we will conduct a global stability analysis of the blade-plate EHD flow to probe its global linear dynamics. This is motivated by the consideration of understanding how and when the steady flow may remain stable or become unstable and thus transition to another flow state \citep{perri2020electrically}. We will primarily show the eigenvectors which can indicate the most important flow region for the perturbative dynamics in this flow. This has become our motivation particularly because we aim to present the results in a more comprehensible manner for the experimentalists working on this flow to explain, e.g., where the disturbances accumulate and develop. There seems no previous work in the literature on the stability analysis of this steady EHD flow.

The remaining part of this paper is organised as follows. In section \ref{problemformulation}, we describe the physical problem, present the nonlinear governing equations with boundary conditions and formulate the framework of the global linear stability analysis. Section \ref{numericalmethod} introduces the numerical methods. In section \ref{results}, we report the numerical results on the Moffatt-like eddies and analyse their global stability. The conclusion is drawn in section \ref{Conclusions} with some discussions on the flow physics. Appendices A and B detail the verification of our numerical simulation and analysis. Appendix C provides a nomenclature of the symbols used in this work.

\section{Problem formulation}\label{problemformulation}
\begin{figure}
	\centering
	\subfigure{
		\begin{minipage}[h]{0.4\textwidth}
			\centering
			\includegraphics[height=3.5cm]{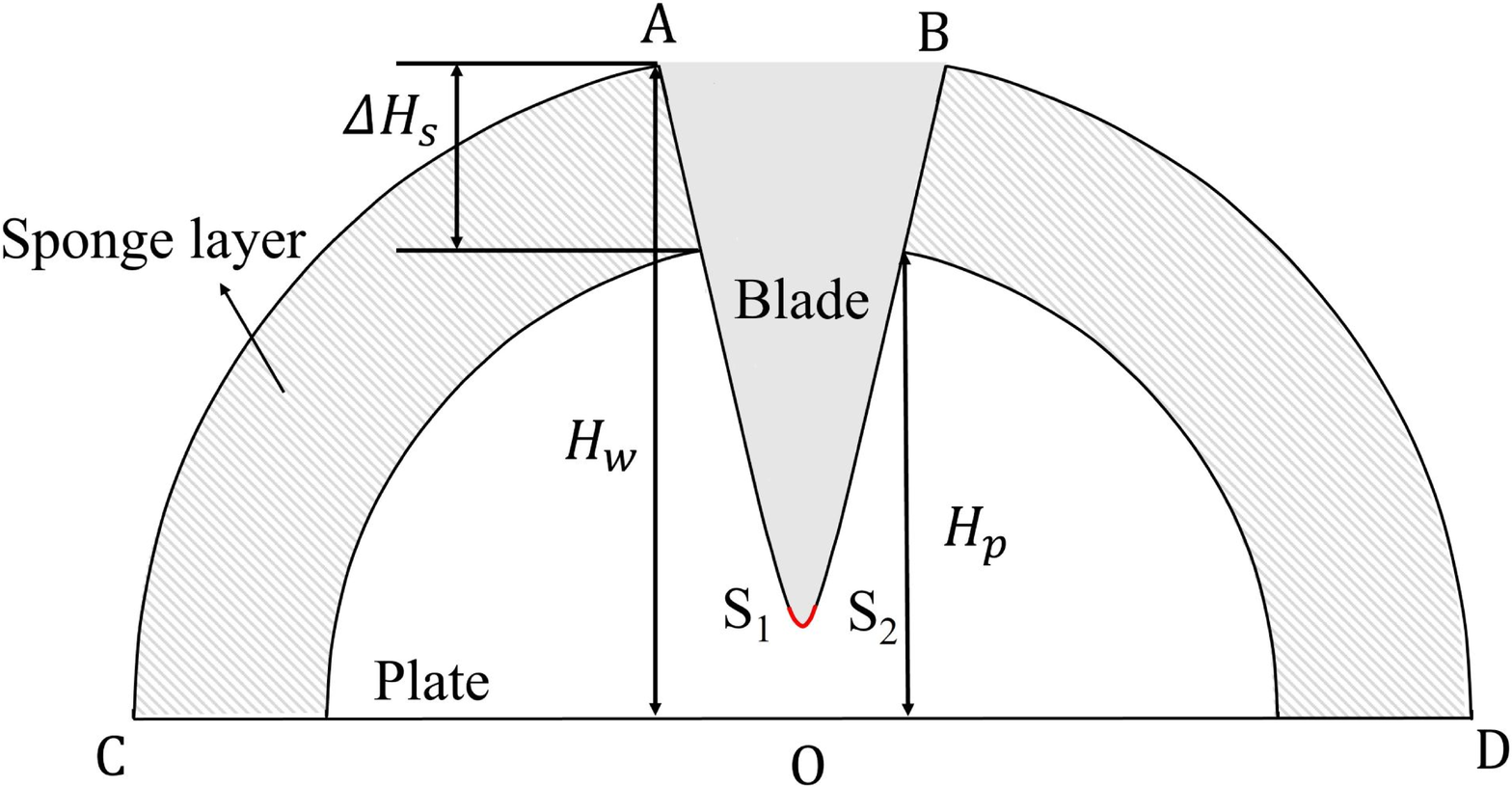}
			\put(-180,90){(a)}
			\label{fig.fig2a}
		\end{minipage}
	}
	\hspace{30pt}
	\subfigure{
		\begin{minipage}[h]{0.4\textwidth}
			\centering
			\includegraphics[height=3.8cm]{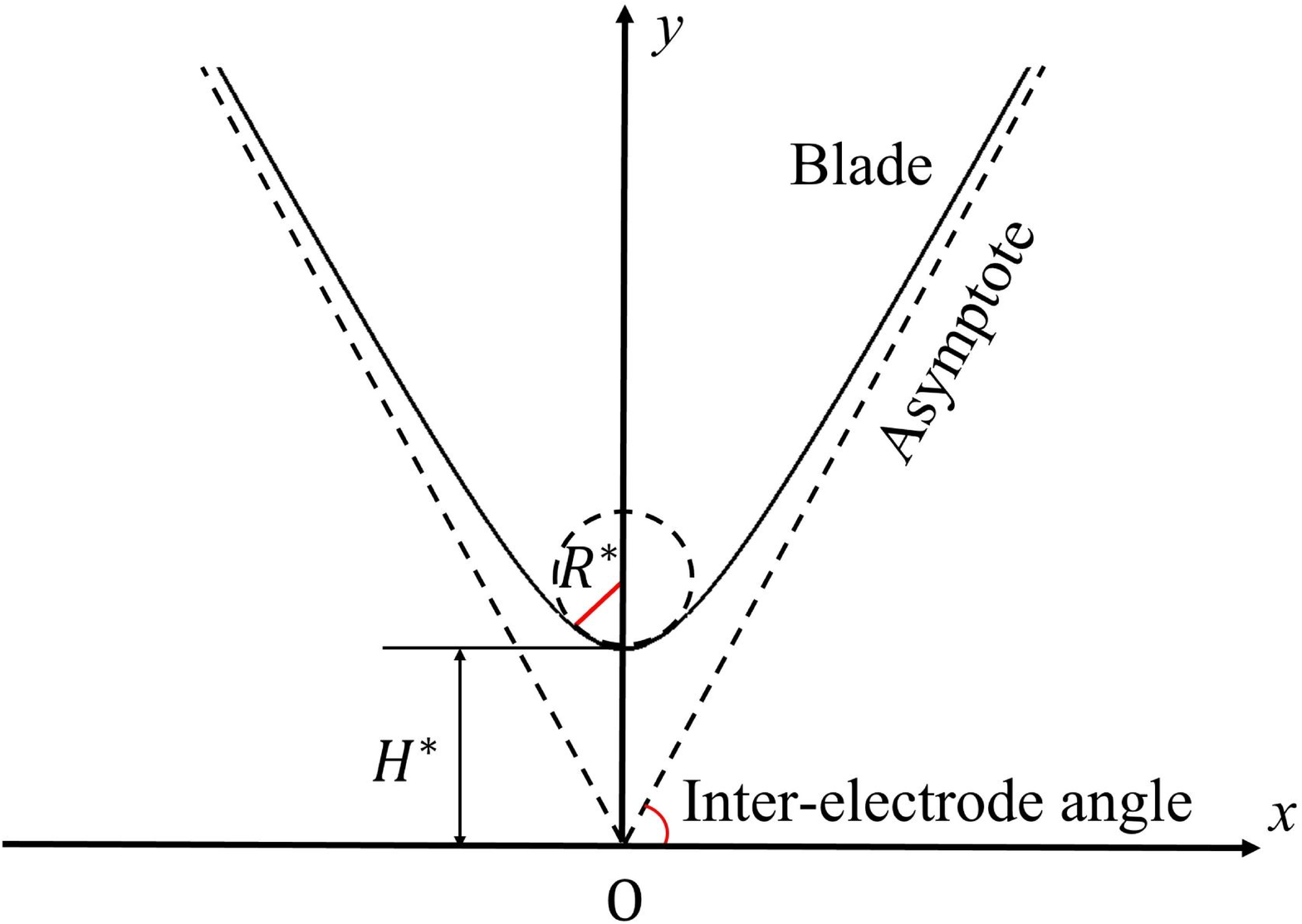}
			\put(-160,90){(b)}
			\label{fig.fig2b}
		\end{minipage}
	}
	\subfigure{
		\begin{minipage}[h]{0.4\textwidth}
			\centering
			\includegraphics[height=4cm]{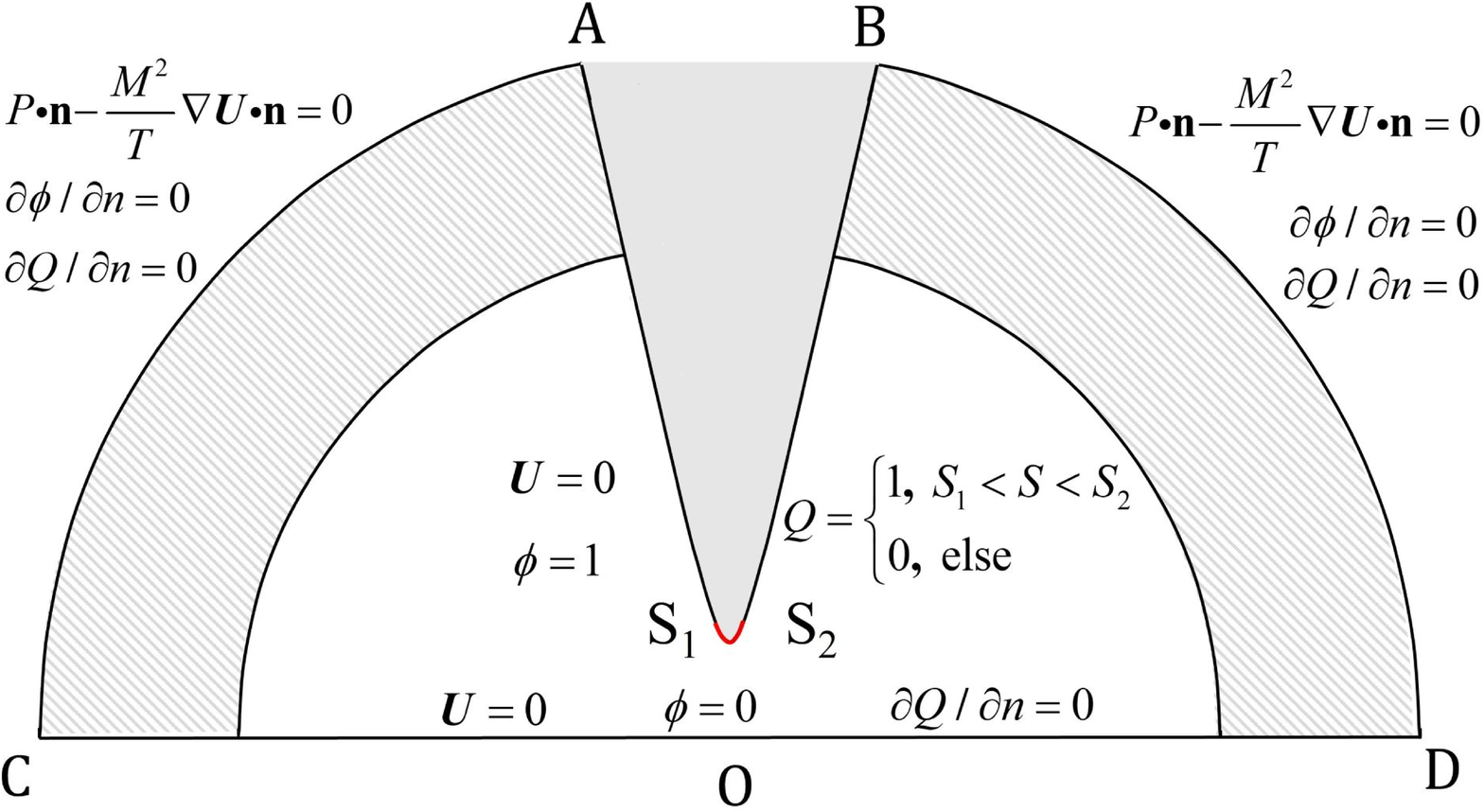}
			\put(-240,100){(c)}
			\label{fig.fig2c}
		\end{minipage}
	}
	\caption{(a) The computational domain in our numerical simulations. The white region is the physical flow domain and the hatching lines denote the sponge layer, which is used to damp the reflection of the outgoing waves. (b) A sketch of the hyperbolic blade electrode. (c) Specification of the boundary conditions. See the paragraph above Eq. \ref{eq.dimenless} for the meanings of the variables.}
	\label{fig.model}
\end{figure}

As shown in figure \ref{fig.fig2a}, we will consider in this work a two-dimensional EHD flow which arises in a dielectric liquid between a flat plate and a blade electrode. The shape of the blade electrode satisfies a hyperbolic equation as follows
\begin{equation}
	\left\{
	\begin{aligned}
		x^*&=\sqrt{R^*H^*}\mathrm{sinh}(\tau^*),\\
		y^*&=H^*\mathrm{cosh}(\tau^*),
	\end{aligned}
	\ \ \ \ \ \  \ \right.\tau^*\in \mathbb{R}.
	\label{eq.blade}
\end{equation}
This is a hyperbola with the its center being the original point $ O $, as shown in figure \ref{fig.fig2b}. In this paper, dimensional variables and parameters are denoted with a superscript $^ \ast $.  In the above equation, $ H^* $ represents the distance from the blade tip to the plate electrode; $ \tau^* $ defines a particular point on the hyperbola; $ R^* $ (red line in figure \ref{fig.fig2b}) is the radius of curvature of the blade tip, which determines the sharpness of the hyperbolic blade. Additionally, we label the angle between the asymptote of the hyperbolic blade and the bottom plate as the inter-electrode angle following \cite{perri2021particle}, as shown in figure \ref{fig.fig2b}.

In the current work, the dielectric liquid is assumed to be incompressible and Newtonian, which is characterised by the constant permittivity $ \varepsilon ^ * $, density $ \rho^*_0 $ and viscosity $  \mu^* $. A constant electric potential $\phi_0^* $ is imposed to the blade electrode, and the plate electrode is grounded, forming a nonuniform electric field between the blade and the plate. Charges are injected from the blade electrode, with the ionic mobility $ \Ky $ and charge diffusion constant $ D_\nu ^ * $. In theory, there are two main mechanisms for charges generation, i.e., conduction mechanism and injection mechanism. In the conduction scenario, charges are generated in the liquid as a result of the dissociation-recombination process of a solute or impurity present in the liquid. When the electric field is stronger than a critical value, charge injection can occur, where charges are produced by the electrochemical reaction between the interface of the electrode and the neutral impurities \citep{atten1996electrohydrodynamic, daaboul2017study}. Thus, in the case of a strong electric field, even though conduction mechanism may also be at work, the charge injection mechanism will play a leading role. Since this paper considers a strong electric field, we assume an injection mechanism only, that is, unipolar charges with strength $\Qy_0$ are issued from a fixed region on the hyperbolic blade between points $ S_1 $ and $ S_2 $, see the red portion in figure \ref{fig.fig2a}. This is also the consideration in the numerical simulations of \cite{perri2020electrically}.

In order to facilitate the subsequent presentation, we nondimensionalise the governing equations by appropriate physical scales. The length is non-dimensionalised by $ H^* $ (distance from the blade tip to the plate electrode), the time $ \ty $ by $ {H^*}^2/(\Ky\Delta\phiy_0) $, where $ \Delta\phiy_0 $ the potential difference applied to the electrodes, the electric potential $ \phiy $ by $ \Delta\phiy_0 $, the electric density $ \Qy $ by $ \Qy_0 $ (injected charge density), the velocity $ \Uy $ by $ \Ky\Delta\phiy_0/H^* $, the pressure $ \Py $ by $\rho^*_0{\Ky}^2\Delta{\phiy_0}^2/{H^*}^2 $, the electric field $ \Ey $ by $ \Delta\phiy_0/H^* $. Therefore, the non-dimensional equations for the EHD flow read \citep{castellanos1998electrohydrodynamics}
\begin{subequations}
	\begin{align}
		\nabla  \bcdot \Ub &= 0,
		\label{eq.dimless1}\\
		\frac{{\partial \Ub}}{{\partial t}} + (\Ub \bcdot \nabla )\Ub &=  - \nabla P + \frac{{{M^2}}}{T}{\nabla ^2}\Ub + C{M^2}Q\Eb+\mathbf{F}_{su},\\
		\frac{{\partial Q}}{{\partial t}} + \nabla  \bcdot [(\Eb + \Ub)Q] &= \frac{1}{{Fe}}{\nabla ^2}Q+F_{sq},
		\label{eq.dimless4}\\
		{\nabla ^2}\phi  &=  - CQ, \\
		\Eb &=  - \nabla \phi,
		\label{eq.dimenless5}
	\end{align}
	\label{eq.dimenless}
\end{subequations}
where
\begin{equation}
	\begin{split}
		M = \frac{\sqrt{\frac{\varepsilon ^ *}{ \rho _0^ * }}}{K^ * },\; \ \ \  T = \frac{{{\varepsilon ^ * }\Delta \phi _0^ * }}{{{K^ * }{\mu ^ * }}},\; \ \ \ C = \frac{{Q_0^ * {H^*}^2}}{{\Delta \phi _0^ * {\varepsilon ^ * }}},\; \ \ \ Fe = \frac{{{K^ * }\Delta \phi _0^ * }}{{D_\nu ^ * }}.
	\end{split} \label{nonnumbers}
\end{equation}
\begin{table}
	\centering
	\begin{tabular}{l c c c}
		Boundary conditions for &  velocity ($ \Ub $) & charge density (Q) & electric potential ($ \phi $)\\
		Plate electrode &   $ \Ub=0 $   & $ \partial Q/\partial y=0 $ & $ \phi=0 $\\
		Blade electrode (excluding injector)  & $ \Ub=0 $     & $ Q=0$  & $ \phi=1 $\\
		Injector ($ S_1\leq S\leq S_2 $)  & $ \Ub=0 $     & $ Q=1$  & $ \phi=1 $\\
		Arc $ \overset{\frown}{AC}$ and arc $\overset{\frown}{BD}$ & $ P\bcdot\mathbf{n}-\frac{M^2}{T}\nabla\Ub\bcdot\mathbf{n}=0 $ & $ \partial Q/\partial n=0 $ & $ \partial \phi/\partial n=0 $\\
	\end{tabular}
	\caption{Boundary conditions for numerical simulation of nonlinear blade-plate EHD flow.}
	\label{table:nonbc}
\end{table}
The forcing terms $ \mathbf{F}_{su} $ and $ F_{sq} $ are considered due to the artificial viscosity in the sponge layer. They damp the waves and their reflections near the outflow boundary, and will not affect the flow dynamics in the physics domain, to be described shortly.
Note that these equations are in principle the same as those used by \cite{perri2020electrically}, except that the geometries are different in the two works and that we additionally consider the charge diffusion effect ($\frac{1}{{Fe}}{\nabla ^2}Q$) in Eq. \ref{eq.dimless4}. More specifically, the first two equations are the continuity equation and Navier-Stokes (NS) equations with additional Coulomb force terms. The last three equations are the Maxwell's equations in the so-called quasielectrostatic limit (which refers to the case where the charge relaxation time or the electromagnetic waves' transition time is much shorter than the flow characteristic time, see Section 1.2 in \cite{castellanos1998electrohydrodynamics}). 

In the above equations, $ M $ is defined as the ratio between the hydrodynamic mobility to the ionic mobility. The electric Rayleigh number $T$ determines the ratio between the Coulomb force and the viscous force, which quantifies the strength of the imposed electric field. The parameter $ C $ measures the ion injection intensity and $ Fe $ is the inverse of the charge diffusion coefficient. The boundary conditions are summarised in figure \ref{fig.fig2c} and table \ref{table:nonbc}. We will conduct DNS of these equations to solve for their steady solutions, which are the Moffatt-like eddies in the EHD flow to be presented.

\subsection{Linearisation}\label{Linearisation}
We are also interested in the perturbative dynamics of the steady solution to the nonlinear equations. We will apply the classical global linear stability theory \citep{Theofilis2003,Theofilis2011} to study the linearised EHD flows in the blade-plate geometry.

In the linear stability analysis, we evoke Reynolds decomposition of the flow state, which writes the total flow state as the sum of a base flow and a perturbative component, that is, $\Ub=\bar {{\Ub}}+\ub , P=\bar{P}+p, Q=\bar{Q}+q, \phi=\bar{\phi}+\varphi, \Eb=\bar{\Eb}+\eb$. The base flow is steady in the current analysis (for example, $\bar {{\Ub}}=\bar {{\Ub}}(x,y)$ not a function of time). Inserting this decomposition into equation (\ref{eq.dimenless}) and subtracting the base flow, we arrive at the linearised equations for the perturbed variables $ \mathbf{f}=(\ub,p,q,\varphi,\eb)^T $
\begin{subequations}
	\begin{align}		
		\label{eq.line1}
		\nabla  \bcdot \ub &= 0,\\
		\label{eq.line2}
		\frac{{\partial \ub}}{{\partial t}} + (\ub \bcdot \nabla )\bar \Ub + (\bar \Ub \bcdot \nabla )\ub & =  - \nabla p +\frac{{{M^2}}}{T}{\nabla ^2}\ub + C{M^2}(q\bar \Eb + \bar Q\eb)+\mathbf{F}_{su},\\
		\label{eq.line3}
		\frac{{\partial q}}{{\partial t}} + \nabla  \bcdot [(\bar\Eb + \bar\Ub)q+(\eb + \ub)\bar Q] &= \frac{1}{{Fe}}{\nabla ^2}q+F_{sq},\\
		\label{eq.line4}
		{\nabla ^2}\varphi & =  - Cq,\\
		\label{eq.line5}
		\eb& =  - \nabla \varphi.
	\end{align}
	\label{eq.linear}
\end{subequations}
\begin{table}
	\centering
	\begin{tabular}{l c c c}
		Boundary conditions for &  velocity ($ \ub $) & charge density (q) & electric potential ($ \varphi $)\\
		Plate electrode &   $ \ub=0 $   & $ \partial q/\partial y=0 $ & $ \varphi=0 $\\
		Blade electrode (excluding injector)  & $ \ub=0 $     & $ q=0$  & $ \varphi=0 $\\
		Injector ($ S_1\leq S\leq S_2 $)  & $ \ub=0 $     & $ q=0$  & $ \varphi=0 $\\
		Arc $ \overset{\frown}{AC}$ and arc $\overset{\frown}{BD}$ & $ p\bcdot\mathbf{n}-\frac{M^2}{T}\nabla\ub\bcdot\mathbf{n}=0 $ & $ \partial q/\partial n=0 $ & $ \partial \varphi/\partial n=0 $\\
	\end{tabular}
	\caption{Boundary conditions for numerical simulation of linear blade-plate EHD flow.}
	\label{table:linbc}
\end{table}
The boundary conditions for the perturbations are listed in table \ref{table:linbc}. The above linearised equations can be written in a compact form
\begin{equation}
	\frac{\partial \mathbf{f}}{\partial t}=\Lh \mathbf{f},
	\label{eq.ml}
\end{equation}
where $ \Lh$ represents the linearised operator in the blade-plate EHD flow. Since we consider the steady state as the base flow, a wave-like solution for the perturbation can be assumed, which reads
\begin{equation}
	\mathbf{f}(x,y,t)=\tilde{\mathbf{f}}(x,y)e^{ \omega t}.
	\label{eq.m2}
\end{equation}
This expression, inserted into the linear equation (\ref{eq.ml}), leads to an eigenvalue problem
\begin{equation}\label{eigenproblem}
	\omega\tilde{\mathbf{f}}=\Lh\tilde{\mathbf{f}},
\end{equation}
where $\omega$ is the complex eigenvalue, with its real part denoting the temporal growth rate of perturbations (i.e., positive $\omega$ means growth of the disturbance and negative $\omega$ decay of the disturbance), and its imaginary part representing the phase speed. Correspondingly, the eigenvector is $\tilde{\mathbf{f}}$.

Waves are generated because of the impingement of the charged flow on the plate electrode and they propagate towards the outer boundary in the computational domain.
In order to minimise the reflections of outgoing disturbances from the boundary, a sponge region is applied, shown as hatching lines in figure \ref{fig.fig2a}. There are different ways to implement the sponge region and we follow the method of \cite{chevalier2007simson}. An additional volume force is added to the governing equations
\begin{equation}
	\mathbf{F}_{su}=-\lambda (r)\mathbf{u}, \ \ \ \  \text{and} \ \ \ \  	F_q=-\lambda (r)q.
\end{equation}
The parameter $ \lambda $ is defined by
\begin{equation}
	\lambda (r)= \lambda_{max}\bcdot S\left( \frac{r-r_{start}}{\Delta_{rise}}\right) 
\end{equation}
where $ \lambda_{max} $ is the maximum strength of the damping, $ r_{start} $ is the radial position where the sponge region starts and $ \Delta_{rise} $ corresponds to the rise distance of the damping function. The smooth step function $S$, using $x$ as the argument, reads
\begin{equation}
	S(x)=\left\{
	\begin{aligned}
		&0, & x\leq 0,\\
		&1/(1+e^{1/(x-1)+1/x}), & 0<x<1, \\
		&1, & x\geq 1.
	\end{aligned}
	\right.
\end{equation}

\section{Numerical methods}\label{numericalmethod}
In this paper, the computational flow solver Nek5000 \citep{Paul2008} is used to perform the numerical simulations, which is based on Legendre polynomial-based spectral-element method (SEM) \citep{patera1984spectral}. This method has the advantages of both geometrical flexibility of finite-element methods and accuracy of spectral methods. The $ P_N-P_{N-2} $ formulation is used for the spatial discretisation, that is, within each element, the velocity is expressed as a linear combination of Lagrangian basis functions of order $ N $ on the Gauss-Lobatto-Legendre (GLL) nodes, whereas the pressure is discretised by Lagrangian interpolants of order $ N-2 $ on the Gauss-Legendre (GL) quadrature points. In the current work, we take $ N=7 $ for most cases. The time discretisation scheme adopted in Nek5000 is the semi-implicit scheme $ BDF_k/EXT_k $, in which the viscous terms are implicitly discretised based on a backward differential formula of order $ k $, and the nonlinear convection term is explicitly advanced by an extrapolation scheme of order $ k $. In this work, $ k=2 $ is applied. 

For the linear stability analysis, the Implicitly Restarted Arnoldi Method (IRAM) \citep{lehoucq1996deflation} is adopted to compute the eigenpairs of the linear system. IRAM is an iterative eigenvalue solver based on the projection of the problem on an orthogonal basis, in which process a Krylov subspace $ \mathbf K_n $ of dimension $ n $ is created. The Krylov subspace of the exponential propagator $ \mathcal{M} $ and the initial vector $ \mathbf{f}_0 $ is defined as $\mathbf{K}_n(\mathcal{M},\mathbf{f}_0)=\{\mathbf{f}_0,\mathcal{M}\mathbf{f}_0,...,\mathcal{M}^{n-1}\mathbf{f}_0\}$.
This Krylov subspace converges at the eigenvector corresponding to the eigenvalue with the largest moduli. This simple iteration is known as the power method, which is simple to perform, but converges slowly and can only obtain the leading eigenpair of the problem. In order to obtain more eigen-information from the iteration, a Gram-Schmidt orthogonalisation process is applied, and the residual information is collected. Eventually, a small-dimensional Hessenberg matrix $ \mathcal{H} $ is formed to approximate the eigen-information of the exponential propagator $ \mathcal{M} $ and its eigenpairs can be calculated easily. We use the LAPACK package \citep{anderson1999lapack} for IRAM.

\section{Result and Discussions} \label{results}

In this section, we present the results of DNS and global stability analysis of EHD flows in the blade-plate configuration. At this point, it is instructive to specify the parameter range considered in this work. The typical value of $ Fe $ (inverse of charge diffusion) for the real dielectric fluids lies within the range of $ 10^3-10^4 $ \citep{P1989Role}, thus we choose an intermediate value $ Fe = 5\times 10^3 $ in this work, except in the section where we study its effect. Previous works often neglected this charge diffusion term \citep{perez2009numerical,wu2013direct,perri2020electrically}, but it has been shown that the charge diffusion has a non-negligible effect on the linear stability and bifurcation in EHD flows \citep{Zhang2015,Zhang2016Weakly,feng2021deterministic}.
Since a strong injection regime has been considered in this work, we take $ C = 5$. The typical value of $ M $ is higher than $ 3 $ for most dielectric liquids \citep{P1989Role} and $ M=50 $ is used in the following. The remaining parameters will be specified in each case to be presented. We also mention that we have used many symbols to denote various flow parameters and define the geometry. For a clearer understanding of the results below, it is useful to consult the table in Appendix C for the nomenclature.

\subsection{Moffatt eddies} \label{moffatt}
Inspired by the experimental work of the EHD flow in the needle-plate configuration, where the Moffatt-like vortices were observed \citep{perri2020electrically,perri2021particle}, we will explore the Moffatt-like eddies in the 2-D blade-plate EHD flow. As our flow is in a Cartesian coordinate, the results can be compared to the theoretical prediction by \cite{moffatt} for the vortices in a wedge formed by two flat plates and induced by a disturbance near the corner. For a better presentation of the results, in the following, we will first summarise the theoretical results of \cite{moffatt}.
\begin{figure}
	\centering
	\subfigure{
		\begin{minipage}[h]{0.4\textwidth}
			\centering
			\includegraphics[height=4.5cm]{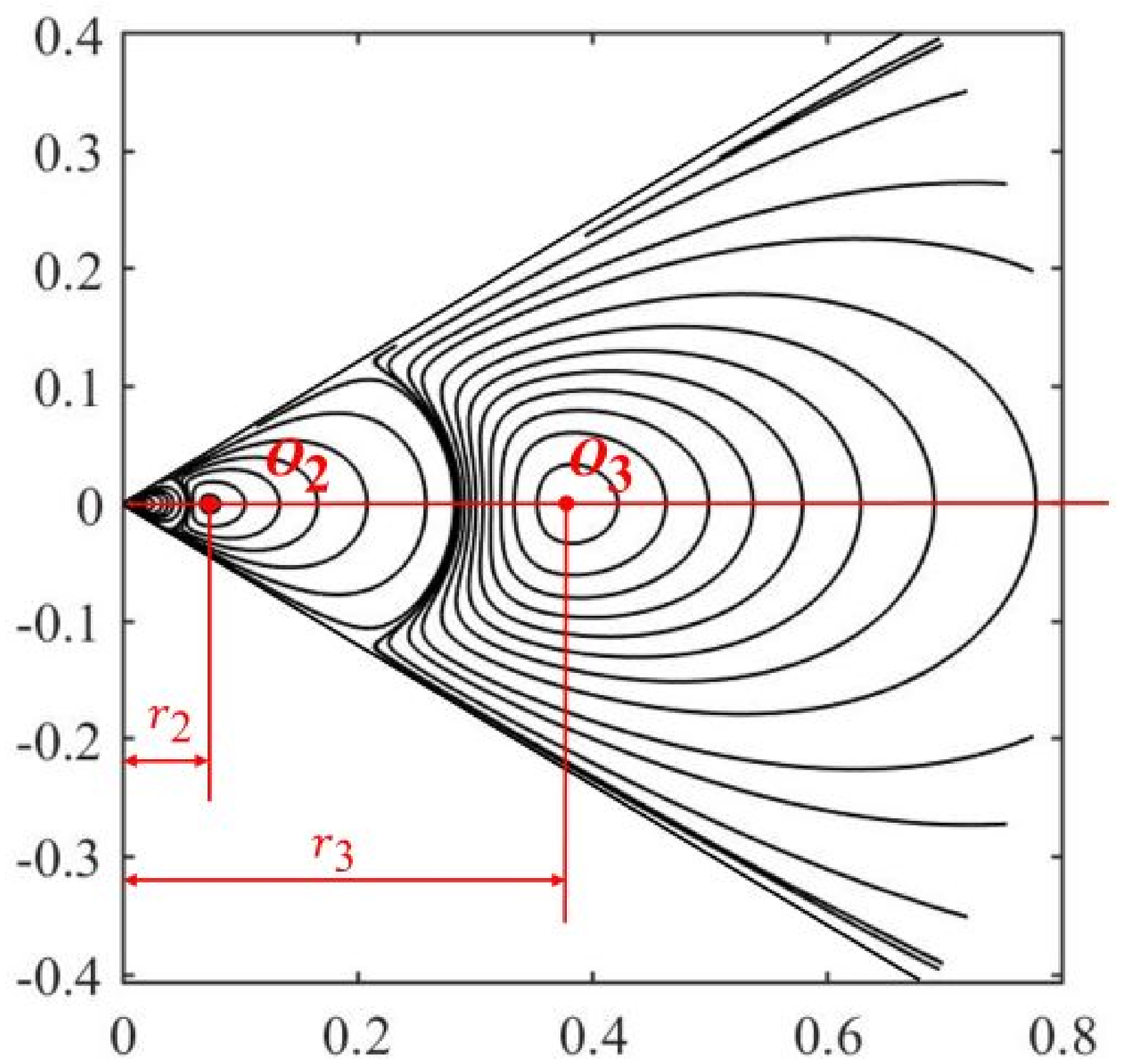}
			\put(-150,120){(a)}
			\label{fig.r03Mat1}
		\end{minipage}
	}
	\hspace{20pt}
	\subfigure{
		\begin{minipage}[h]{0.4\textwidth}
			\centering
			\includegraphics[height=4.5cm]{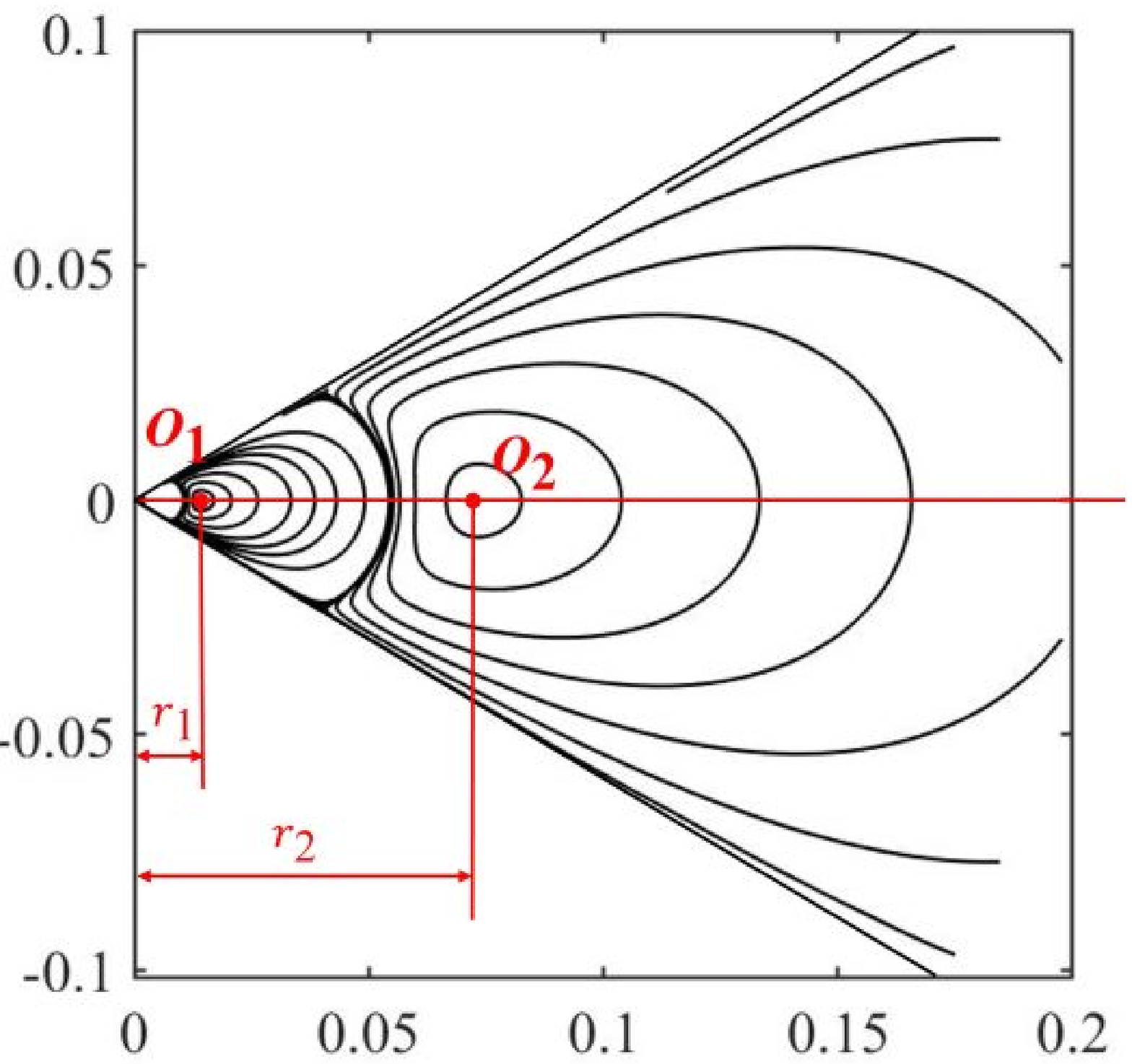}
			\put(-150,120){(b)}
			\label{fig.r03Mat2}
		\end{minipage}
	}
	\caption{(a) Theoretically predicted  streamlines in corner eddies for the corner angle $ 2\alpha=61.3^\circ $, based on the theory of \cite{moffatt}. The flow is induced by a disturbance near the corner. (b) A close-up view of the corner.}
	\label{fig.r03Mat}
\end{figure}

According to the theory developed in \cite{moffatt}, the flow field in the case of corner angle $ 2\alpha=61.3^\circ $ is plotted in figure \ref{fig.r03Mat}. A sequence of geometrically and dynamically similar vortices are formed, and their centers all fall on the corner bisector. We mark their centers as $ O_1, O_2, O_3 $ (counted from the corner) and denote the distance from the center to the corner as $ r_1, r_2, r_3 $, respectively. Theoretically, the radial coordinates of the vortex centers of Moffatt are in an equal ratio sequence and satisfy the following relationship \citep{moffatt}
\begin{equation}
	\frac{r_{n+2}}{r_{n+1}}=\frac{r_{n+1}}{r_n}=\frac{r_{n+2}-r_{n+1}}{r_{n+1}-r_n}=\rho, \label{rratio}
\end{equation}
where $ \rho $ is the ratio only dependent on the angle $ 2\alpha $; $ r_n $ is the distance between the corner and the center of the $n$-th eddy, counting from the corner.  For the case of $ 2\alpha=61.3^\circ $, it can be calculated from equation (3.6) in \cite{moffatt} that $ \rho=5.22 $. Therefore, we have in this case $\frac{r_{3}}{r_2}=\frac{r_{2}}{r_1}=\frac{r_{3}-r_{2}}{r_{2}-r_1}=5.22$.

Besides, the intensity of two successive eddies follows a constant ratio as well, which according to \cite{moffatt} reads
\begin{equation}
	\frac{|v_{\theta}|_{n+1/2}}{|v_{\theta}|_{n+3/2}}=\Omega,
\end{equation}
where $ \Omega $ is also only dependent on the angle $ 2\alpha $ and $ v_{\theta} $ is the azimuthal velocity. We use $ |v_{\theta}|_{n+1/2} $ to denote the absolute value  of the local maximum azimuthal velocity of $n$-th vortex (which can represent the intensity of the vortex). Similarly, the value of $\Omega$ can be obtained theoretically \citep{moffatt} and it is equal to $710.56 $ in the case of $ 2\alpha=61.3^\circ $, that is $\frac{|v_{\theta}|_{1+1/2}}{|v_{\theta}|_{2+1/2}}=\frac{|v_{\theta}|_{2+1/2}}{|v_{\theta}|_{3+1/2}}=710.56$.
These equations summarise the flow quantities that we will probe and compare to in our numerical simulations of EHD flows.

\subsubsection{Formation of the eddies in the blade-plate EHD flow}
In the following, we present and characterise the Moffatt-like eddies in the blade-plate EHD flow. Appendix A furnishes a detailed verification step of the domain size and grid resolution in our computations. We focus on characterising the generation and evolution of the Moffatt eddies in this subsection. The parameters in this subsection are $ T=500 $, $ C=5 $, $ M=50 $, $ Fe=5\times 10^3, R=0.05 $.

\begin{figure}
	\centering
	\subfigure{
		\begin{minipage}[h]{0.4\textwidth}
			\centering
			\includegraphics[height=4cm]{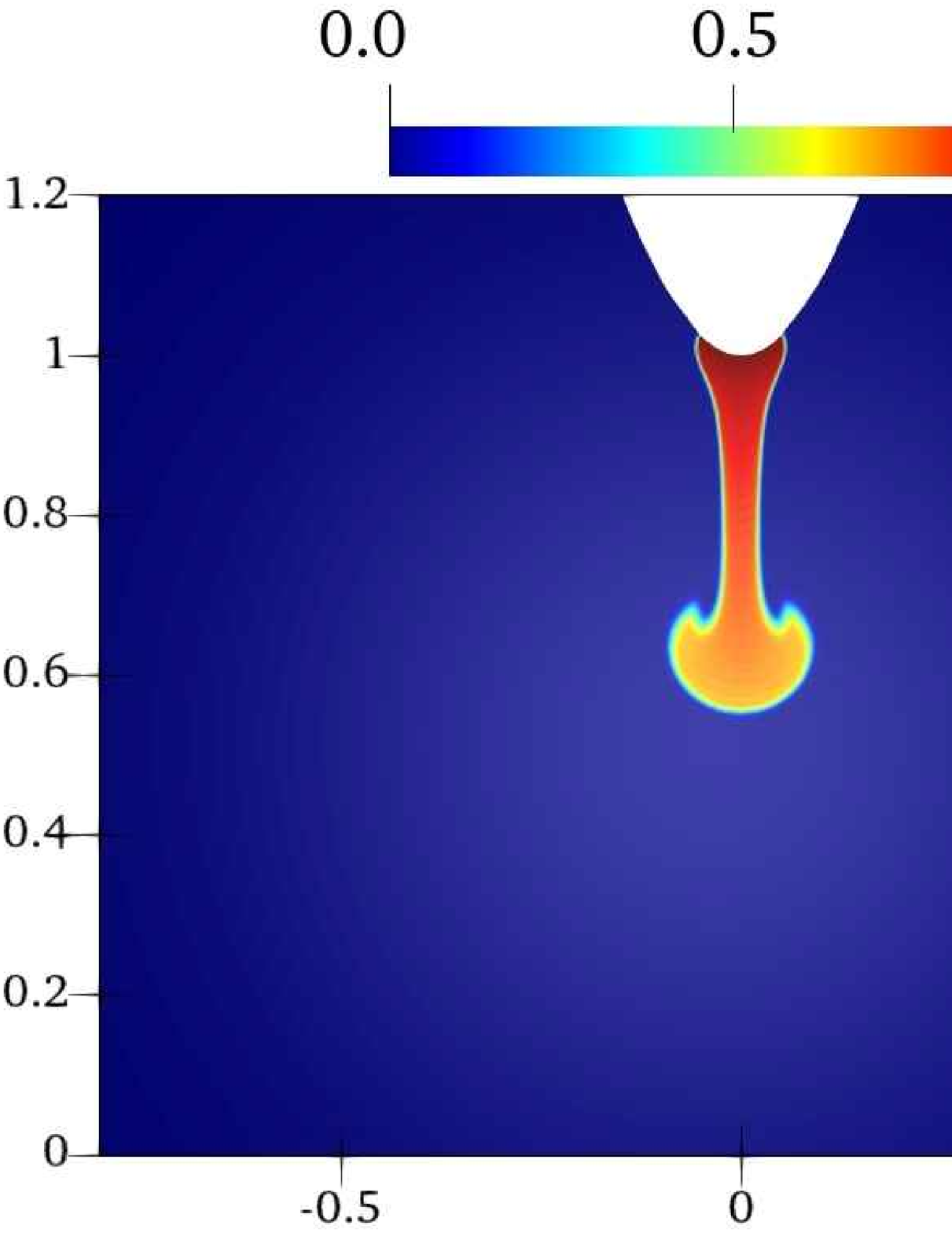}
			\put(-150,90){(a)}
			\put(-140,49){$ y $}
			\put(-64,-5){$ x $}
			\label{fig.r03charge01}
	\end{minipage}}
	\hspace{20pt}
	\subfigure{
		\begin{minipage}[h]{0.4\textwidth}
			\centering
			\includegraphics[height=4cm]{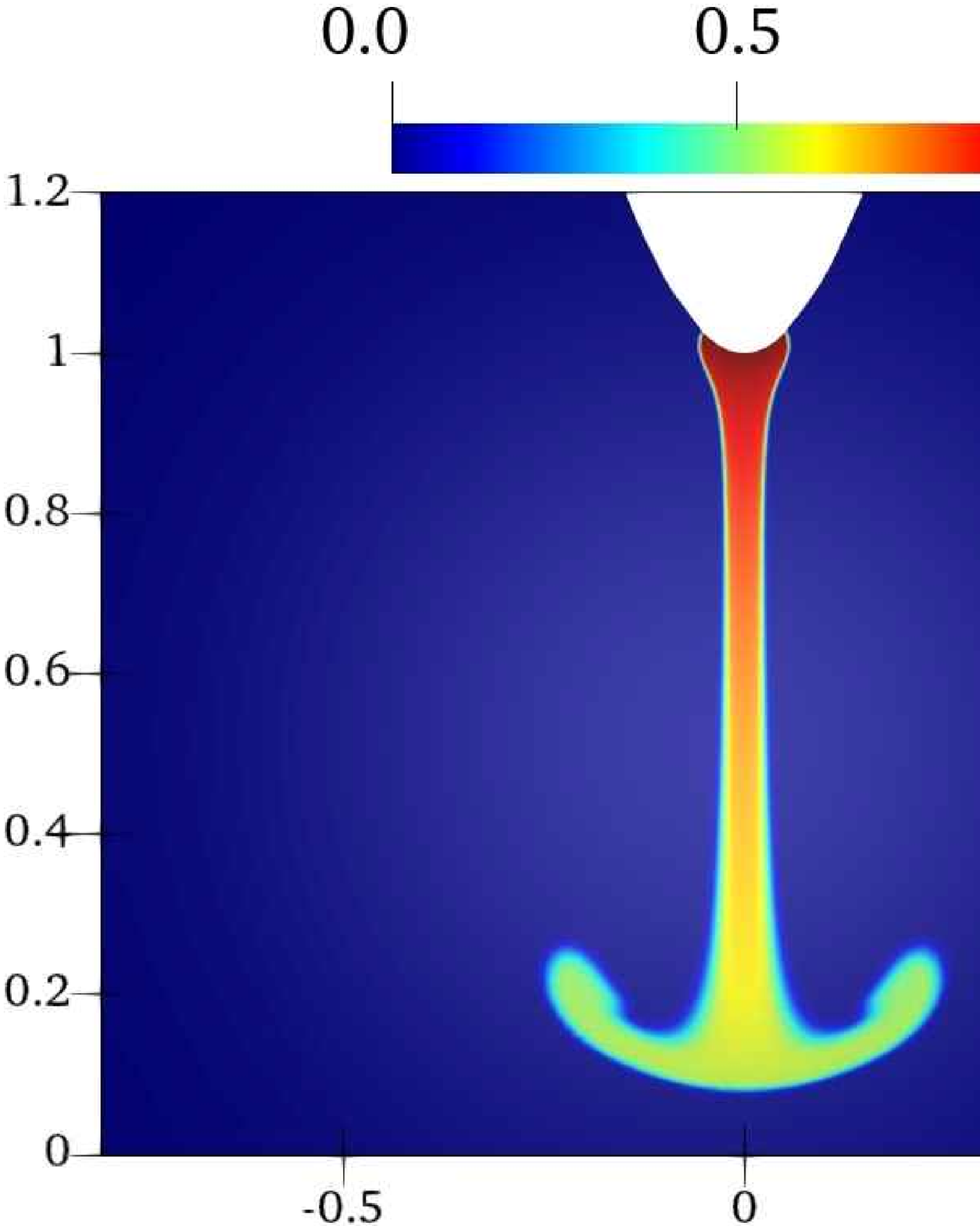}
			\put(-150,90){(b)}
			\put(-140,49){$ y $}
            \put(-63,-5){$ x $}			
			\label{fig.r03charge02}
	\end{minipage}}
	\hspace{20pt}
	\subfigure{
		\begin{minipage}[h]{0.4\textwidth}
			\centering
			\includegraphics[height=4cm]{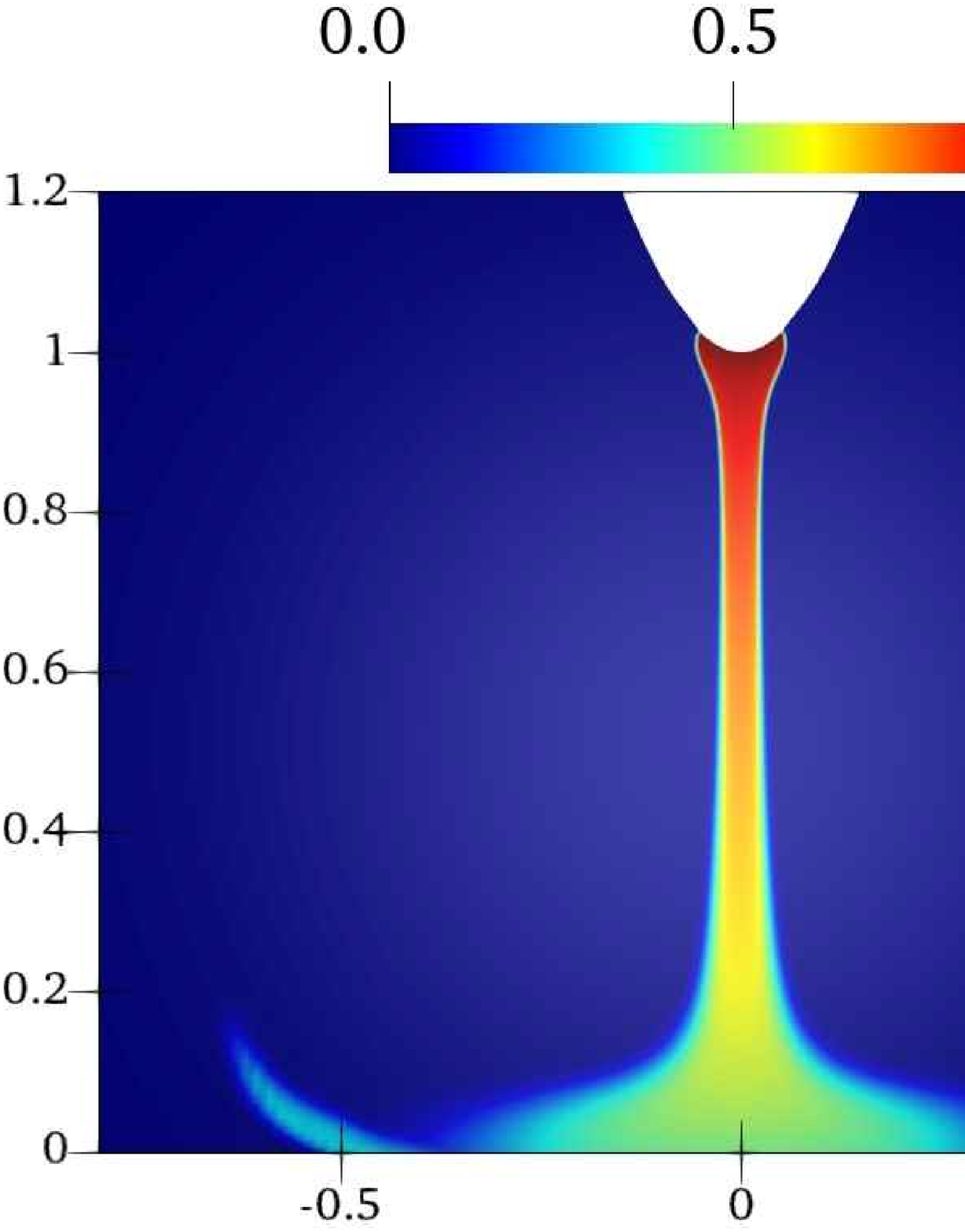}
			\put(-150,90){(c)}
		    \put(-140,49){$ y $}
			\put(-63,-5){$ x $}
			\label{fig.r03charge04}
	\end{minipage}}
	\hspace{20pt}
	\subfigure{
		\begin{minipage}[h]{0.4\textwidth}
			\centering
			\includegraphics[height=4cm]{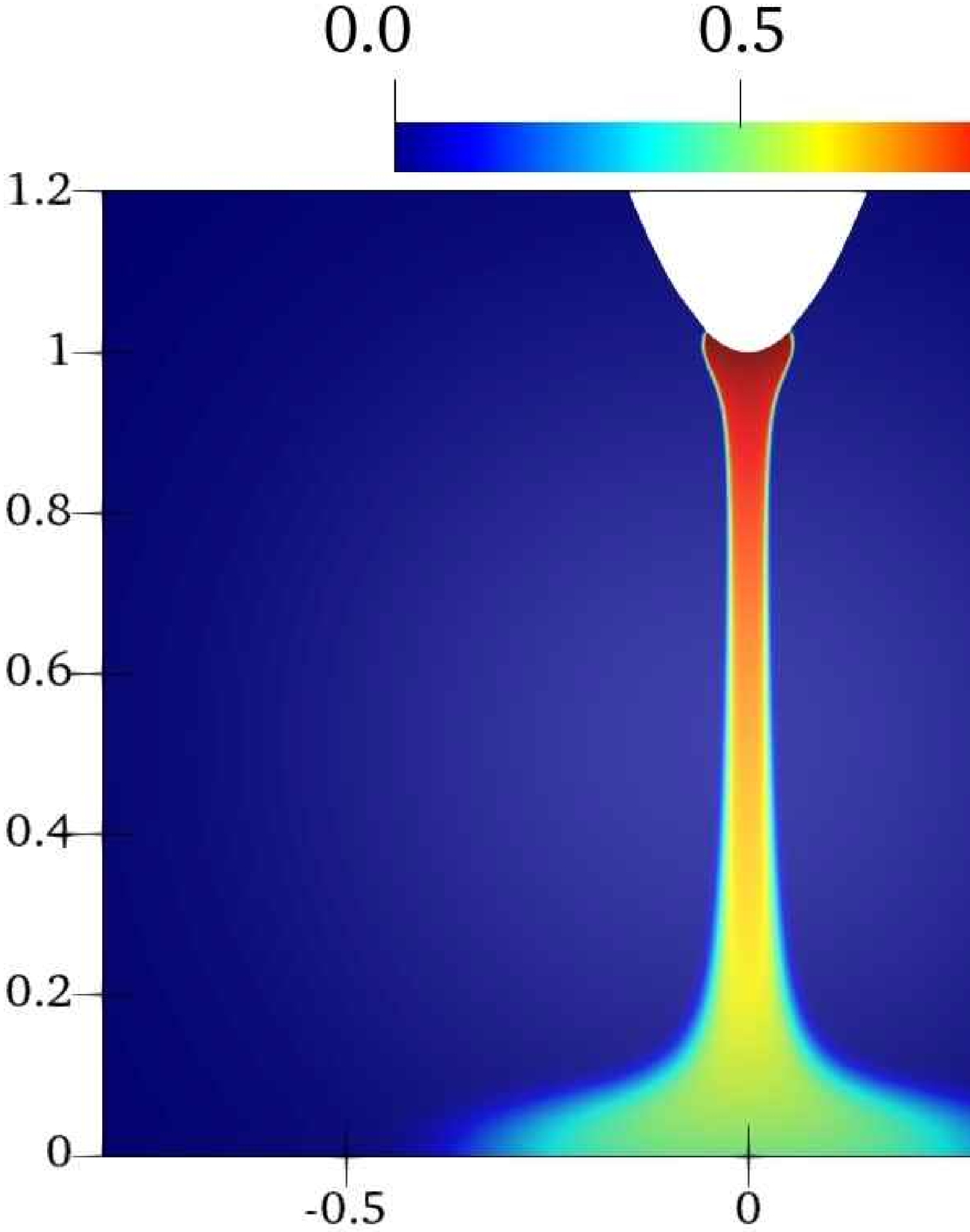}
			\put(-150,90){(d)}
			\put(-140,49){$ y $}
			\put(-63,-5){$ x $}
			\label{fig.r03charge1}
	\end{minipage}}
	\subfigure{
	\begin{minipage}[h]{0.4\textwidth}
		\centering
		\includegraphics[height=4cm]{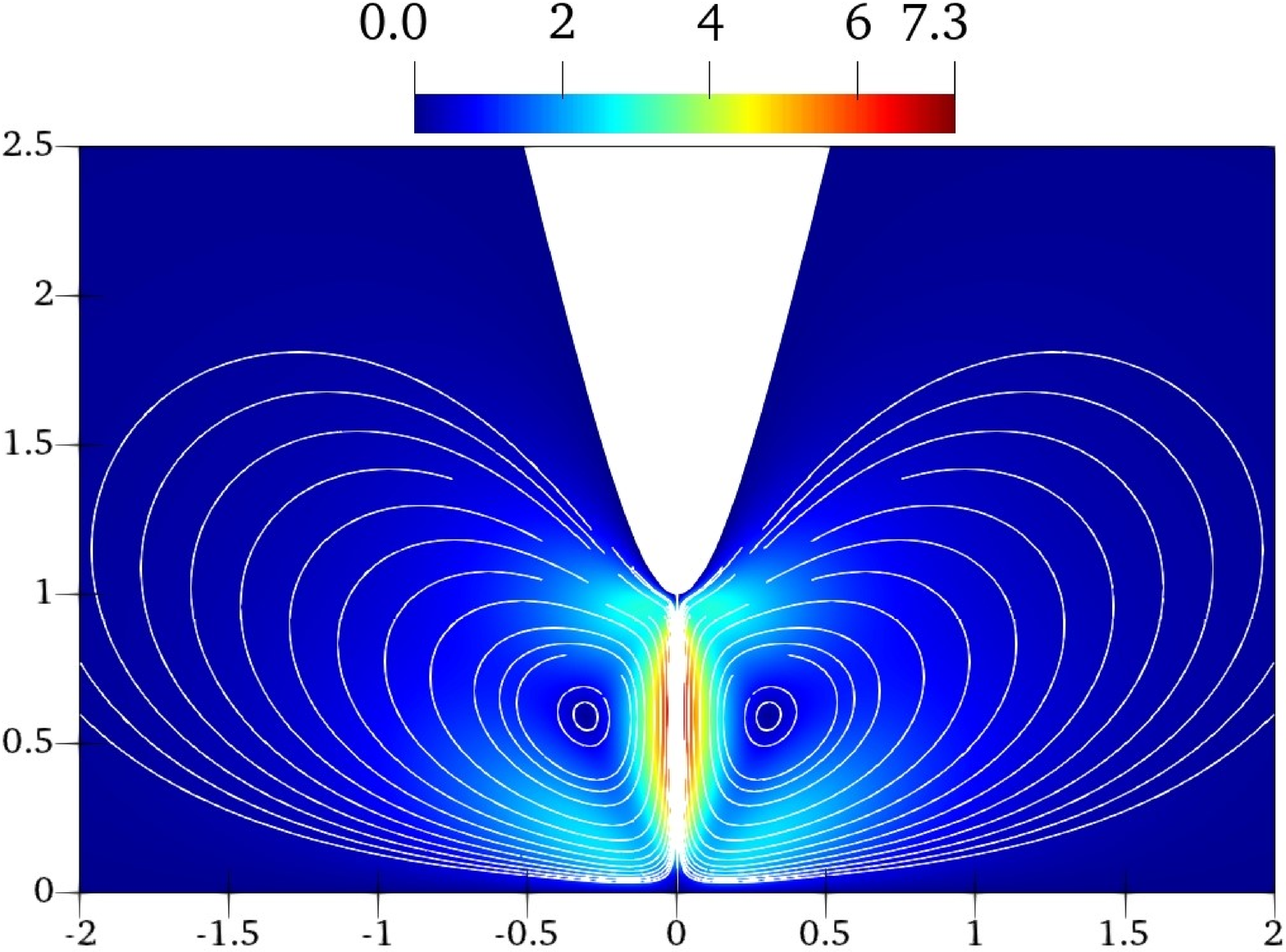}
		\put(-170,90){(e)}
		\put(-160,49){$ y $}
		\put(-74,-5){$ x $}
		\label{fig.v1small}
\end{minipage}}
	\caption{Evolution of the charge density distribution between the blade injector and the plate electrode at (a) $ t=0.1 $; (b) $ t=0.2 $; (c) $ t=0.4 $ (d) $ t=1.0 $. (e) The velocity field at final steady state and the streamlines. The white region in the middle is due to the clustering of the streamlines. The parameters are $ T=500 $, $ C=5 $, $ M=50 $, $ Fe=5\times 10^3, R=0.05 $.}
	\label{fig.charge}
\end{figure}

Figures \ref{fig.r03charge01}-\ref{fig.r03charge1} show that charges are injected from the blade tip and move towards the plate electrode driven by the potential difference, then impinge on the flat plate, leading to the formation of two steady symmetrical vortices in the central region, see figure \ref{fig.v1small}. This result is similar to previous numerical simulations in \cite{wu2013direct,pan2021energy}. It is noted that the flow pattern of the EHD flow here resembles the thermal plume \citep{vazquez1996thermal,lesshafft2015linear} and the impinging jet flow \citep{park2004numerical,meliga2011global}.

\begin{figure}
	\centering
	\subfigure{
		\begin{minipage}[h]{0.4\linewidth}
			\centering
			\includegraphics[height=3.2cm]{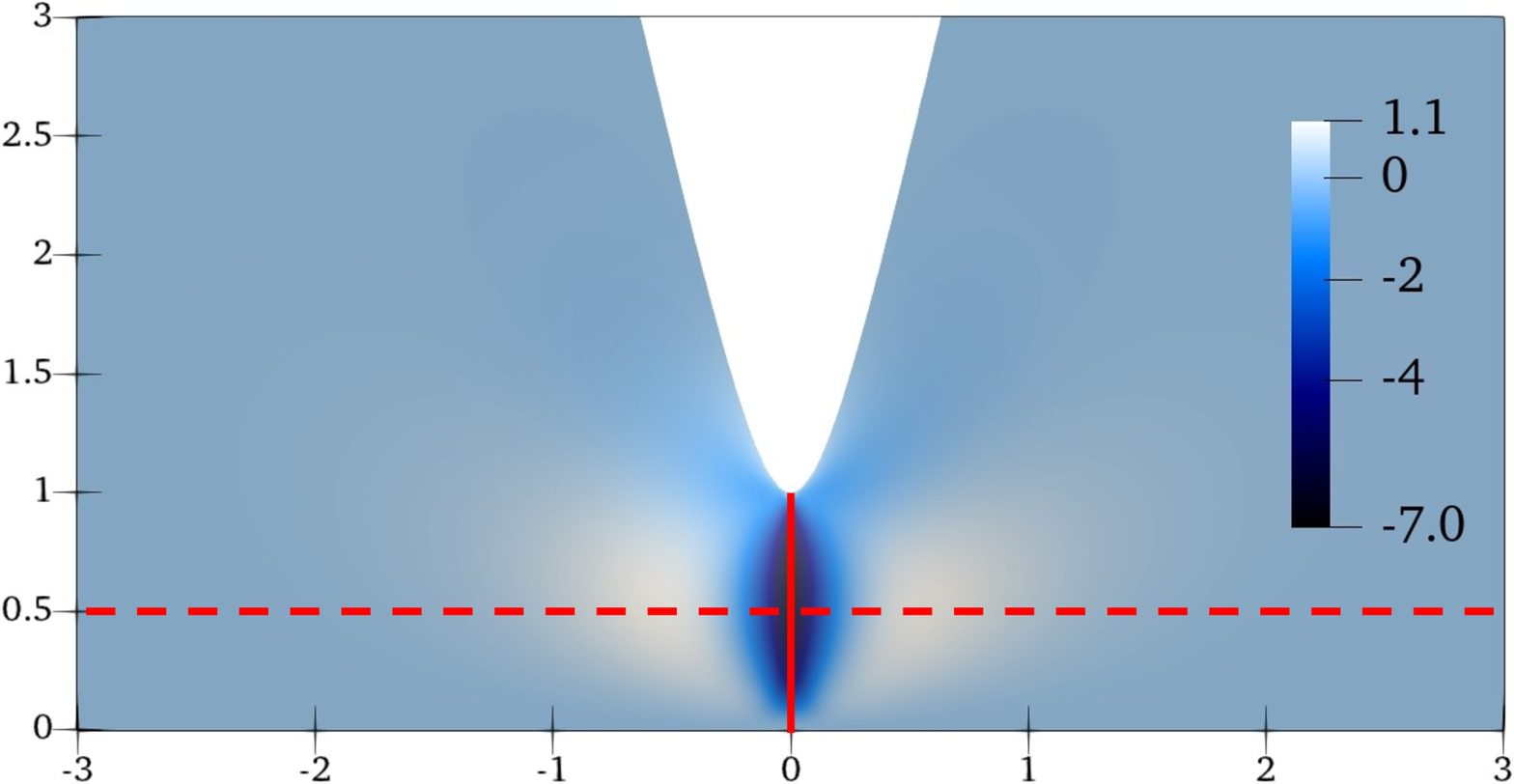}
			\put(-190,80){(a)}
			\put(-180,49){$ y $}
			\put(-85,-5){$ x $}
			\label{fig.uycon}
	\end{minipage}}
	\hspace{100pt}
	\subfigure{
		\begin{minipage}[h]{0.4\linewidth}
			\centering
			\includegraphics[height=4.5cm]{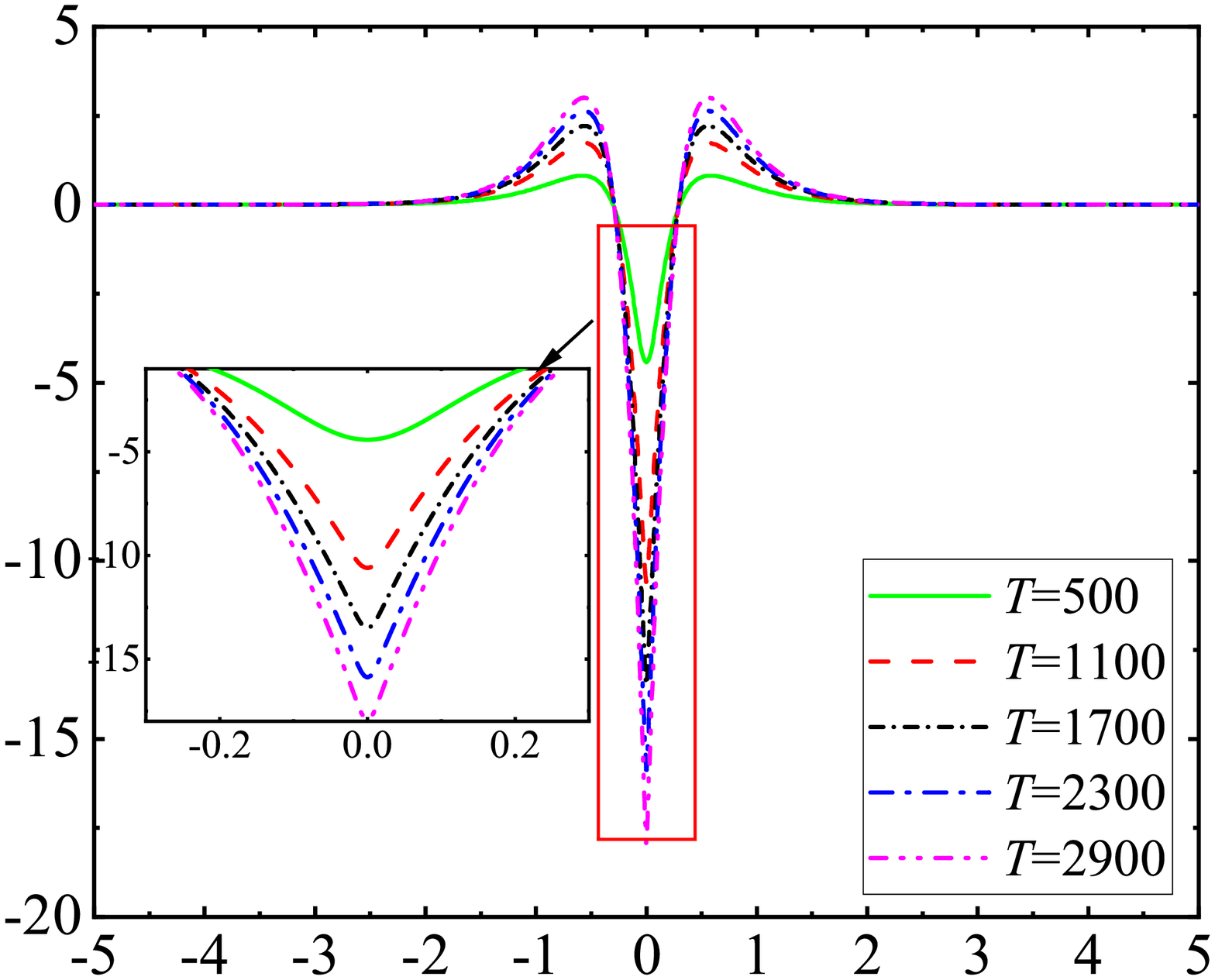}
			\put(-170,120){(b)}
			\put(-170,60){$ V $}
			\put(-75,-7){$ x $}
			\label{fig.uyy05}
	\end{minipage}}
	\hspace{20pt}
	\subfigure{
		\begin{minipage}[h]{0.4\linewidth}
			\centering
			\includegraphics[height=4.5cm]{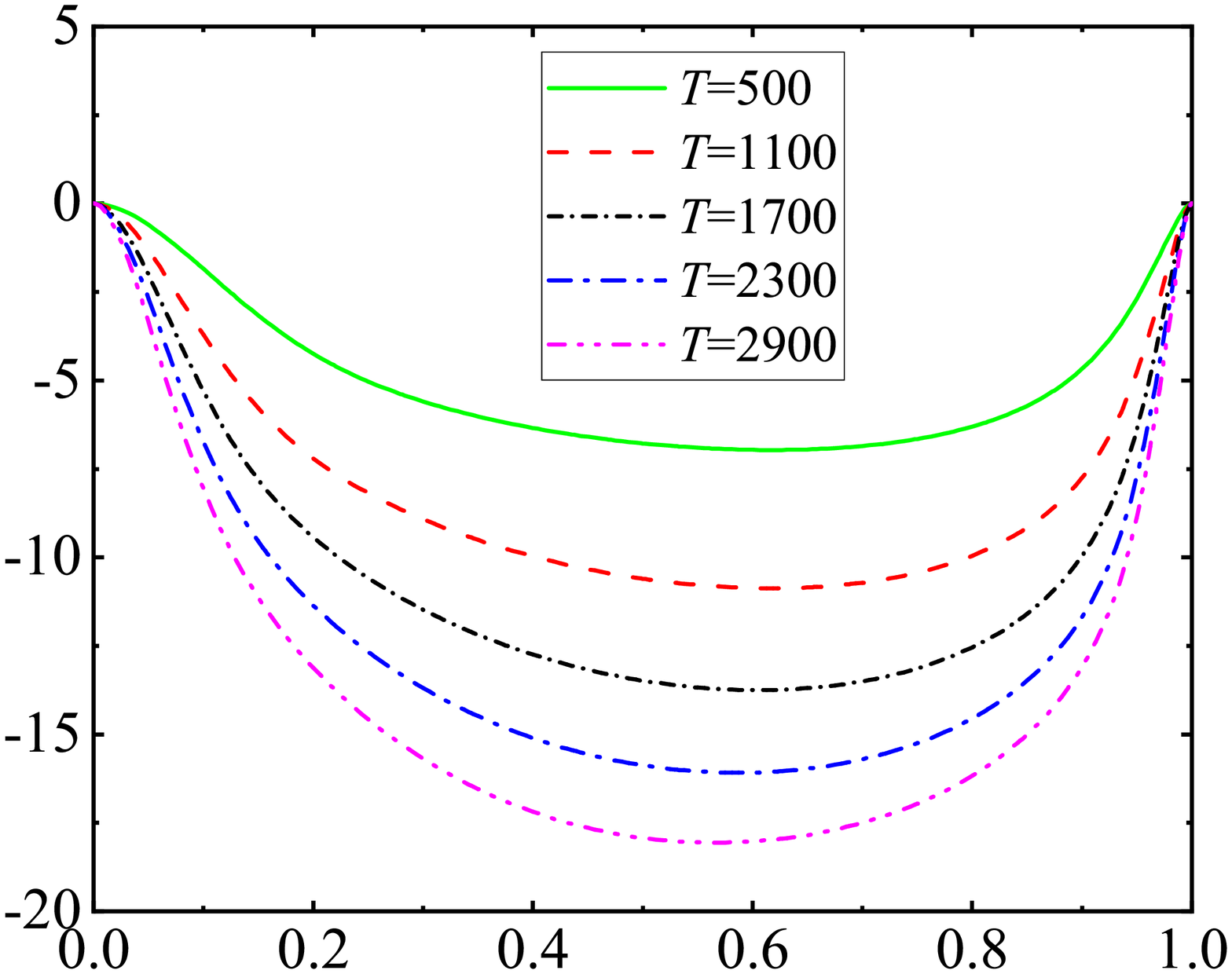}
			\put(-170,120){(c)}
			\put(-170,60){$ V $}
			\put(-75,-7){$ y $}
			\label{fig.uyx0}
	\end{minipage}}
	\caption{ (a) Contour of vertical velocity at final steady state ($ V $) at $ T=500 $. Vertical velocity profiles at different $ T $ versus (b) $ x $ at $ y=0.5 $; (c) $y$ at $ x=0 $. The other parameters are the same as figure \ref{fig.charge}.}
	\label{fig.uy}
\end{figure}

Figure \ref{fig.uycon} shows the distribution of the vertical velocity $V$ at the final steady state in the case of $ T=500 $. Figure \ref{fig.uyy05} displays the profile of $V$ for different electric Rayleigh numbers $ T $ probed at the middle horizontal line between the blade and the plate, as shown by the red dashed line in panel \ref{fig.uycon}. It indicates that the vertical velocity is symmetrical with respect to the central vertical axis. Its absolute value is largest at the central vertical axis and then rapidly decreases towards both sides. Somewhere around $x=0.5$ at the center of the vortices, the velocity amplitude reaches another local maximum and then gradually decreases to zero with increasing $x$. In figure \ref{fig.uyx0}, the vertical velocity $ V $ along the central vertical axis (the solid red line in panel \ref{fig.uycon}) is plotted. We find that the fluid accelerates rapidly near the blade injector ($ y=1 $) and decelerates due to the impingement on the plate. In addition, as expected, increasing $ T $ increases the absolute vertical velocity because Coulomb force is proportional to the potential difference, leading to a larger velocity, as shown in both panels (b) and (c). 

\begin{figure}
	\centering
	\subfigure{
		\begin{minipage}[h]{0.4\linewidth}
			\centering
			\includegraphics[height=4.5cm]{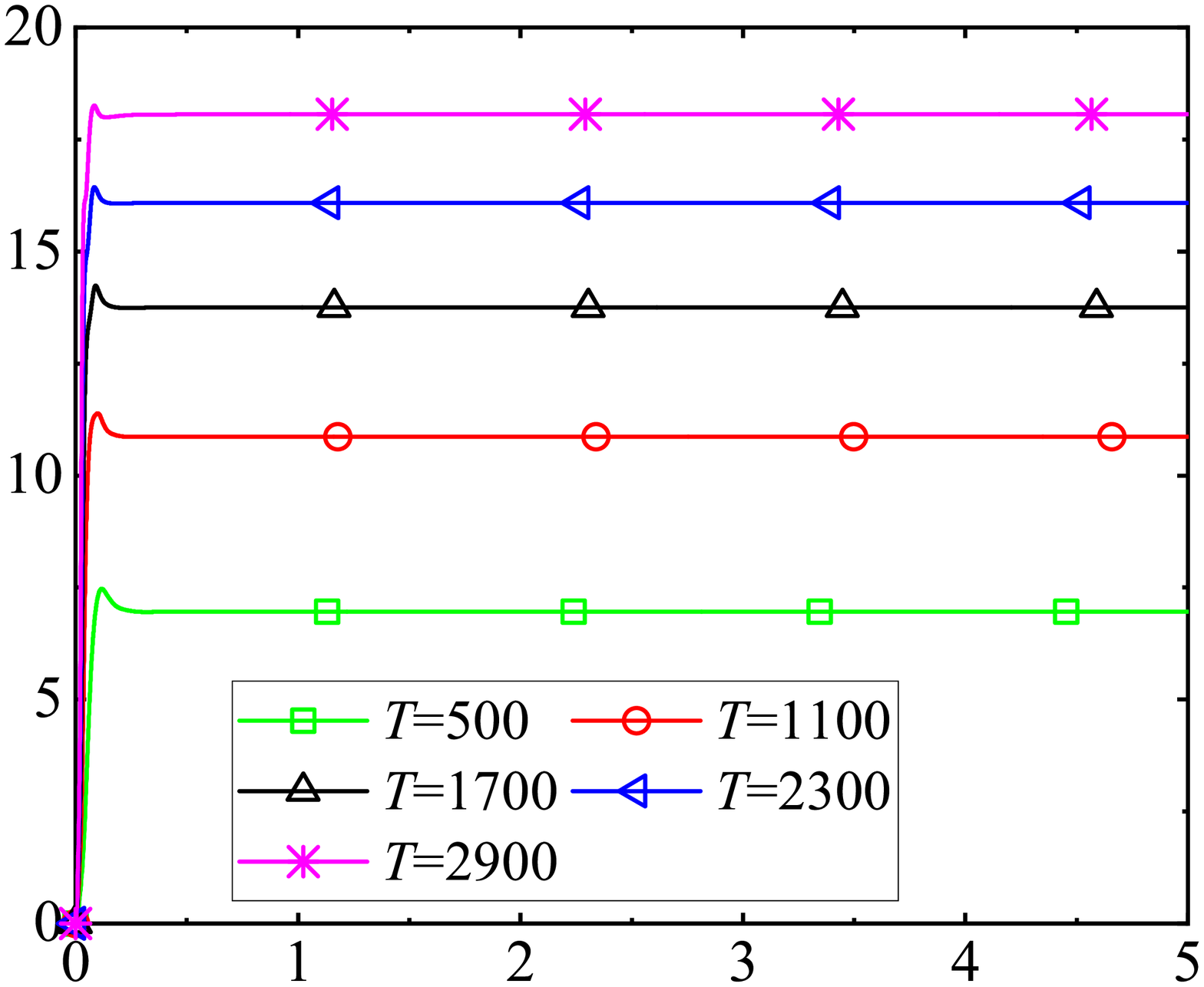}
			\put(-180,120){(a)}
			\put(-185,65){$ |\Ub|_{max} $}
			\put(-80,-10){$ t $}
			\label{fig.Umax}
	\end{minipage}}
	\hspace{20pt}
	\subfigure{
		\begin{minipage}[h]{0.4\linewidth}
			\centering
			\includegraphics[height=4.5cm]{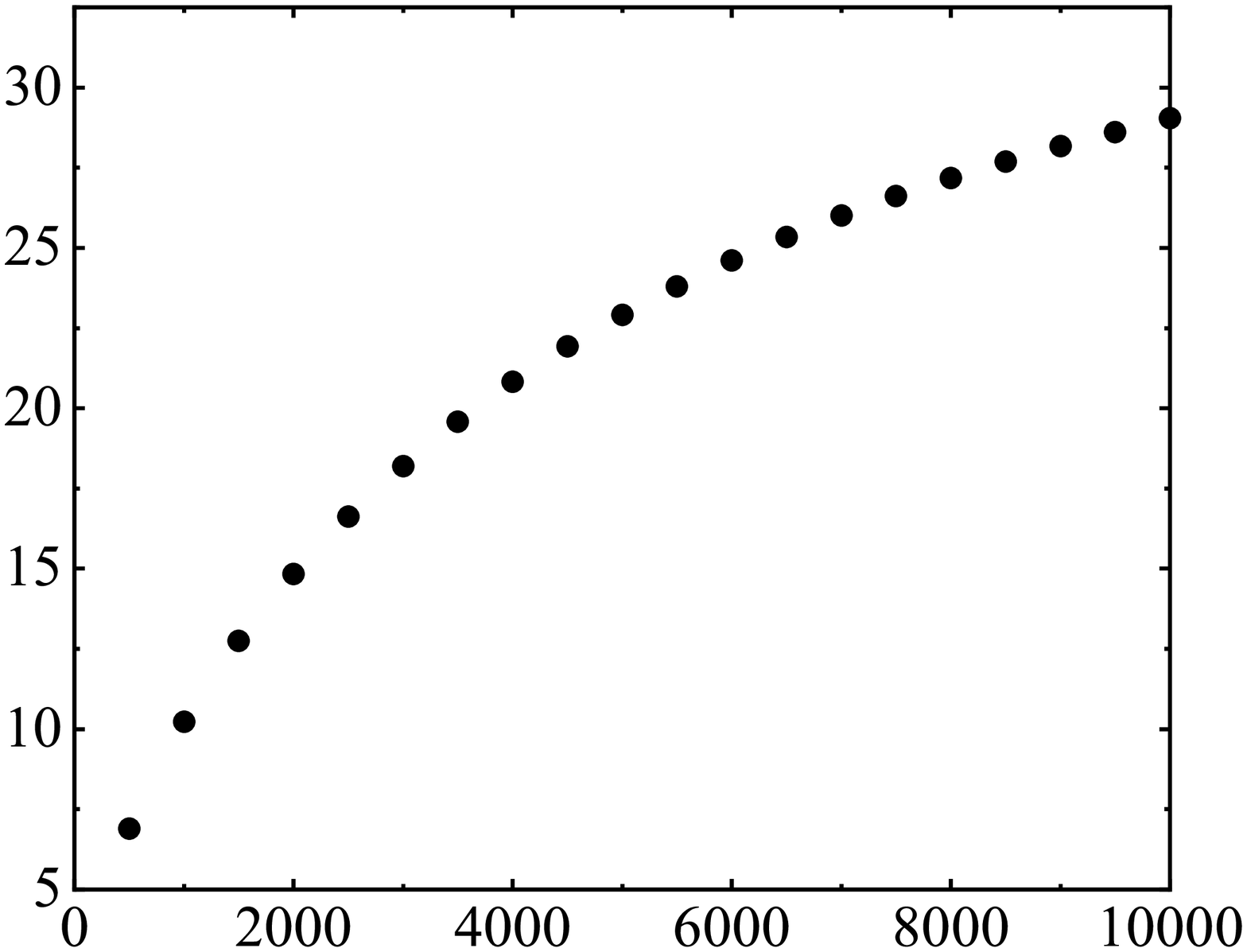}
			\put(-190,120){(b)}
			\put(-195,65){$ |\Ub|_{max}^{s} $}
			\put(-80,-10){$ T $}
			\label{fig.UmaxT}
	\end{minipage}}
	\caption{(a) Evolution of the maximum velocity norm, and (b) saturated maximum velocity magnitude at the steady state versus $ T $. The other parameters are the same as figure \ref{fig.charge}.}
	\label{fig.T}
\end{figure}

In order to obtain a global view of the velocity amplitude in the flow, figure \ref{fig.Umax} presents the evolution of the maximum velocity magnitude $ |\Ub|_{max} $ in the whole domain. As we can see, its value also increases with increasing $ T \in[500,2900]$. The typical characteristics of the EHD flow structure discussed above are consistent with those obtained from previous numerical simulations \citep{park2004numerical,wu2013direct} and experiments in the injection regime \citep{yan2013velocity,sun2020experimental,daaboul2017study}. Figure \ref{fig.UmaxT} depicts $ |\Ub|_{max}^{s} $ (where the superscript 's' represents the saturated $ |\Ub|_{max} $ of the final steady state, see panel a) as a function of $ T $ from 500 to 10000. We find that with the increase of $ T $, $ |\Ub|_{max}^s $ monotonically increases in this range of $T$. The growth of $ |\Ub|_{max}^s $ gradually slows down when $T$ is large.

\begin{figure}
	\centering
	\subfigure{
		\begin{minipage}[h]{0.4\textwidth}
			\centering
			\includegraphics[height=4cm]{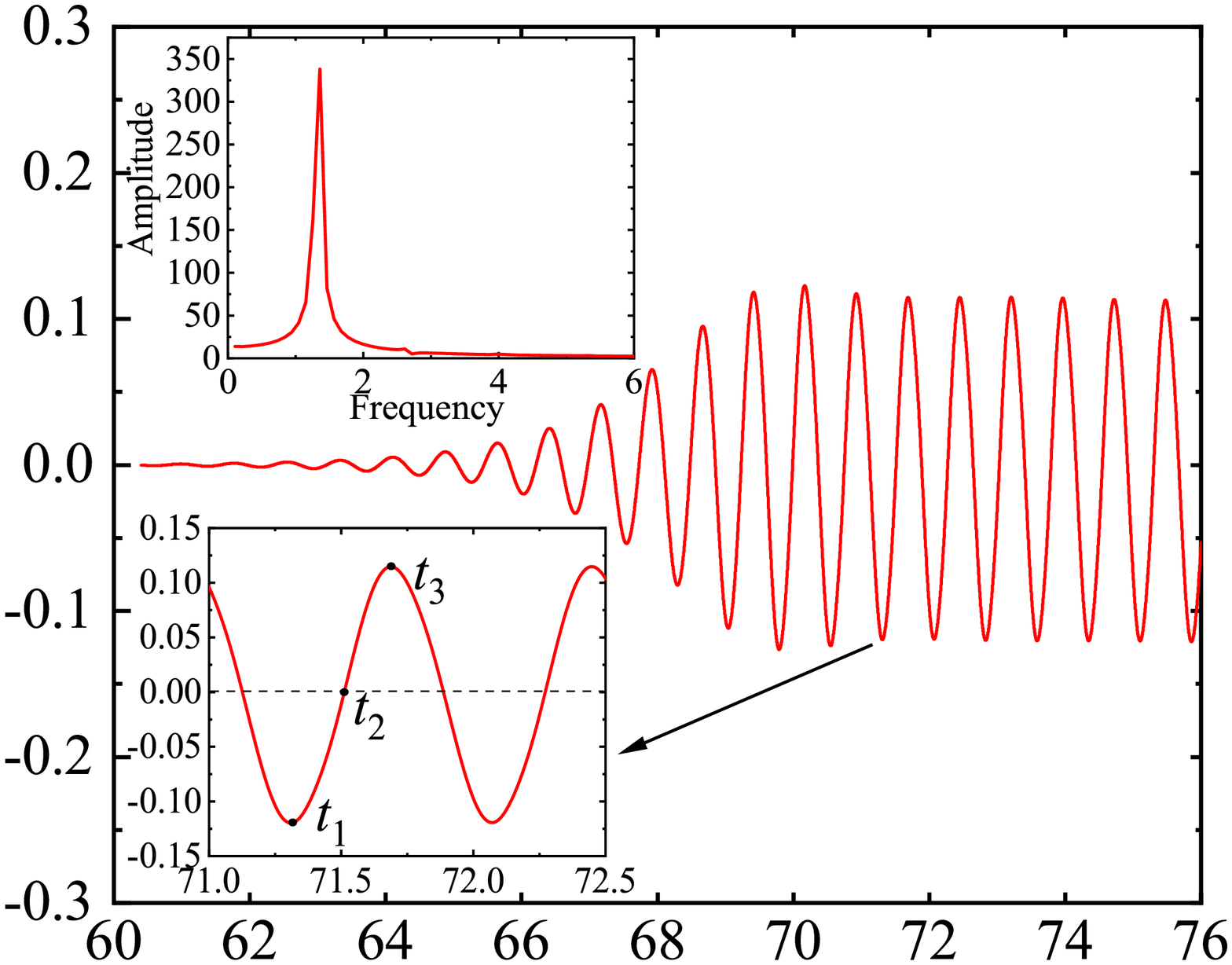}
			\put(-150,100){(a)}
			\put(-90,115){$ T=3\times10^4 $}
			\put(-150,60){$ U_x $}
			\put(-70,-5){$ t $}
			\label{fig.T3e4}
	\end{minipage}}
	\hspace{20pt}
	\subfigure{
		\begin{minipage}[h]{0.4\textwidth}
			\centering
			\includegraphics[height=4cm]{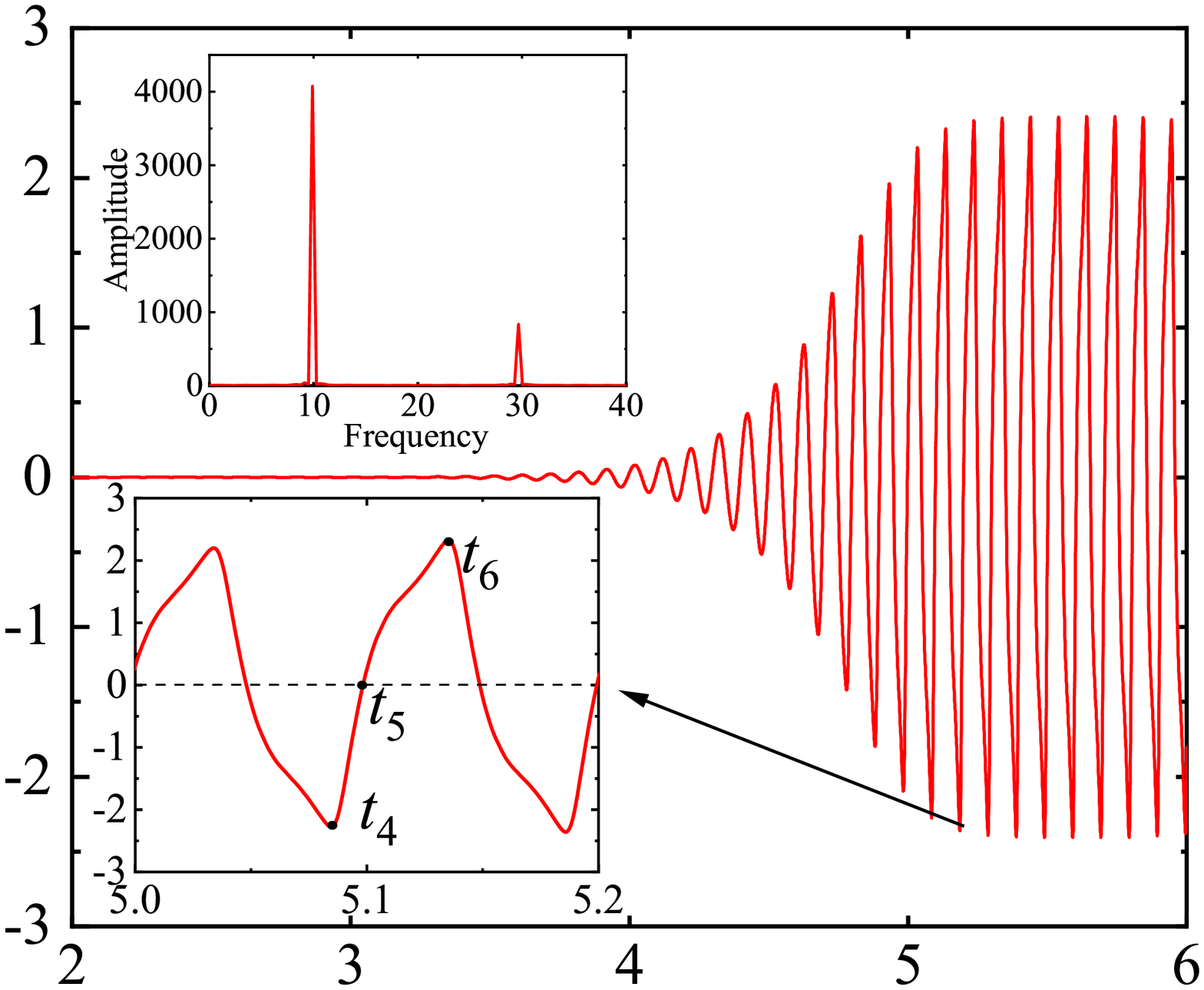}
			\put(-150,100){(e)}
			\put(-90,115){$ T=4\times10^4 $}
			\put(-150,60){$ U_x $}
			\put(-70,-5){$ t $}
			\label{fig.T4e4}
	\end{minipage}}
	\hspace{20pt}
	\subfigure{
		\begin{minipage}[h]{0.4\textwidth}
			\centering
			\includegraphics[height=4cm]{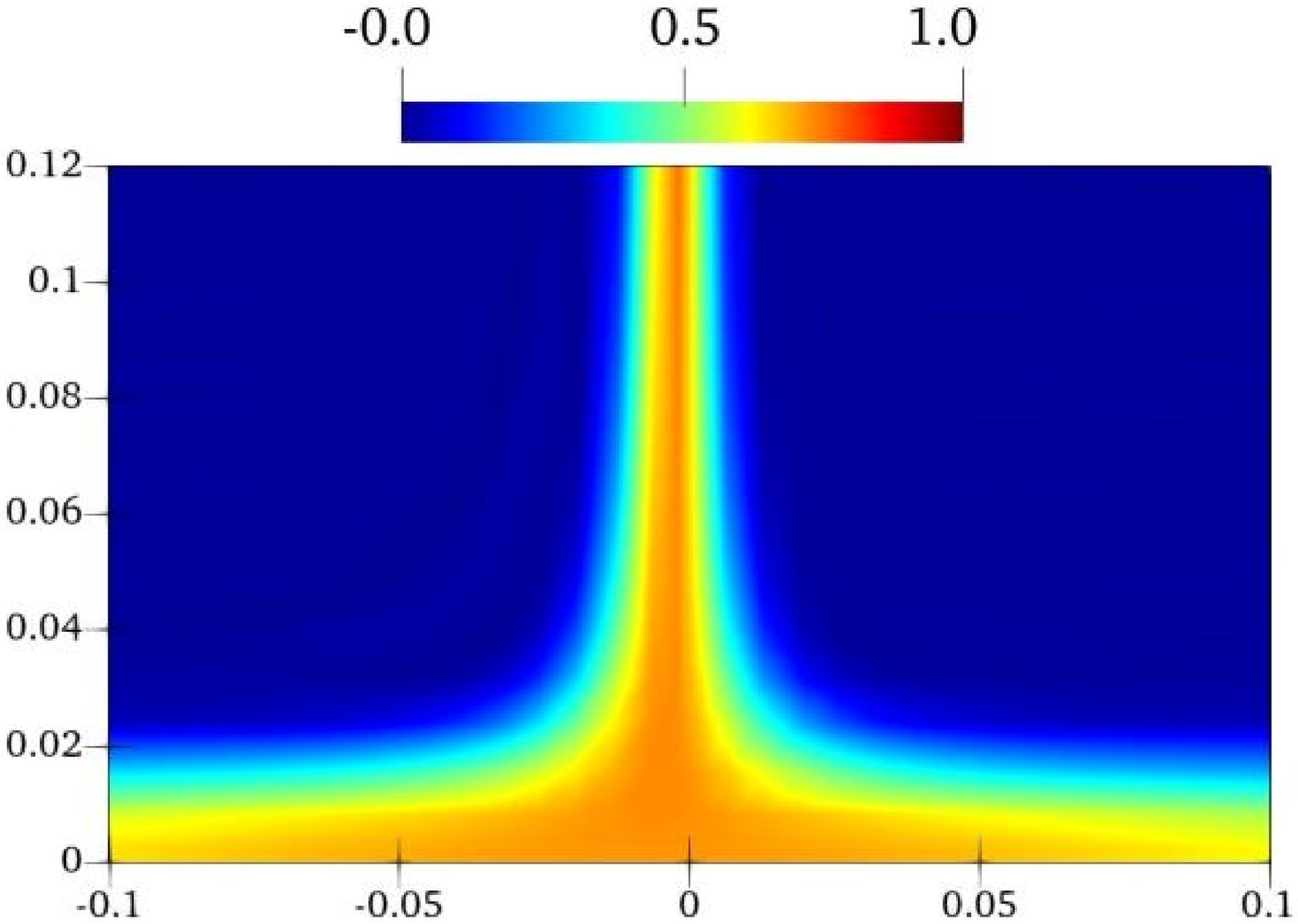}
			\put(-175,90){(b)}
			\put(-170,49){$ y $}
			\put(-82,-5){$ x $}			
			\label{fig.qt1}
	\end{minipage}}
	\hspace{20pt}
	\subfigure{
		\begin{minipage}[h]{0.4\textwidth}
			\centering
			\includegraphics[height=4cm]{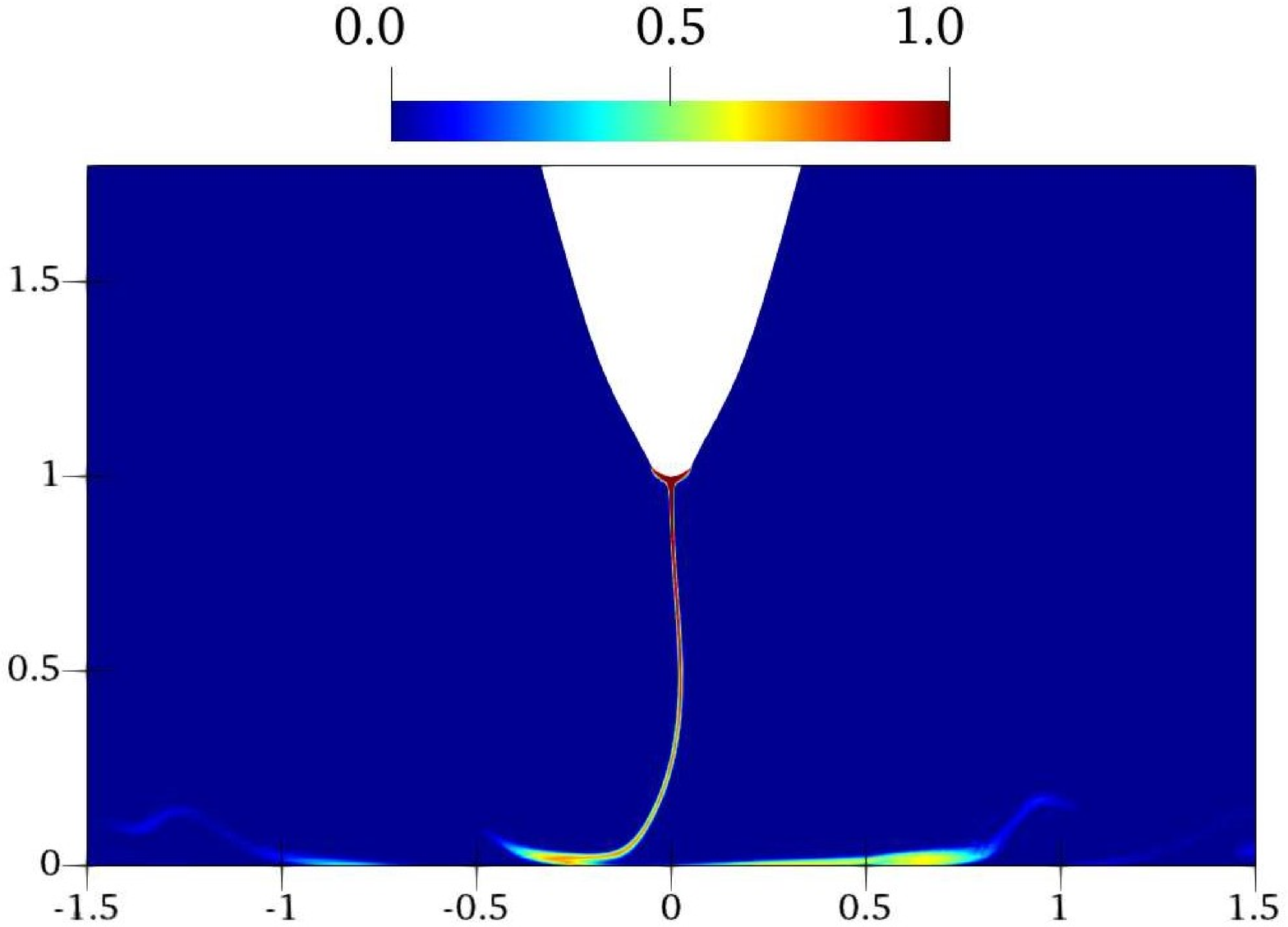}
			\put(-170,90){(f)}
			\put(-165,49){$ y $}
			\put(-82,-5){$ x $}			
			\label{fig.t1}
	\end{minipage}}
	\hspace{20pt}
	\subfigure{
		\begin{minipage}[h]{0.4\textwidth}
			\centering
			\includegraphics[height=4cm]{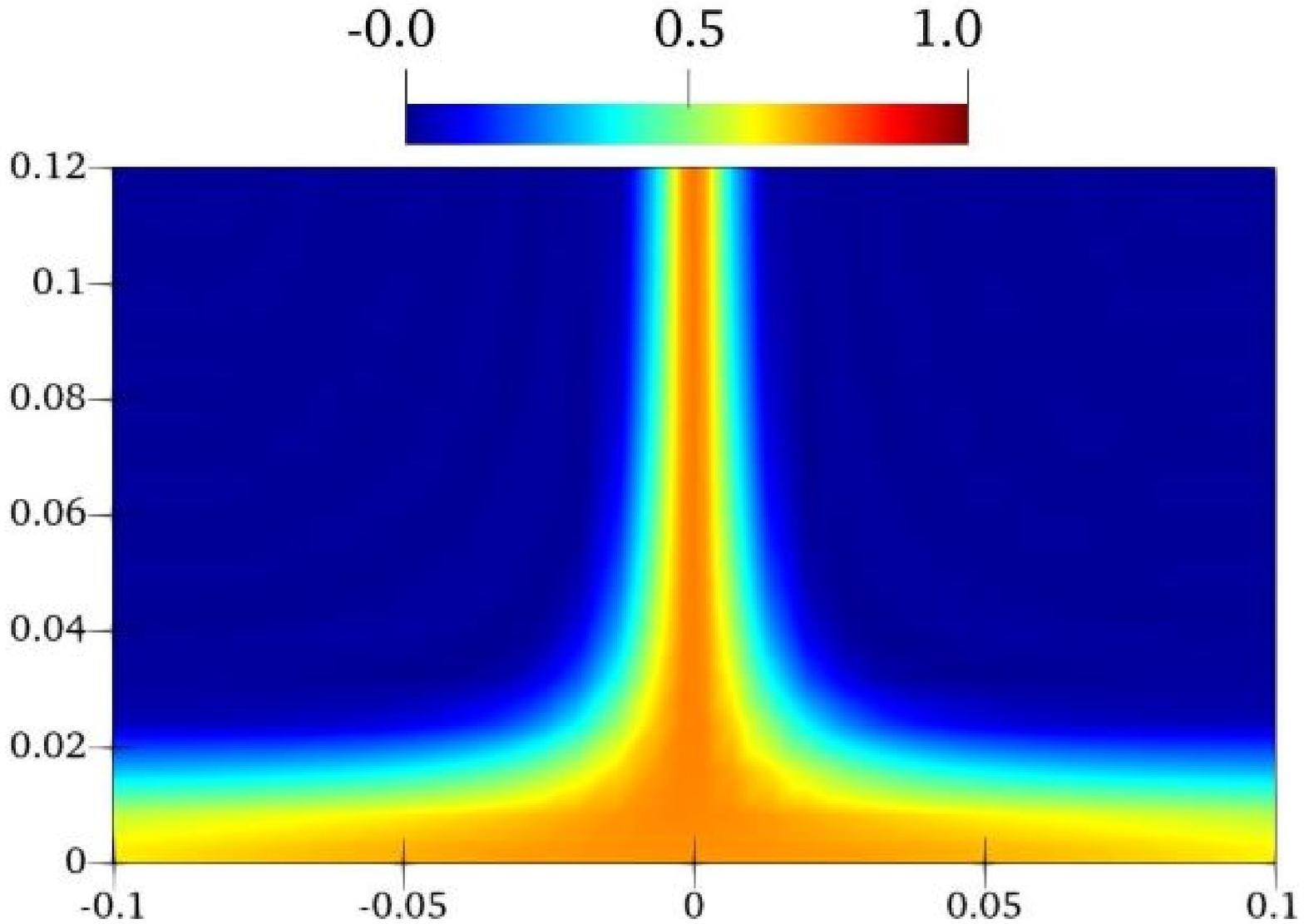}
			\put(-175,90){(c)}
			\put(-170,49){$ y $}
			\put(-79,-5){$ x $}	
			\label{fig.qt2}
	\end{minipage}}
	\hspace{20pt}
	\subfigure{
		\begin{minipage}[h]{0.4\textwidth}
			\centering
			\includegraphics[height=4cm]{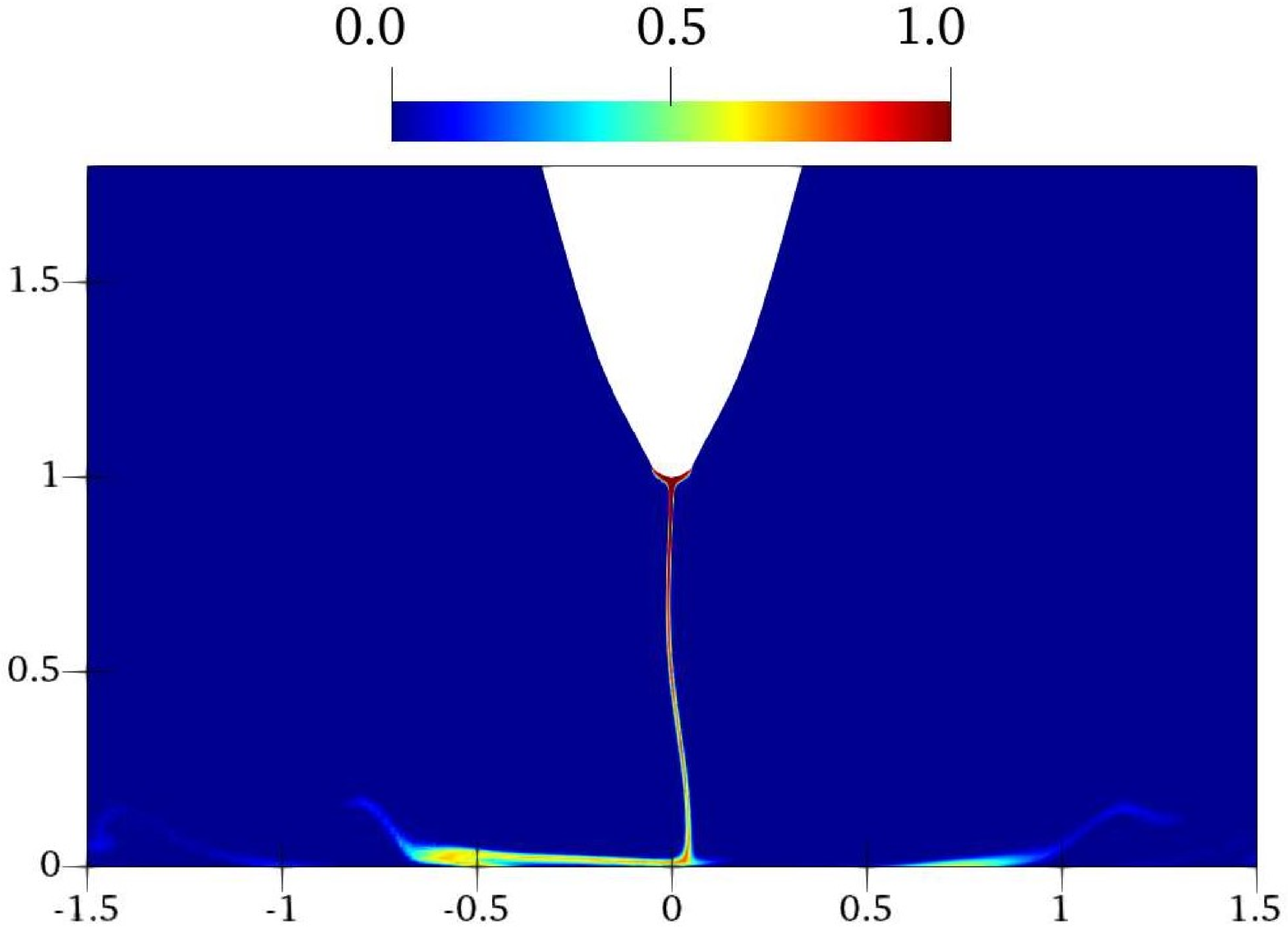}
			\put(-170,90){(g)}
			\put(-165,49){$ y $}
			\put(-79,-5){$ x $}	
			\label{fig.t2}
	\end{minipage}}
	\hspace{20pt}
	\subfigure{
		\begin{minipage}[h]{0.4\textwidth}
			\centering
			\includegraphics[height=4cm]{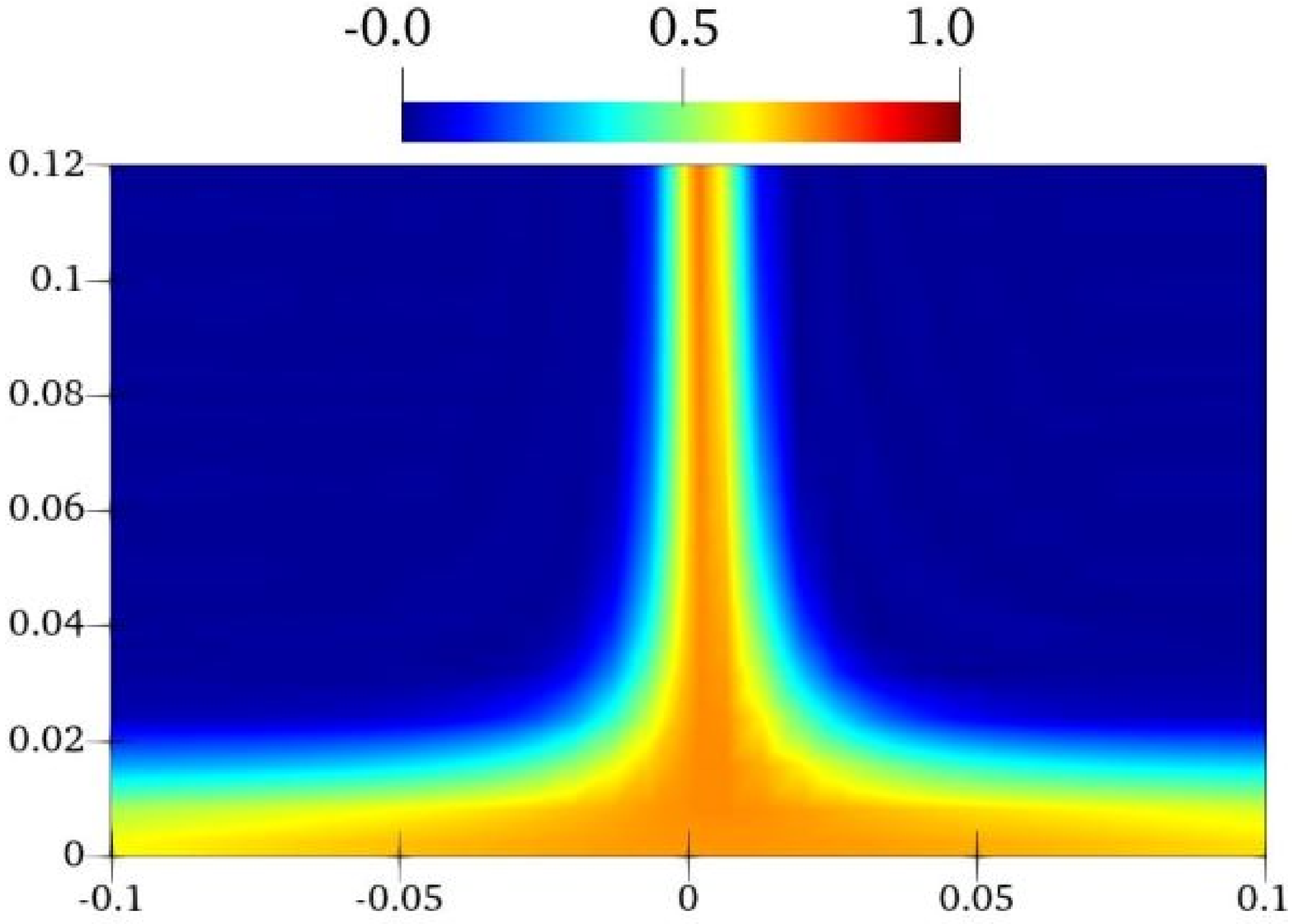}
			\put(-175,90){(d)}
			\put(-170,49){$ y $}
			\put(-79,-5){$ x $}	
			\label{fig.qt3}
	\end{minipage}}
	\hspace{20pt}
	\subfigure{
		\begin{minipage}[h]{0.4\textwidth}
			\centering
			\includegraphics[height=4cm]{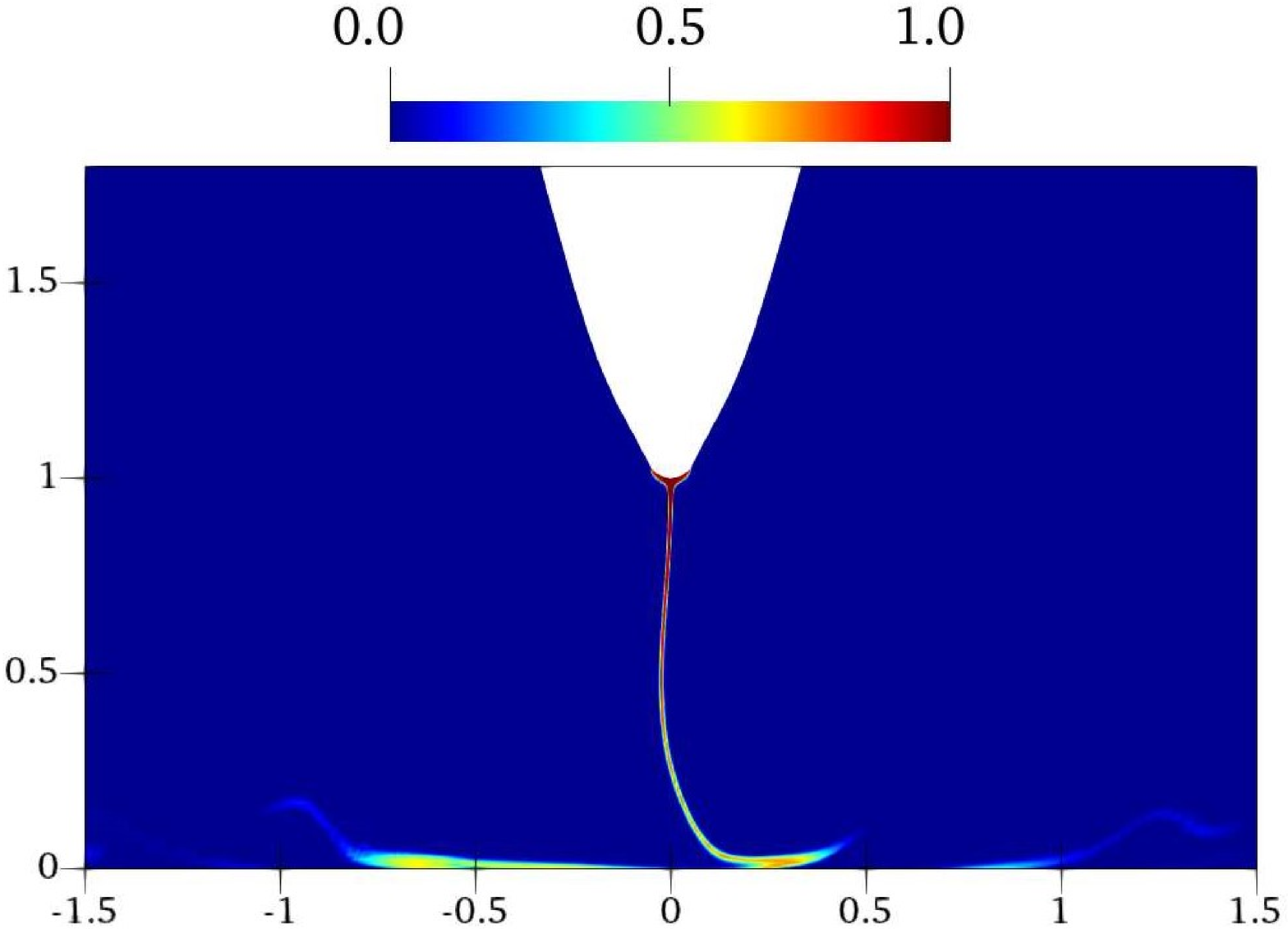}
			\put(-170,90){(h)}
			\put(-165,49){$ y $}
			\put(-79,-5){$ x $}	
			\label{fig.t3}
	\end{minipage}}
	\caption{Evolution of the $x$-velocity of the point (0,0.5) (panels (a) and (e)) and charge density distribution at different times in blade-plate EHD flow at $ T=3\times10^4 $ (left column) and $ T=4\times10^4 $ (right column). Among the two insets in each panel, the bottom one zooms in the oscillation period and the top one shows the Fast Fourier Transform (FFT) of the oscillation signal when its amplitude is stable. Distribution of charge density at (b) $ t=t_1 $; (c) $ t=t_2 $; (d) $ t=t_3 $ at $ T=3\times10^4 $ and at (f) $ t=t_4 $; (g) $ t=t_5 $; (h) $ t=t_6 $ at $ T=4\times10^4 $. ($ t_1-t_6$ are denoted in the inset of panel (a) and (e)). One has to look very closely to notice the small oscillation in panels (b)-(d). The other parameters are the same as those in figure \ref{fig.charge}.}
	\label{fig.osc}
\end{figure} 

In \cite{perri2020electrically}, the authors observed transient EHD flows, suggestive of flow bifurcation to another state. In our numerical simulations, we can also observe an unsteady flow when increasing $T$ in the blade-plate EHD flow. In figure \ref{fig.osc}, we present the oscillation behavior at $ T=3\times10^4 $ (left column) and $ T=4\times10^4 $ (right column). As shown in figure~\ref{fig.T3e4}, the time evolution of $ U_x $ sampled at a point (0,0.5) at a large $ T=3\times10^4 $ transitions from stable to periodic oscillation. The corresponding charge density distributions at different times in half period, namely $ t_1, t_2$ and $ t_3 $ (see the inset of panel (a)), are depicted in panels (b)-(d). Note that the magnitude of the oscillating $U_x$ is small in this case. The panels (b)-(d) at the first sight look the same. Nevertheless, one has to look at the figures closely to observe that the vertical structure of the charge jet swing from left to right by scrutinising its relative position to the central axis. At larger $ T=4\times10^4 $, the oscillation becomes more violent. It can be seen from panel (e) that, the amplitude of the oscillating $ U_x $ at the point (0,0.5) increases (note the range of $y$-axis). The oscillation also seems to deviate from a single-frequency behaviour (unlike the smaller $T=3\times10^4$ in panel (a)) due to the stronger nonlinearity at the larger $T$. In addition, the swing of the charge jet is more obvious, which can be seen in panels (f)-(h). Additionally, we display the result of Fast Fourier Transform (FFT) of the $ U_x $ in the stable oscillation stage, as shown in the inset of panel (a) and (e). One can see that the dominant frequencies are 1.36 and 9.89 for $ T=3\times10^4 $ and  $ T=4\times10^4 $, respectively. In the latter case, there is another spike at the frequency $\approx 30$. Note that when we calculated the FFT of the velocity signal at (0.5,0.1) for $T=4\times10^4$, we observed two frequencies at around 10 and 20 (not shown), which is more consistent with the weakly nonlinear phenomenon (that the dominant frequency $f$ interacts with itself to generate $2f$).

\subsubsection{Characteristics of the Moffatt-like eddies in the blade-plate EHD flow, compared to \cite{moffatt} } 

\begin{figure}
	\centering
	\subfigure{
	\begin{minipage}[h]{0.4\linewidth}
		\centering
		\includegraphics[height=5.5cm]{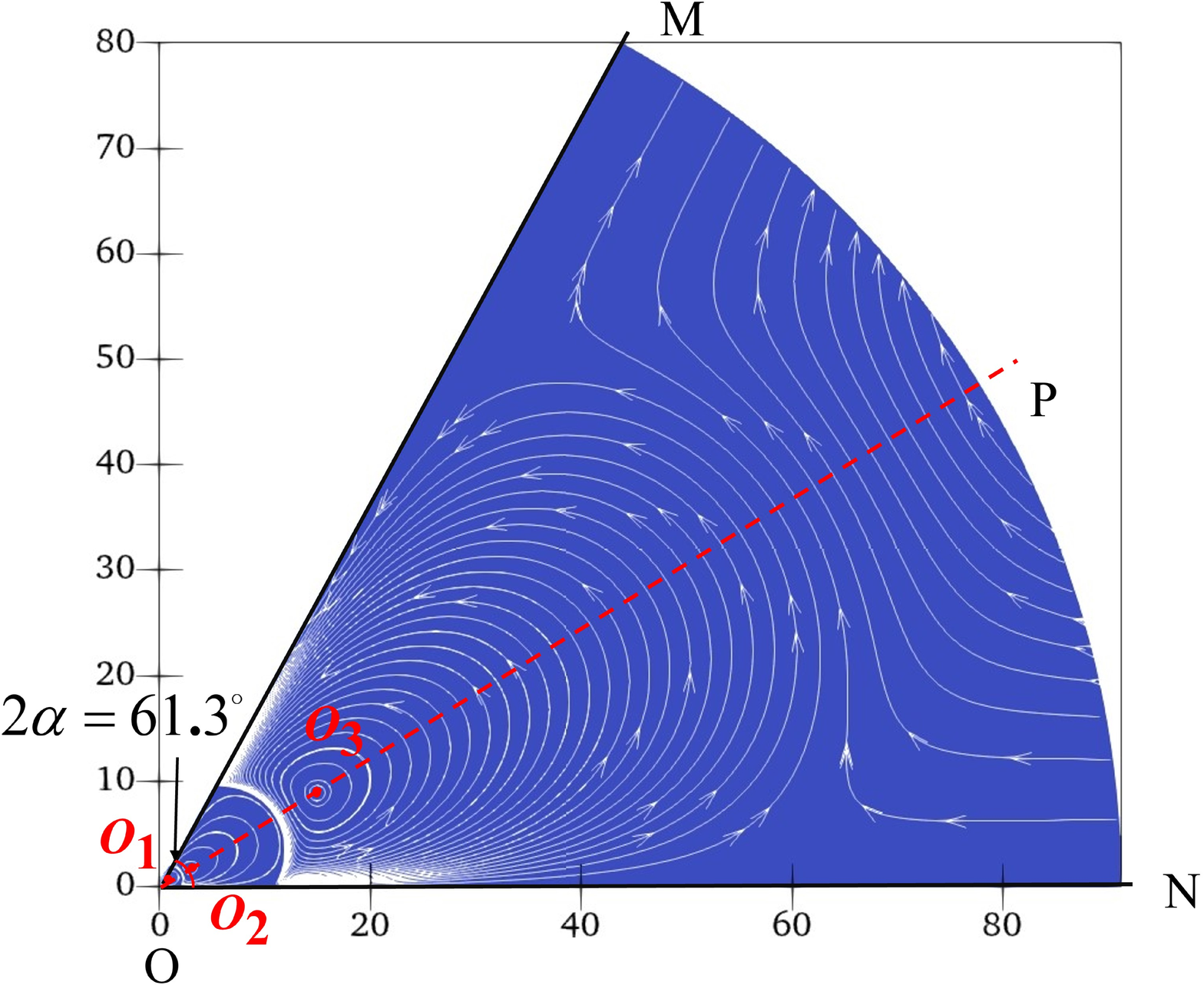}
		\put(-190,130){(a)}
		\label{fig.r03}
	\end{minipage}}
	\hspace{20pt}
	\subfigure{
	\begin{minipage}[h]{0.4\linewidth}
		\centering
		\includegraphics[height=5.5cm]{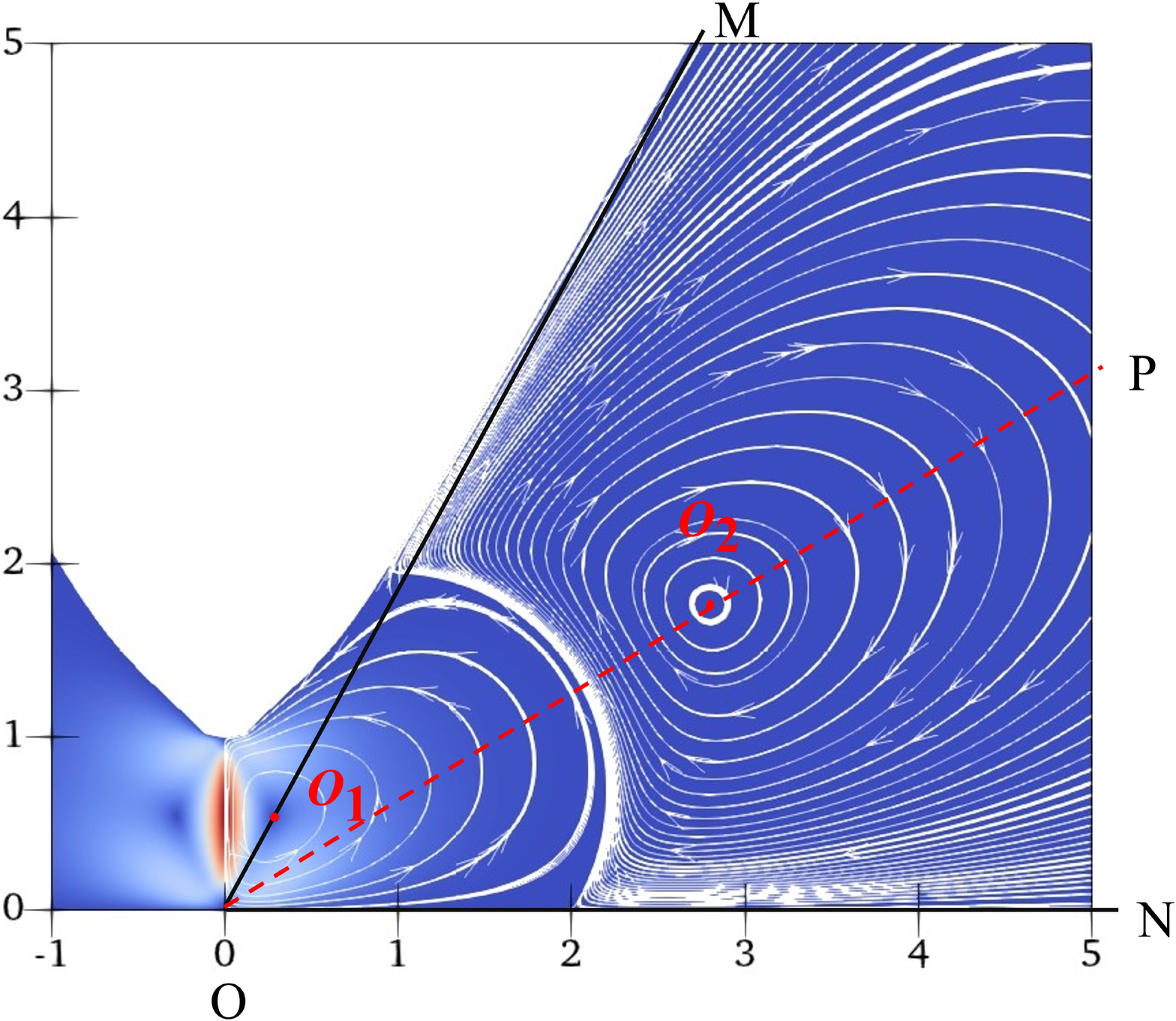}
		\put(-190,130){(b)}
		\label{fig.r03small}
	\end{minipage}}
	\caption{Streamlines in blade-plate EHD flows with $ R=0.3 $ ($ 2\alpha=61.3^\circ $). (a) The whole flow domain. (b) Zoom-in around the near field. The other parameters are $ T=500$, $Fe=5\times10^3$, $C=5$, $M=50$. Points $O_1,O_2,O_3$ are the centers of the three vortex structures. The red dashed half-line $OP$ connects points $O$ and $O_3$. }
	\label{fig.r03s}
\end{figure}

After a pair of small vortices are formed near the tip due to the charge injection (in the range of $x\in[1,2]$), two larger pairs of vortices are formed further away from the corner region driven by the viscous force (note that the viscous force includes both flow viscosity and charge diffusion, to be discussed shortly). Now we take the case of $ R=0.3, T=500, Fe=5\times10^3$ as an example to illustrate the properties of the Moffatt-like eddies in our blade-plate EHD flow. Figure \ref{fig.r03s} shows the streamline patterns in the nonlinear simulations of the steady EHD flow. It can be observed from the velocity vectors on the streamlines that the adjacent vortices are counter-rotating.  We draw the asymptote of the hyperbola passing the origin $O (0,0) $, labeled as $OM$ (more clearly in panel (b)). The half-line denoting the plate electrode is labelled as $ON$. We connect the origin $O$ to the center of the third vortex $O_3$ and mark the half-line $OP$. With these notations, $ \angle MON=61.3^\circ $ is the inter-electrode angle (as denoted in figure \ref{fig.fig2b}), which can be calculated exactly by $ \angle MON=\arctan(1/\sqrt{R}) $. To some extent, these vortices resemble those in \cite{moffatt} with the included angle $ 2\alpha=61.3^\circ $ between two rigid boundaries, but differences exist, especially, in the 'corner' area around the original point, due to the specific configuration of the blade-plate electrodes. 

It can be seen from panel \ref{fig.r03small} that the center $ O_2 $ of the second vortex falls almost on the line $OP$ with a slight deviation and the center of the vortex 1 (point $O_1$) does not locate at the line $OP$, but above it. It can be extracted that the coordinates of the three vortex centers are $ O_1(0.278,0.544)$, $ O_2(2.794,1.764) $, $ O_3(14.908,8.853) $ in this case. In our blade-plate EHD flow, the distances between the centers of the three vortices and the original point $O$ are respectively denoted as $ r_1 $, $ r_2 $ and $ r_3 $ (counted from the corner). Even though these notions are the same as those in equation (\ref{rratio}) for the Moffatt eddies, confusion can be easily dispelled within the context. We obtain their values as $ r_1=0.611, r_2=3.304, r_3=17.339 $. We find that $ r_3/r_2=5.25 $, which is close to the theoretical prediction 5.22. However, $ r_2/r_1=5.41 $, discernibly different from the theoretical solution. This is mainly because of the geometrical differences mentioned above in the 'corner' region. Besides, it is noted that in the theory of Moffatt eddies \citep{moffatt}, the calculated flow is a solution to the Navier-Stokes equations in the absence of volume forces. However, in the EHD case, there is a Coulomb force concentrated around the axis of symmetry of the geometry. The off-axis extension of the Coulomb force by diffusion and Coulomb repulsion generates a finite region where the volume force is non-zero. These are not the assumptions of the solution of Moffatt eddies, which may also contribute to the discrepancy of the first vortex between our result and that in Moffatt eddies.

\begin{figure}
	\centering
	\subfigure{	
	\begin{minipage}[h]{0.4\linewidth}
		\raggedleft
		\includegraphics[height=4.5cm,width=5.5cm]{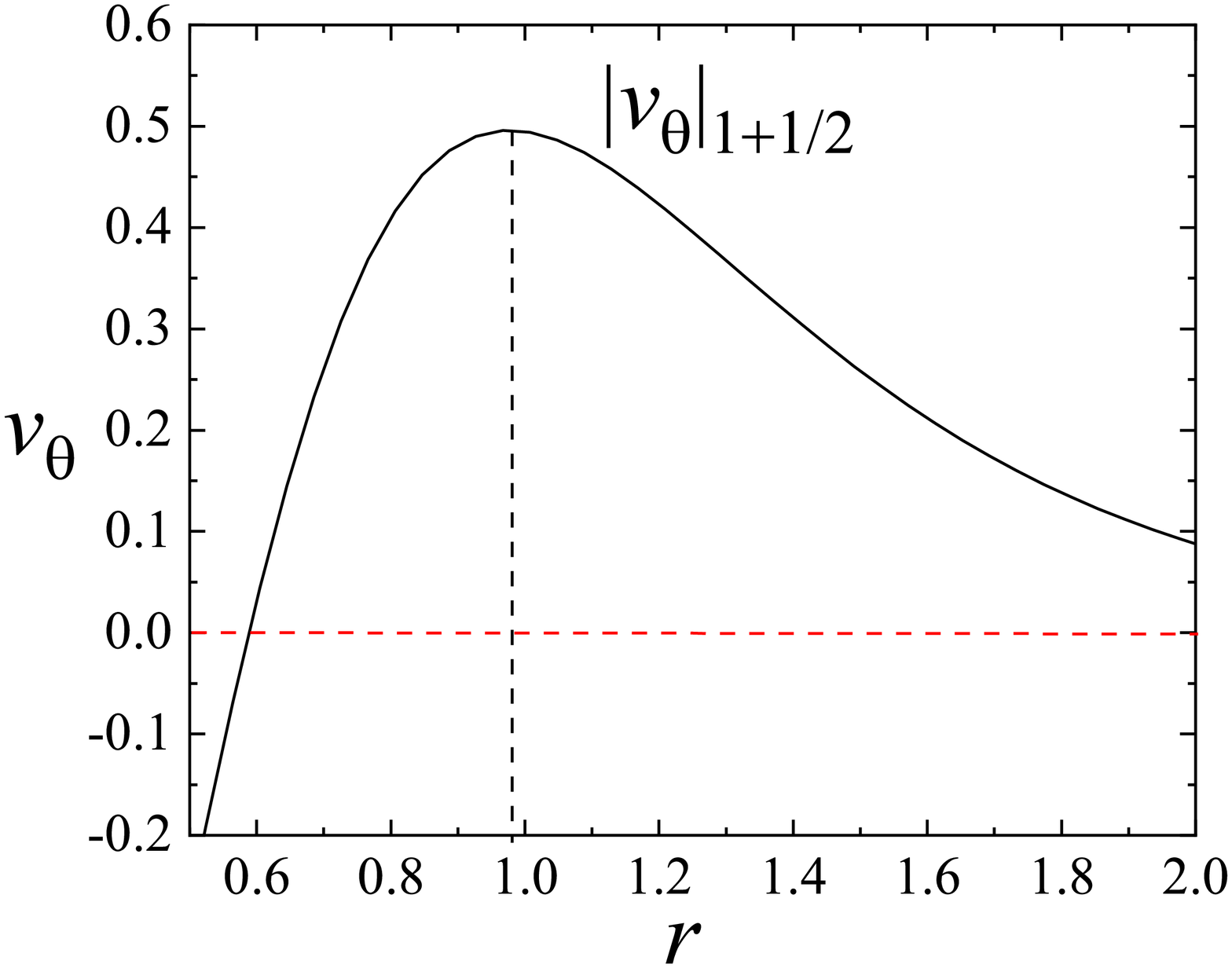}
		\put(-170,120){(a)}
		\label{fig.vor161}
	\end{minipage}}
	\hspace{20pt}
	\subfigure{
	\begin{minipage}[h]{0.4\linewidth}
		\flushright
		\includegraphics[height=4.5cm,width=5.5cm]{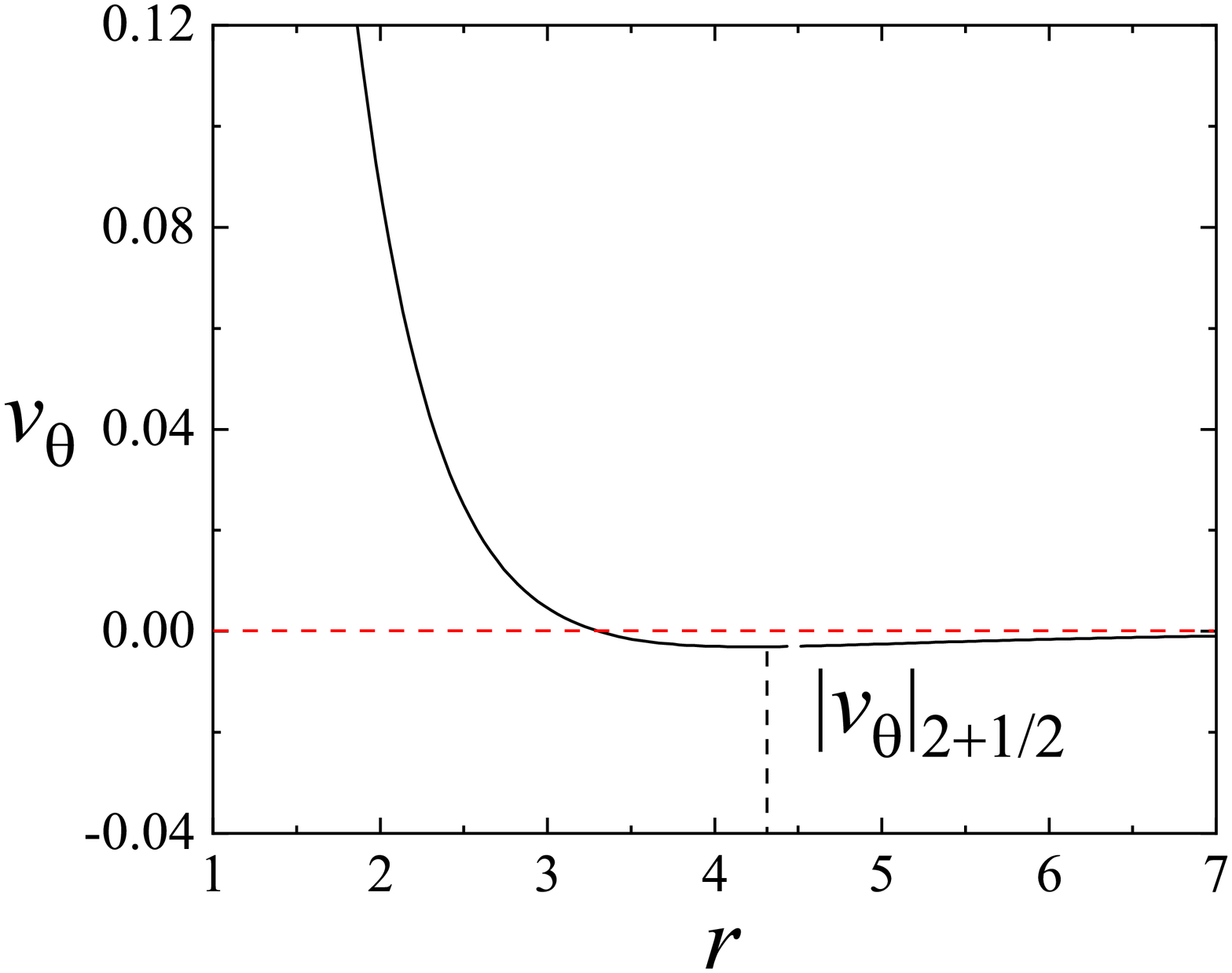}
		\put(-170,120){(b)}
		\label{fig.vor261}
	\end{minipage}}
	\subfigure{
	\begin{minipage}[h]{0.4\linewidth}
		\raggedleft
		\includegraphics[height=4.5cm,width=5.5cm]{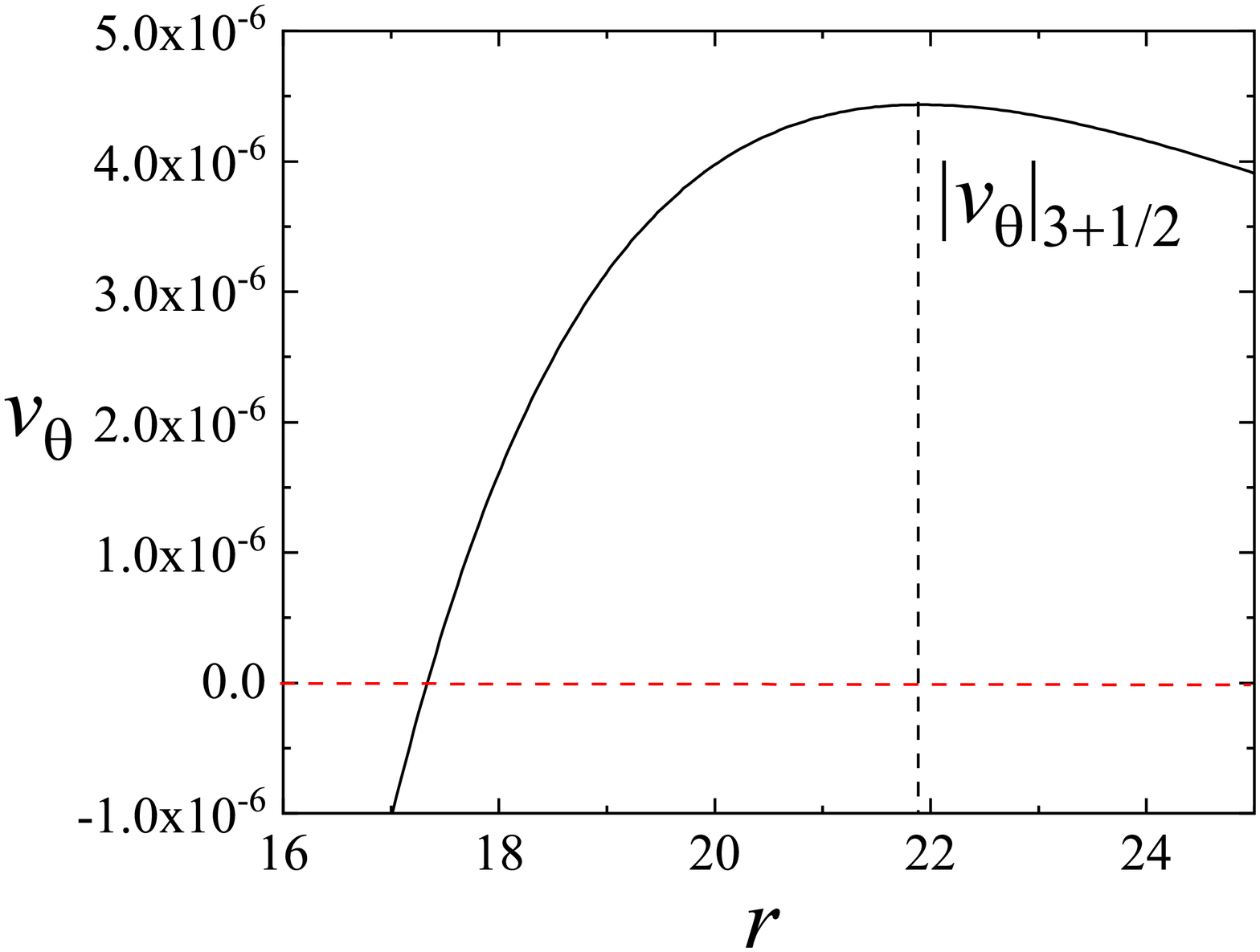}
		\put(-165,120){(c)}
		\label{fig.vor361}
	\end{minipage}}
	\hspace{30pt}
    \subfigure{
	\begin{minipage}[h]{0.4\linewidth}
		\flushright
		\includegraphics[height=4.5cm,width=5.5cm]{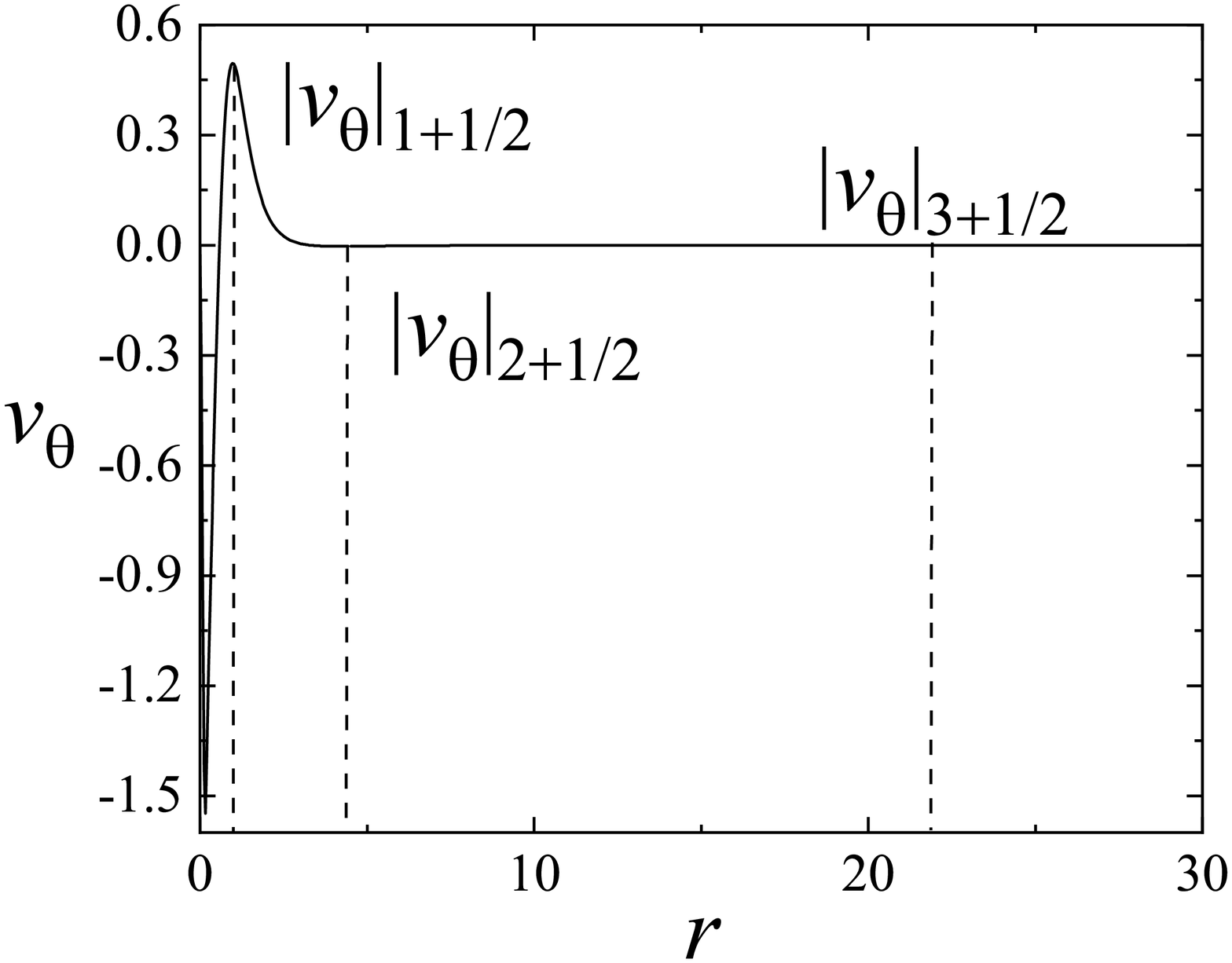}
		\put(-170,120){(d)}
		\label{fig.vor36}
    \end{minipage}}
	\caption{The azimuthal velocity distribution along the line $ OP $ in the section near the local maximum value. The inter-electrode angle is $ 61.3^\circ $. (a) Vortex 1; (b) vortex 2; (c) vortex 3; (d) a full view of $ v_\theta $ along the line $ OP $. The parameters are the same as figure \ref{fig.r03s}.}
	\label{fig.vortex61}
\end{figure}

As shown in figures \ref{fig.vor161}-\ref{fig.vor361}, we present the values of the azimuthal velocity $v_\theta$ on the line $ OP $. Note that the direction of $v_\theta$ is perpendicular to $OP$. The local maximum of the amplitude of $v_\theta$ is denoted as $|v_\theta|_{n+1/2}$, where $n$ means the number of the vortex counting from the corner.
The $r$ axis in these figures measures the radial distance from the original point on the $OP$ line. We can find the ratios of eddy intensity as (the theoretical value is $ 710.56 $):
\begin{equation}
	\frac{|v_{\theta}|_{1+1/2}}{|v_{\theta}|_{2+1/2}}=\frac{0.496}{0.00306}=162.09, \ \ \ \  \frac{|v_{\theta}|_{2+1/2}}{|v_{\theta}|_{3+1/2}}=\frac{0.00306}{4.436\times10^{-6}}=689.81.
\end{equation}
Similar to the size ratio, the intensity ratio of vortex 2 and 3 is closer to the theoretical solution, while the ratio of vortex 1 and 2 is noticeably different. It is also noted that in figure \ref{fig.vor361}, the position corresponding to $ v_{\theta} $ = 0 is the position of the center of vortex 3. Panel (d) presents the whole view of $ v_\theta $ along $ OP $. The value of $ v_{\theta} $ first decreases (and its amplitude increases) due to the potential difference between the plate and the blade electrodes, then increases and reaches the first local maximum in the first vortex. The remaining two local maxima can also be similarly understood with the help of the flow field shown in figure \ref{fig.r03s}.

\subsubsection{Effects of parameters: inter-electrode angle, intensity of the electric field, charge diffusion} 
In order to characterise the Moffatt-like eddies in the EHD flow in detail, we study the effects of several parameters in this subsection. 

We first change the inter-electrode angle to see the change of the Moffatt-like eddies in the blade-plate EHD flows. The inter-electrode angle can be adjusted by changing the radius of the curvature $ R $. The investigated values of the inter-electrode angles are $ 77.4^\circ, 72.5^\circ $ and $ 66^\circ $, which correspond to the radius of the hyperbolic function $ R=0.05, R=0.1 $ and $ R=0.2 $ (as in Eq. \ref{eq.blade}), respectively. The comparison between the results of numerical simulations and theoretical analyses is presented in  table \ref{table:r}. It can be seen that with the decrease of the inter-electrode angle (or larger $R$), $ r_1 $, $ r_2 $ and $ r_3 $ all decrease, indicating that the sizes of vortices shrink. 
Besides, the local maximum azimuthal velocities of vortex 2 and 3 ($ |v_{\theta}|_{2+1/2} $, $ |v_{\theta}|_{3+1/2}$, respectively) increase with decreasing $ 2\alpha $ and that of vortex 1 ($ |v_{\theta}|_{1+1/2} $) decreases as $ 2\alpha $ decreases. We suspect that these trends of the local maximum azimuthal velocity may not be generalisable since the results may depend on the specific setting in our flow configuration; especially we fix the $y$-coordinates of $ S_1 $ and $ S_2 $ (depicted in figure \ref{fig.model}) for the ion injection. From the table, we can also see that the maximum vertical velocity $ |U_y|_{max}^s $ increases as $ R $ increases. In addition, the results show that the size and intensity ratio of the 2nd vortex over the 3rd vortex at different inter-electrode angles have a good agreement with the theoretical solutions, which are at a large distance from the corner (so less affected by the corner geometry), indicating that the Moffatt-like eddies observed in the (relative) farfield of blade-plate EHD flow closely obey the similarity solutions presented by \cite{moffatt}. Furthermore, $ \angle PON$ is close to $ \alpha $ for all $R$, meaning that the center of the vortex 3 falls almost on the bisector of the inter-electrode angle.

\begin{table}
	\def~{\hphantom{0}}
	\begin{center}
		\begin{tabular}{l c c c c c}
			&& $ R=0.05 $& $ R=0.1$ & $ R=0.2 $ & $ R=0.3 $ \\ 
			&& $ (2\alpha=77.4^\circ) $& $ (2\alpha=72.5^\circ$) & $(2\alpha=66^\circ) $ & $(2\alpha=61.3^\circ) $ \\ 
			\hline
			\cite{moffatt}&$ \rho $&  ~9.37~ & ~7.73~ & ~6.12~ & ~5.22~ \\
						& $ \Omega $&~$ 1195.39 $~& ~$ 995.71 $~ & ~$ 808.76 $~& ~$ 710.56 $~\\	
						&$ \angle PON$&~$ 38.7^\circ $~ & ~$ 36.25^\circ $~ & ~$ 33^\circ $~& ~$ 30.65^\circ $~\\
		    \hline			
			Present&$ |U_y|_{max}^s $   &   7.067     &    7.319&   7.488&   7.611\\
			&$ O_1 $   &   (0.312,0.599)     &(0.295,0.574)&(0.278,0.555)&(0.278,0.544)\\
			&$ O_2 $   &  (4.181,3.433)      &(3.638,2.756)&(3.003,2.049)&(2.794,1.764)\\
			&$ O_3 $   &   (39.358,31.475)     &(28.381,20.83)&(18.631,12.103)&(14.908,8.853)\\						
			&$ r_1 $ &   ~0.675~ & ~0.645~ & ~0.621~ &   ~0.611~ \\
			&$ r_2 $  & ~5.410~ & ~4.564~  & ~3.635~ &~3.304~ \\
			&$ r_3 $	&  ~50.384~ &  ~35.205~ & ~22.217~&~17.339~\\		
			&$ r_2/r_1 $&  ~8.01~ & ~7.08~ & ~5.85~ & ~5.41~\\
			&$ r_3/r_2 $&  ~9.31~ & ~7.71~ & ~6.11~ & ~5.25~\\
			&$ |v_{\theta}|_{1+1/2} $& ~0.563~ & ~0.556~ &~0.528~   & ~0.496~\\
			&$ |v_{\theta}|_{2+1/2} $&~0.00137~&~0.00186~  &~0.00266~   & ~0.00306~ \\
			&$ |v_{\theta}|_{3+1/2} $&~$ 1.051\times10^{-6} $~ &~$ 1.903\times10^{-6} $~  &~$ 3.370\times10^{-6} $~ &~$ 4.436\times10^{-6} $~ \\
			&$ |v_{\theta}|_{1+1/2}/|v_{\theta}|_{2+1/2}$ &~$ 410.95 $~& ~$ 298.92 $~ & ~$ 198.90 $~& ~$ 162.09 $~\\			
			&$ |v_{\theta}|_{2+1/2}/|v_{\theta}|_{3+1/2}$ & ~$ 1303.5 $~  & ~$ 928.39 $~& ~$ 789.31 $~ & ~$ 689.81 $~\\
			&$ \angle PON$ &~$ 38.6^\circ $~ & ~$ 36.2^\circ $~ & ~$ 33^\circ $~& ~$ 30.7^\circ $~\\
		\end{tabular}
		\caption{Properties of Moffatt vortices at different $ R $ for $ T=500, Fe= 5\times10^3 $. The other parameters are $ C=5, M=50$.}	
		\label{table:r}
	\end{center}
\end{table}

\begin{table}
	\def~{\hphantom{0}}
	\begin{center}
		\begin{tabular}{l c c c c}
		      &$ T=500 $    &$ T=1000 $& $ T=1500 $& \cite{moffatt} \\
            $ |U_y|_{max}^s $   &   7.611     &11.373& 14.241 &\\		       
			$ O_1 $      &(0.278,0.544)  &(0.276,0.542)&(0.274,0.533)&\\
			$ O_2 $       &(2.794,1.764) &(2.793,1.748)&(2.785,1.726)&\\
			$ O_3 $      &(14.908,8.853)&(14.846,8.807)&(14.780,8.775)&\\
			$ r_1 $  & ~0.611~& ~0.608~& ~0.599~  &\\
			$ r_2 $  &~3.304~&~3.295~&~3.276~   & \\
			$ r_3 $	  &~17.339~&~17.262~&~17.189~  &\\
			$ r_2/r_1 $ & ~5.41~& ~5.42~ & ~5.47~& ~5.22~\\
			$ r_3/r_2 $  & ~5.25~ & ~5.24~& ~5.25~ & ~5.22~\\
			$ |v_{\theta}|_{1+1/2} $& ~0.496~ & ~0.720~& ~0.909~&\\
			$ |v_{\theta}|_{2+1/2} $&~0.00306~&~0.00441~&~0.00545~ &\\
			$ |v_{\theta}|_{3+1/2} $ &~$ 4.436\times10^{-6} $~ &~$ 6.386\times10^{-6} $~&~$ 7.942\times10^{-6} $~   &\\
			$ |v_{\theta}|_{1+1/2}/|v_{\theta}|_{2+1/2}$ & ~$ 162.09 $~ & ~$ 163.27 $~& ~$ 166.79 $~& ~$ 710.56 $~\\
			$ |v_{\theta}|_{2+1/2}/|v_{\theta}|_{3+1/2}$ & ~$ 689.81 $~ & ~$ 690.57 $~ & ~$ 686.23 $~ & ~$ 710.56 $~\\
			$ \angle PON$ & ~$ 30.7^\circ $~& ~$ 30.7^\circ $~ & ~$ 30.5^\circ $~& ~$ 30.65^\circ $~\\
		\end{tabular}
		\caption{Properties of Moffatt vortices at different $ T $ for $ Fe= 5\times10^3 $ and $ R=0.3 $ ($ 2\alpha=61.3^\circ $). The other parameters are $ C=5, M=50$.}	
		\label{table:61T}
	\end{center}
\end{table}

\begin{table}
	\def~{\hphantom{0}}
	\begin{center}
		\begin{tabular}{l c c c c c}
			&   $ Fe=100 $&   $ Fe=200 $   &$ Fe=500 $    &$ Fe=5\times10^3 $&  \cite{moffatt} \\ 
			 $ |U_y|_{max} $   &   7.532 &   7.590     &7.615& 7.611 &\\			
			$ O_1 $    &(0.296,0.544) &(0.289,0.544) &(0.283,0.544)     &(0.278,0.544)&\\
			$ O_2 $    &(5.001,3.027)&(3.160,1.980)   &(2.811,1.74)     &(2.794,1.764)&\\
			$ O_3 $     &(26.263,15.550)&(16.767,9.890) &(15.088,8.938)     &(14.908,8.853)&\\
			$ r_1 $   & ~0.686~& ~0.616~& ~0.613~& ~0.611~  &\\
			$ r_2 $   &~5.846~&~3.729~&~3.306~&~3.304~   & \\
			$ r_3 $	 &~30.521~ &~19.466~&~17.532~&~17.339~  &\\
			$ r_2/r_1 $     & ~8.51~& ~6.05~& ~5.39~& ~5.41~ & ~5.22~\\
			$ r_3/r_2 $     &~5.22~& ~5.22~& ~5.27~ & ~5.25~ & ~5.22~\\
			$ |v_{\theta}|_{1+1/2} $& ~0.439~& ~0.537~& ~0.545~ & ~0.496~&\\
			$ |v_{\theta}|_{2+1/2} $&~0.000578~&~0.00222~&~0.00311~&~0.00306~ &\\
			$ |v_{\theta}|_{3+1/2} $&~$ 8.339\times10^{-7} $~ &~$ 3.210\times10^{-6} $~ &~$ 4.501\times10^{-6} $~ &~$ 4.436\times10^{-6} $~   &\\
			$ |v_{\theta}|_{1+1/2}/|v_{\theta}|_{2+1/2}$ & ~$ 759.52$~& ~$ 241.89 $~&~$ 175.24 $~ & ~$ 162.09 $~ & ~$ 710.56 $~\\
			$ |v_{\theta}|_{2+1/2}/|v_{\theta}|_{3+1/2}$ & ~$ 693.13 $~& ~$ 691.59 $~& ~$ 690.95 $~ & ~$ 689.81 $~ & ~$ 710.56 $~\\
			$ \angle PON$ & ~$ 30.63^\circ $~& ~$ 30.5^\circ $~& ~$ 30.7^\circ $~& ~$ 30.7^\circ $~ & ~$ 30.65^\circ $~\\
		\end{tabular}
		\caption{Properties of Moffatt vortices at different $ Fe $ for $ T= 500 $ and $ R=0.3 $ ($ 2\alpha=61.3^\circ $). The other parameters are $ C=5, M=50$.}	
		\label{table:61Fe}
	\end{center}
\end{table}

Next, we study the effect of the electric Rayleigh number $T$, which quantifies the strength of the electric field. 
In table \ref{table:61T}, we summarise the numerical results of $ R=0.3 $ and $ T=500,1000,1500 $, respectively. We can observe that $ r_1 $, $ r_2 $ and $ r_3 $ all decrease as $T$ increases, meaning that the vortex centers are approaching the corner as we increase the intensity of the electric field. In addition, $ |v_{\theta}|_{1+1/2} $, $ |v_{\theta}|_{2+1/2} $, $ |v_{\theta}|_{3+1/2} $ all increase  with the increase of $ T $, which has the same trend as $ |U_y|_{max}^s $, indicating that a higher electric field leads to stronger intensity of the eddies. Furthermore, one can see that the size and intensity ratio of vortex 2 and vortex 3 do not change significantly and conform to the law of Moffatt eddies as we change $T$ from 500 to 1500. Also, in all the cases, the angle $\angle PON$ approaches half of the inter-electrode angle $ \alpha $.

\begin{figure}
	\centering
	\subfigure{
		\begin{minipage}[h]{0.4\textwidth}
			\centering
			\includegraphics[height=4cm]{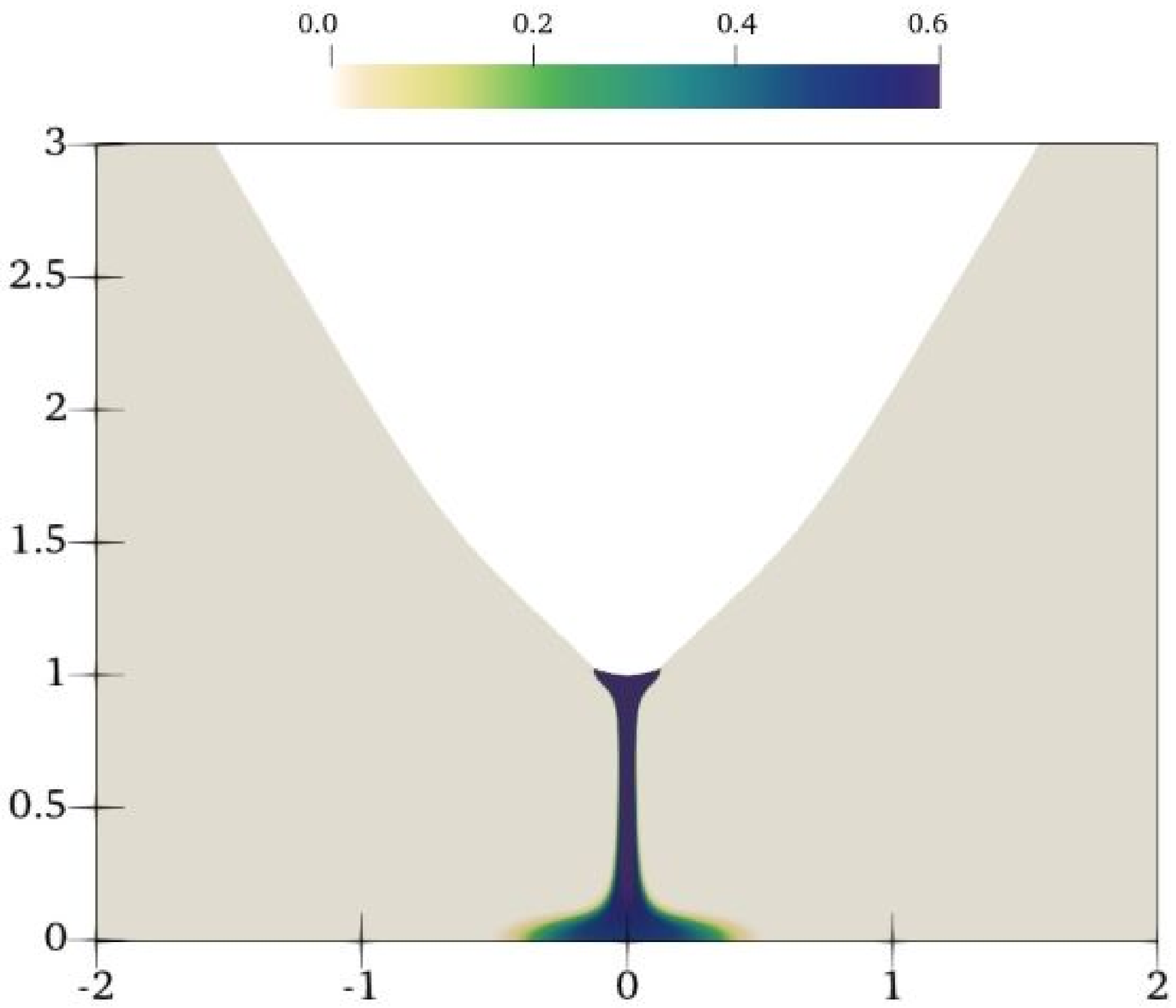}
			\put(-150,90){(a)}
			\put(-142,52){$ y $}
            \put(-65,-5){$ x $}		
            \put(-90,72){$ Fe=5\times10^3 $}	            	
			\label{fig.r03q}
	\end{minipage}}
	\hspace{20pt}
	\subfigure{
		\begin{minipage}[h]{0.4\textwidth}
			\centering
			\includegraphics[height=4cm]{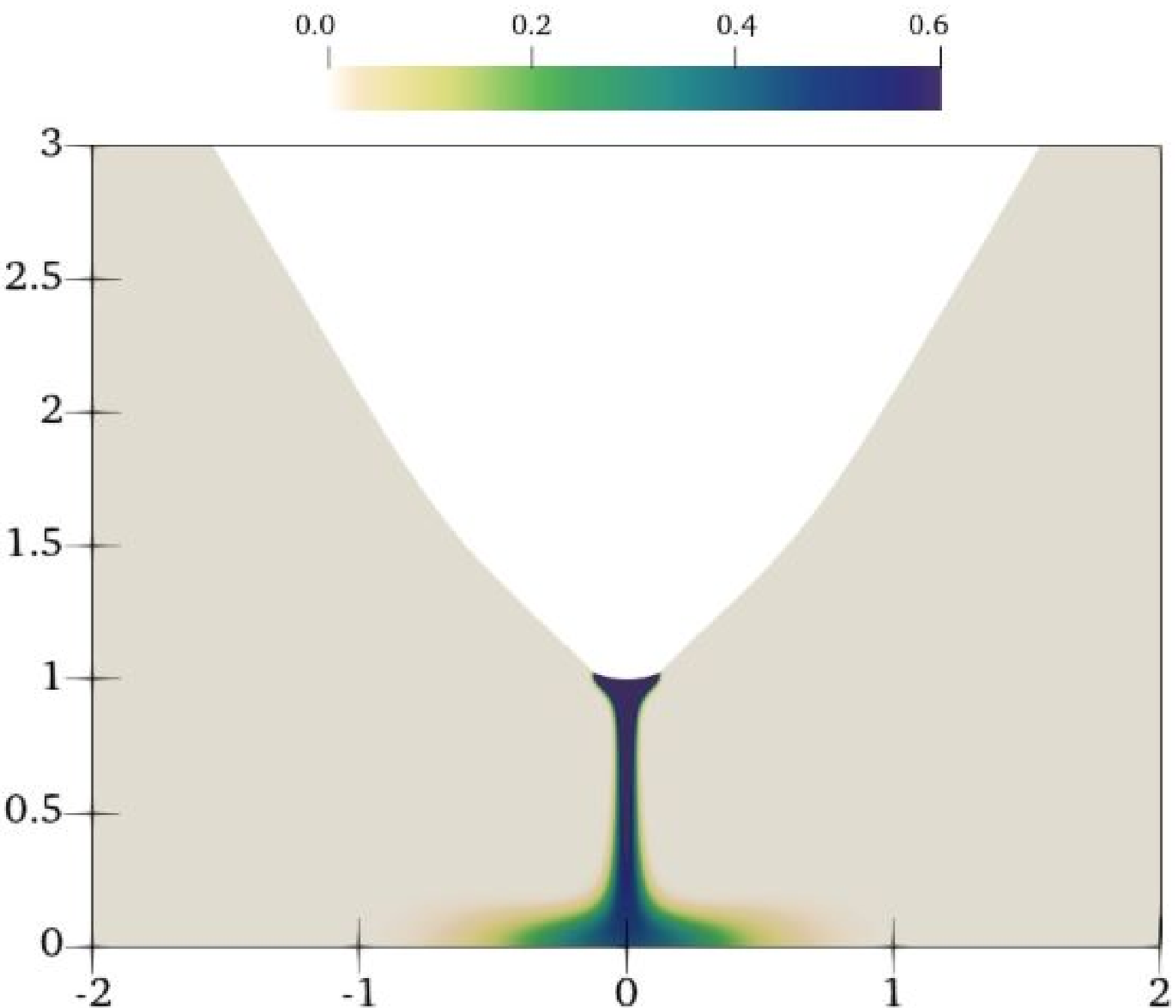}
			\put(-150,90){(b)}			
			\put(-142,52){$ y $}
			\put(-65,-5){$ x $}
            \put(-80,72){$ Fe=500 $}			
			\label{fig.r03qFe500}
	\end{minipage}}
	\hspace{20pt}
	\subfigure{
		\begin{minipage}[h]{0.4\textwidth}
			\centering
			\includegraphics[height=4cm]{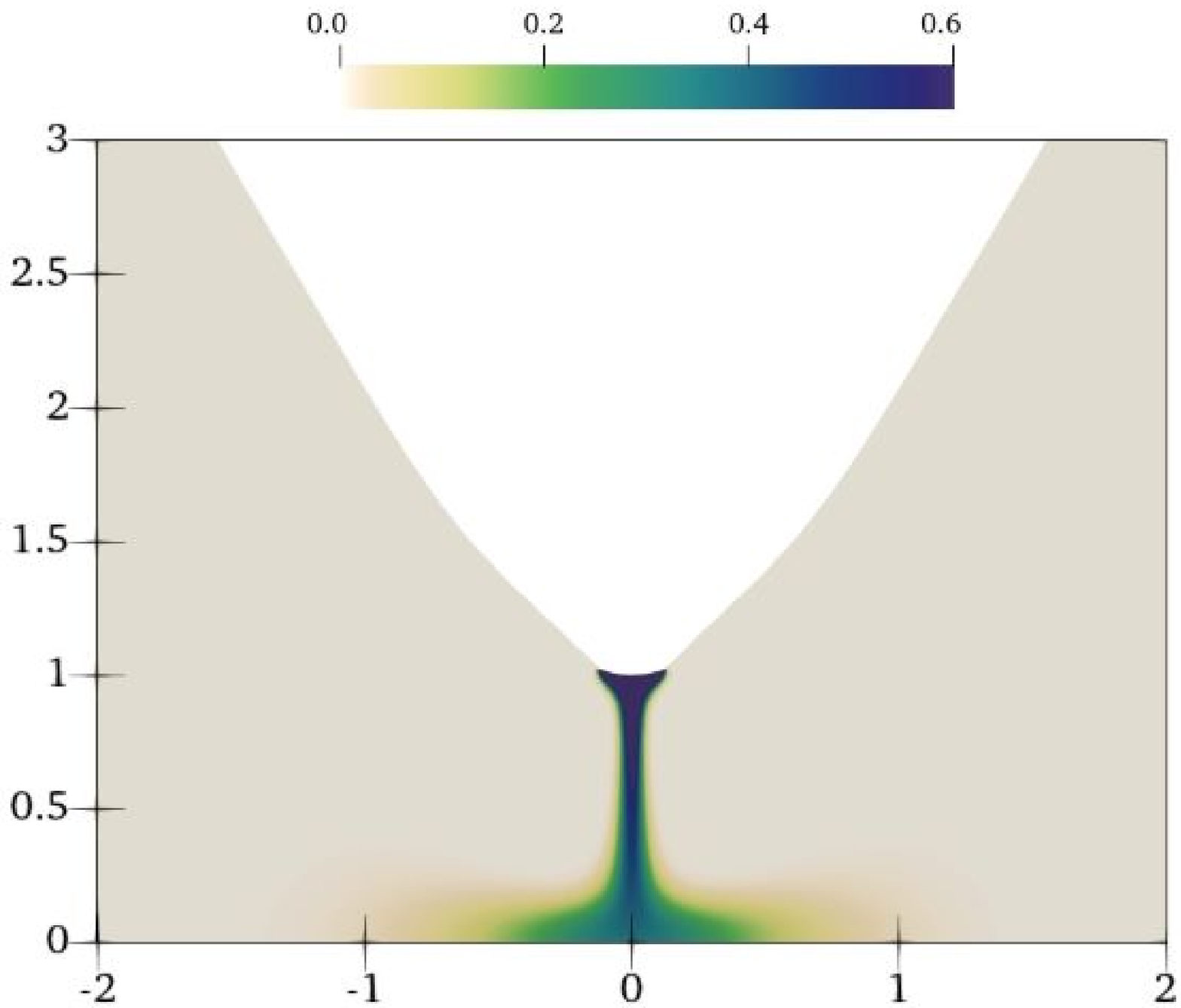}
			\put(-150,90){(c)}
			\put(-142,52){$ y $}
			\put(-65,-5){$ x $}
			\put(-80,72){$ Fe=200 $}
			\label{fig.r03qFe200}
	\end{minipage}}
	\hspace{20pt}
	\subfigure{
	\begin{minipage}[h]{0.4\textwidth}
		\centering
		\includegraphics[height=4cm]{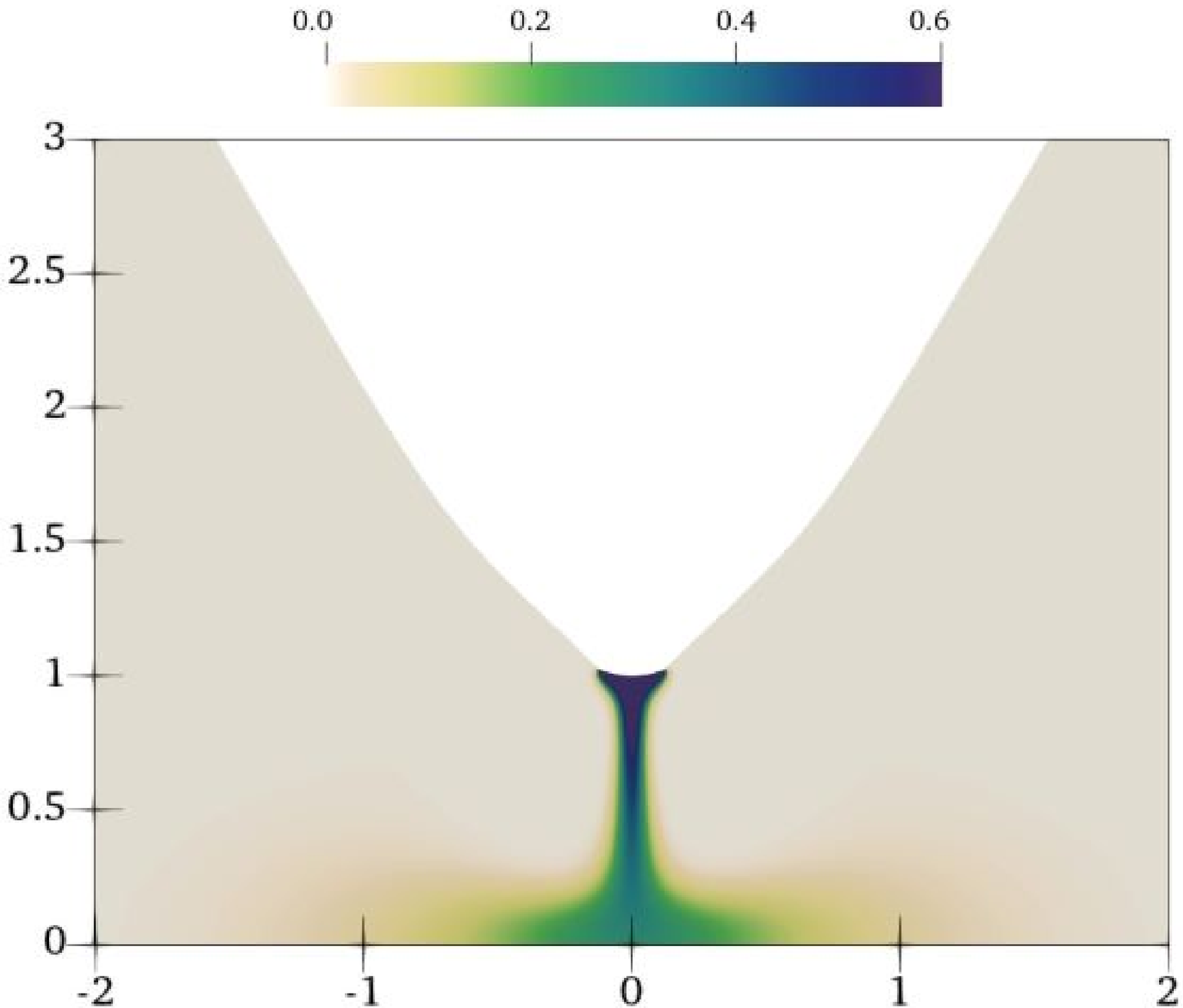}
		\put(-150,90){(d)}
		\put(-142,52){$ y $}
		\put(-65,-5){$ x $}
		\put(-80,72){$ Fe=100 $}
		\label{fig.r03qFe100}
    \end{minipage}}
    \hspace{20pt}
	\subfigure{
		\begin{minipage}[h]{0.4\textwidth}
			\centering
			\includegraphics[height=4cm]{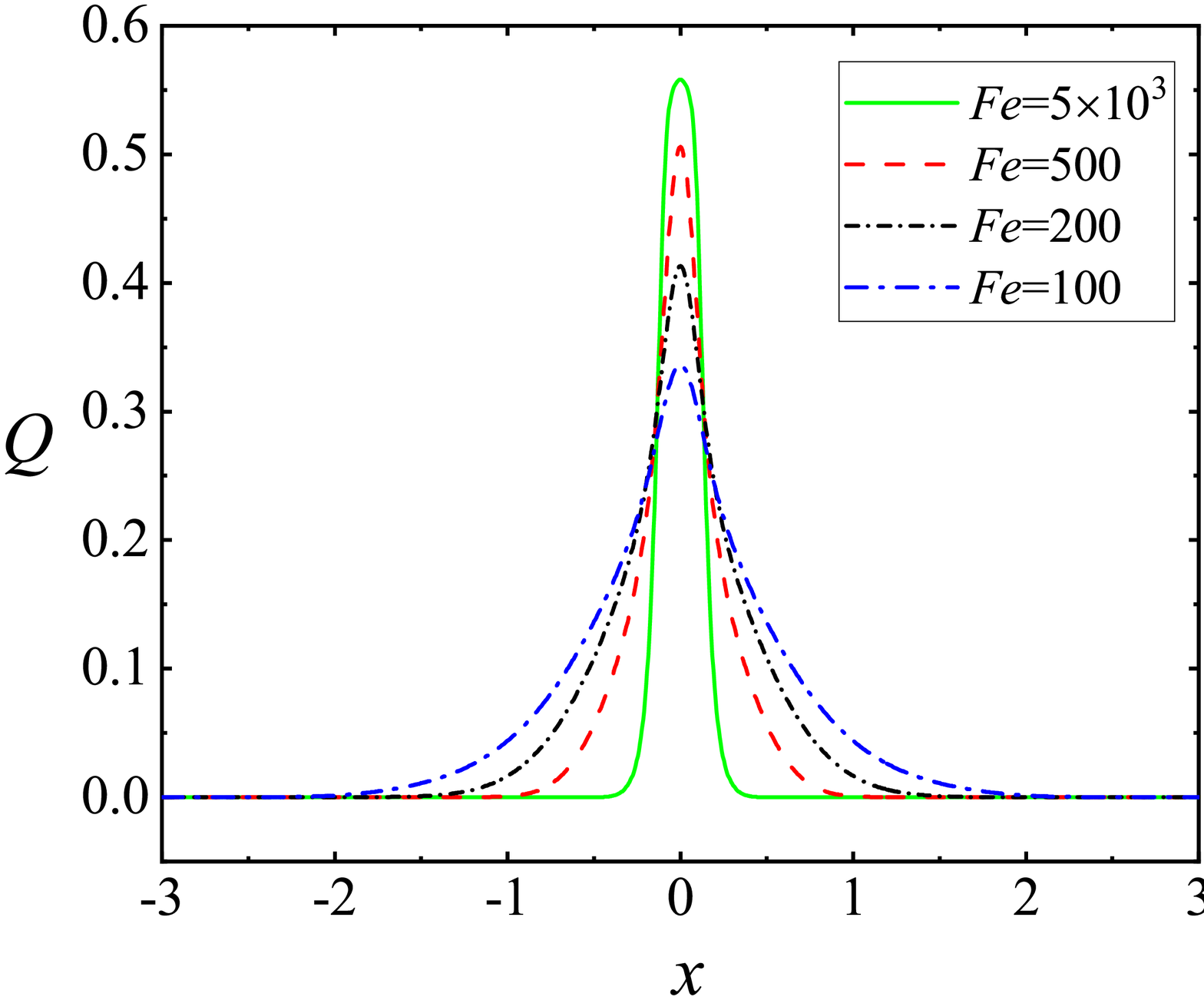}
			\put(-150,98){(e)}
			\label{fig.61q}
	\end{minipage}}
    \hspace{20pt}
    \subfigure{
	\begin{minipage}[h]{0.4\textwidth}
		\centering
		\includegraphics[height=3.7cm]{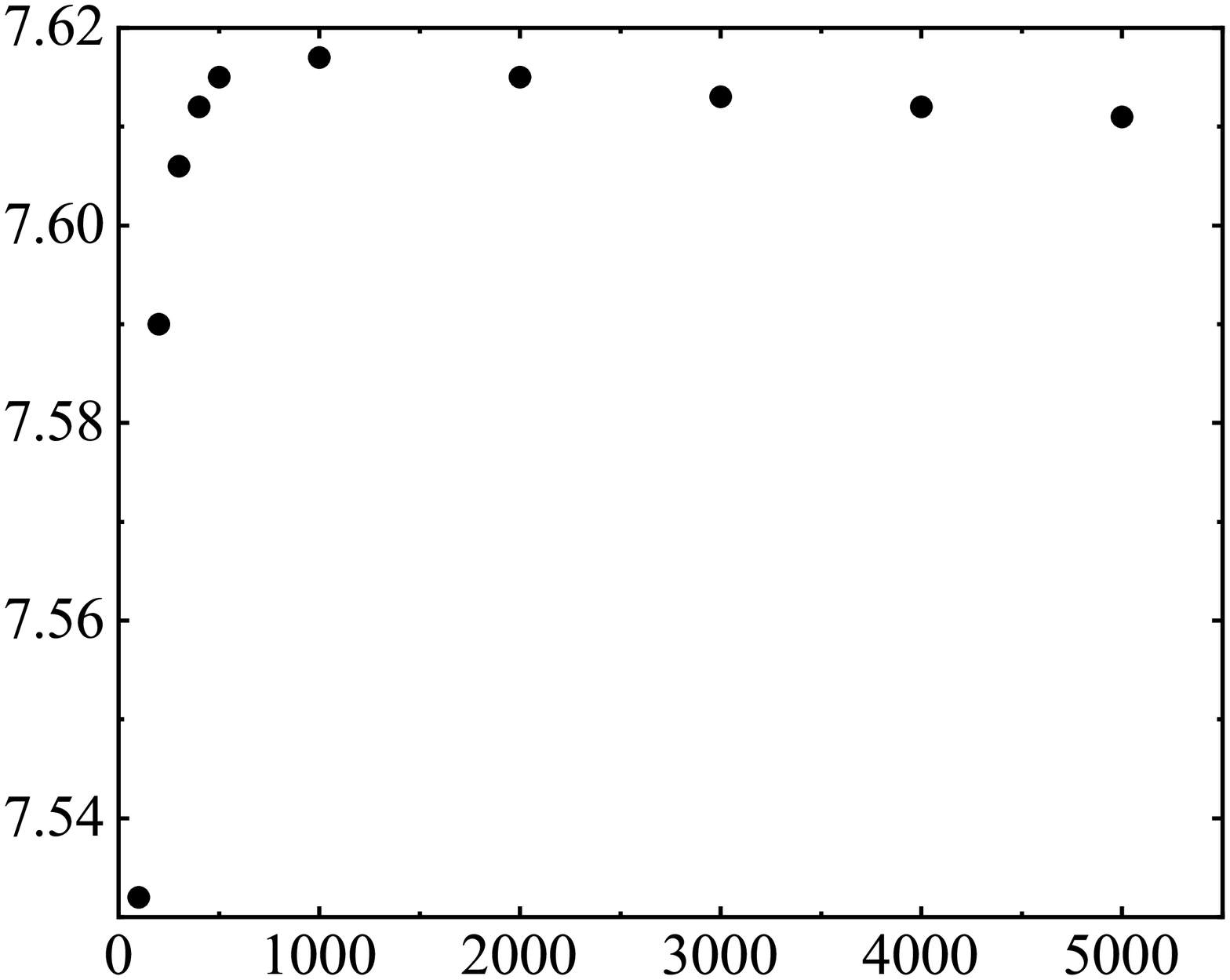}
		\put(-150,90){(f)}
    	\put(-165,60){$ |U_y|_{max}^s $}
        \put(-62,-10){$ Fe $}		
		\label{fig.FeUy}
    \end{minipage}}
	\caption{Distribution of the charge density of final steady state between the blade injector and the plate electrode at $ T=500, R=0.3 $, and (a) $ Fe=5\times10^3 $; (b) $ Fe=500 $; (c) $ Fe=200 $; (d) $ Fe=100 $; (e) along the horizontal line $ y=0.1 $. (f) Maximum vertical velocity magnitude $ |U_y|_{max}^s $ at the steady state versus $ Fe $. The other parameters are $ C=5, M=50$.}
	\label{fig.r03Q}
\end{figure}

Finally, we investigate the influence of charge diffusion on the viscous eddies by changing $Fe$ (note that smaller $Fe$ means stronger charge diffusion effect). In the above sections, we take $ Fe= 5\times10^3 $, corresponding to a relatively small charge diffusion effect. Three additional values of $ Fe $ ($ Fe=500, 200, 100 $) are considered to observe more clearly the effect of charge diffusion. As shown in table \ref{table:61Fe}, from the coordinates of the vortex centers in the table, we find that, even though the $y$-coordinates of the centers of the vortices remain almost the same, the $x$-abscissas of the centers of the vortices tend to move away from the original point, when we decrease the value of $Fe$. It can be explained by the fact that the stronger diffusion motion of the charges from the center towards the two sides drives the fluid to move to both sides (to be discussed with figure \ref{fig.r03Q}), so that the vortex extends farther away, resulting in a larger vortex size. Similar to the observation in the previous cases, the ratios of size and intensity between vortex 2 and vortex 3 are in good agreement with those in \cite{moffatt}, especially when the charge diffusion effect is strong (or $Fe$ is small).We present the distribution of charge density in figure \ref{fig.r03Q}, where the diffusion motion of charges can be observed clearly. From panel (a)-(d), we find that at smaller $ Fe $, more charges diffuse from the center to both sides. Panel (e) displays the charge density distribution at the horizontal line $ y=0.1 $, and it further illustrates that with the decrease of $ Fe $, the charge density in the center reduces and it expands towards both sides.

In terms of vortex intensity $v_{\theta}$ in the table, it seems that the value of $v_{\theta}$ first increases from $ Fe=100 $ to $ Fe=500 $ and then decreases from $ Fe=500 $ to $ Fe= 5\times10^3 $, which has the similar trend with the maximum vertical velocity $ |U_y|_{max}^s $. We then plot the maximum vertical velocity $ |U_y|_{max}^s $ versus $ Fe $ in a larger range of $ Fe $ at $ T=500 $, as shown in figure \ref{fig.FeUy}. We can observe from the figure that, when $ Fe $ is large (or small charge diffusion effect), $ Fe $ has an insignificant effect on the velocity. When $Fe$ is small (say from $ Fe = 100 $ to $ Fe = 1000 $), $ |U_y|_{max}^s $ increases with increasing $ Fe $ corresponding to the decreasing charge diffusion effect. Physically, since charge diffusion is a diffusive force after all, a small charge diffusion effect will hinder less the flow motion. So we observe a large $ |U_y|_{max}^s $ when the charge diffusion is small (or large $Fe$). On the other hand, charge diffusion can also contribute to the movement of the charges in the flow domain. A stronger charge diffusion effect will dispense the charged ions more to the other parts of the flow domain, leaving less charges in the center region where the $ |U_y|_{max}^s $ happens. That is probably why we observe a smaller $ |U_y|_{max}^s $ when $Fe$ is small (or the charge diffusion effect is strong). Thus, the effect of charge diffusion is complex and we indeed observe a maximal effect at an intermediate value of $Fe$ for $ |U_y|_{max}^s $. In general, the variation of $ |U_y|_{max}^s $ (from 7.53 to 7.62) in a large range of $Fe$ (from 100 to 5000) is small.

In a nutshell, the effect of charge diffusion can influence the formation and evolution of the Moffatt eddies. This is an additional factor compared to the situation in the original Stokes flows. The charge diffusion term was neglected in the numerical simulations of \cite{perri2021particle}. Our results indicate the generic existence of Moffatt-like eddies in multi-physical flows (EHD flow in our case) and their complex dependence on those parameters.

\subsection{Linear global stability analysis of the blade-plate EHD flows}\label{stability}

The above section details the numerical results of Moffatt-like eddies in the blade-plate EHD flow using DNS. Next, we will conduct global stability analyses of these steady solutions. Since the main flow motion occurs in the central region near the blade tip, where the eigenmodes also concentrate (to be presented), a small computational domain is adopted in this section in order for a higher computational efficiency. More specifically, we set $  H_p=5 $ and $  \Delta H_s=2 $ for $ T\in[500,2900]$, $  H_p=20 $ and $  \Delta H_s=10 $ for $ T\in[10000,40000]$, see figure \ref{fig.fig2a} for the definitions of $H_p,\Delta H_s$. Additionally, in the linear stability analysis, the disturbance generated near the blade tip and the original point will propagate to the farfield. Because of the finite size of the domain and the imposition of the boundary conditions, reflection of the linear waves at the domain boundary is inevitable. In order to minimize the reflection effect, we consider a sponge zone in the farfield, see again figure \ref{fig.fig2a}, following \cite{chevalier2007simson,appelquist2015global}. A series of validation cases have been carried out and are shown in Appendix B, including verifying the mesh independence and ensuring that the computational domain and the sponge region are large enough to not affect the results in the physical region around the origin. In this section, the parameters are $ R=0.05 $, $ C=5, M=50$ and $ Fe=5\times10^3 $ for most cases.

\begin{figure}
	\centering
	\subfigure{
		\begin{minipage}[h]{0.4\linewidth}
			\includegraphics[height=4cm,width=5cm]{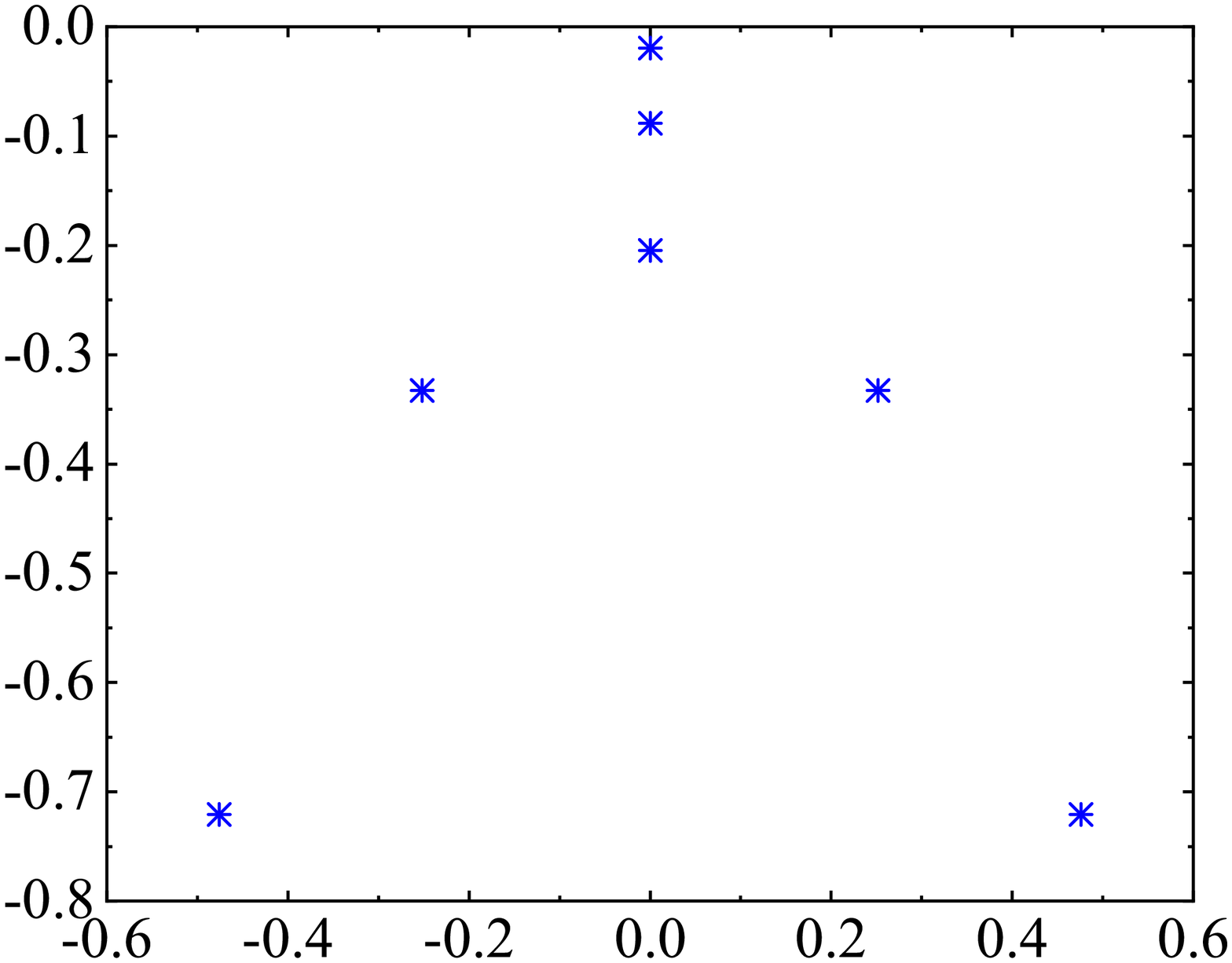}
			\put(-160,65){$ \omega_r $}
			\put(-65,-10){$ \omega_i $}
			\put(-165,105){(a)}
			\label{fig.spectra}
	\end{minipage}}
	\hspace{20pt}
	\subfigure{
		\begin{minipage}[h]{0.4\linewidth}
			\includegraphics[height=4cm,width=5cm]{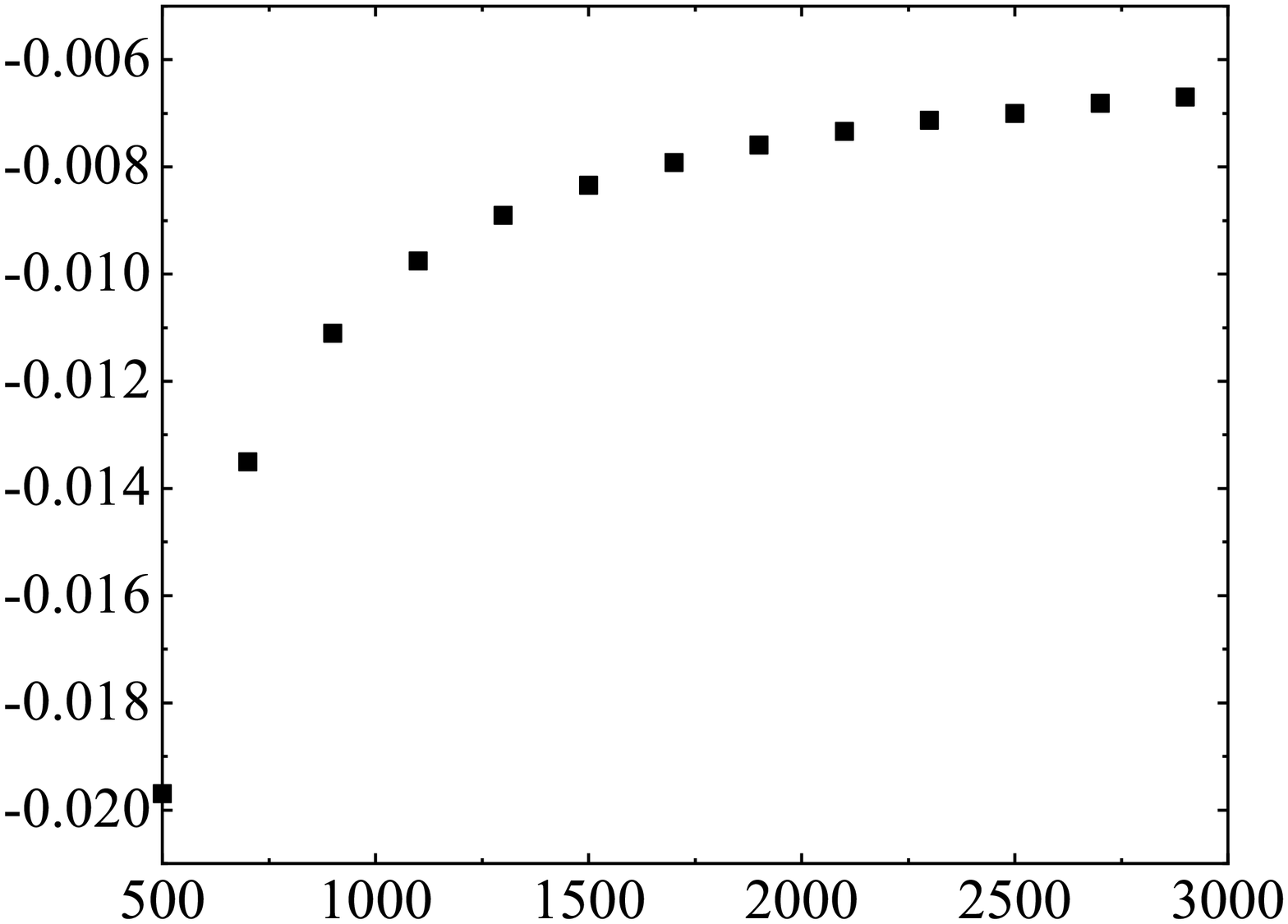}
			\put(-160,65){$ \omega_r $}
			\put(-65,-10){$ T $}
			\put(-160,105){(b)}
			\label{fig.gt}
	\end{minipage}}
	\hspace{20pt}
    \subfigure{
	\begin{minipage}[h]{0.4\linewidth}
		\includegraphics[height=4cm,width=5cm]{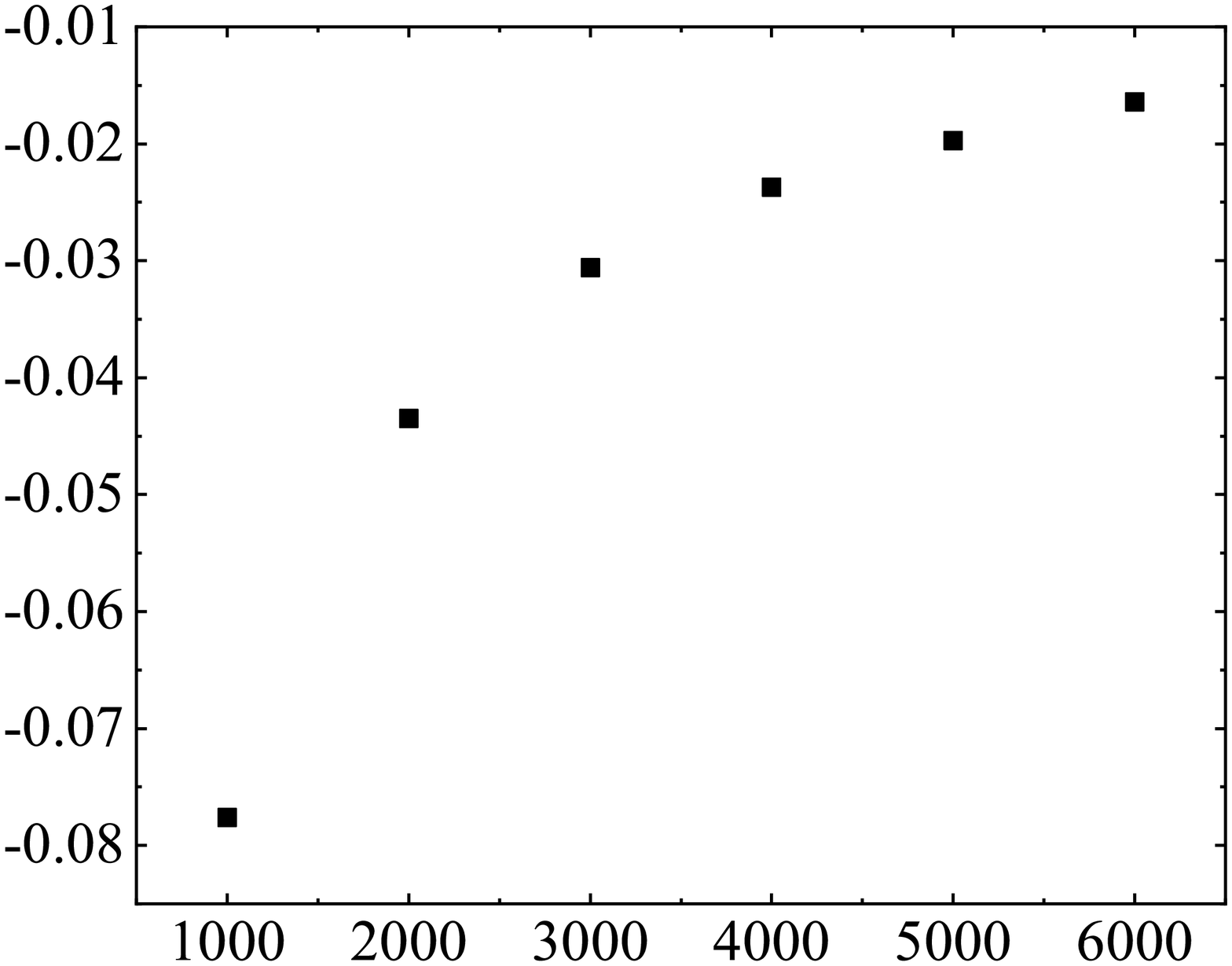}
		\put(-160,65){$ \omega_r $}
		\put(-65,-10){$ Fe $}
		\put(-160,105){(c)}
		\label{fig.Fe}
    \end{minipage}}
	\caption{(a) Least stable eigenvalues of the blade-plate EHD flow at $ T=500, Fe=5\times10^3 $. Growth rate of leading global mode in the blade-plate EHD flow versus (b) $ T $ at $ Fe=5\times10^3 $; (c) $ Fe $ at $ T=500 $. The other parameters are $ C=5, M=50, R=0.05$.}	
	\label{fig.arn}
\end{figure}

We use the steady state of the nonlinear simulation of the blade-plate EHD flow discussed above as the base flow $(\bar {{\Ub}},\bar{Q}, \bar{\phi}, \bar{\Eb}$), and carry out its global linear stability analyses by adding small disturbances on the base flow (section \ref{Linearisation}). We will present below the eigenvalues $\omega$ and eigenvectors $\tilde{\mathbf{f}}$, introduced in equation \ref{eigenproblem}.
We first plot the least stable eigenvalues at $ T=500 $ obtained using IRAM, as shown in figure \ref{fig.spectra}. The first seven least stable modes are presented and they are symmetric with respect to the line $ \omega_i=0 $. Figure \ref{fig.gt} presents the growth rate ($ \omega_r $) of least stable eigenvalue as a function of $ T $. All the growth rates are negative, indicating that perturbations decay  within the range of $ T=500\sim 2900 $ for the steady base state. In this range, the imaginary part of the least stable eigenvalue is zero, indicating that this mode is non-oscillating. It can also be seen from the figure that with the increase of $ T $, the decay rate decreases. It means that increasing $ T $ renders the linear system less stable, but the destabilising effect of increasing $T$ becomes weaker at larger $T$. 
Moreover, we note that in the global stability analysis of the confined impinging jet reported by \cite{meliga2011global}, the variation of the disturbance growth rate by increasing Reynolds number (their governing parameter) follows a similar trend as the one we observe here. This may be related to the similar flow phenomenon (impingement of the flow field) in the two cases. Then, we investigate the influence of $ Fe $ (the inverse of the charge diffusion coefficient) on the global stability of the EHD plume. As shown in figure \ref{fig.Fe}, increasing $ Fe $ (decreasing charge diffusion) destabilizes the linearised flow in the blade-plate EHD geometry. We mention in passing that for each data point in panels (b,c), a nonlinear DNS was first performed to obtain the base state and then a global linear stability analysis of the base state is conducted. This is because the base state is dependent on the parameters.

\begin{figure}
	\centering
	\subfigure{
		\begin{minipage}[h]{0.4\linewidth}
			\centering
			\includegraphics[height=3cm]{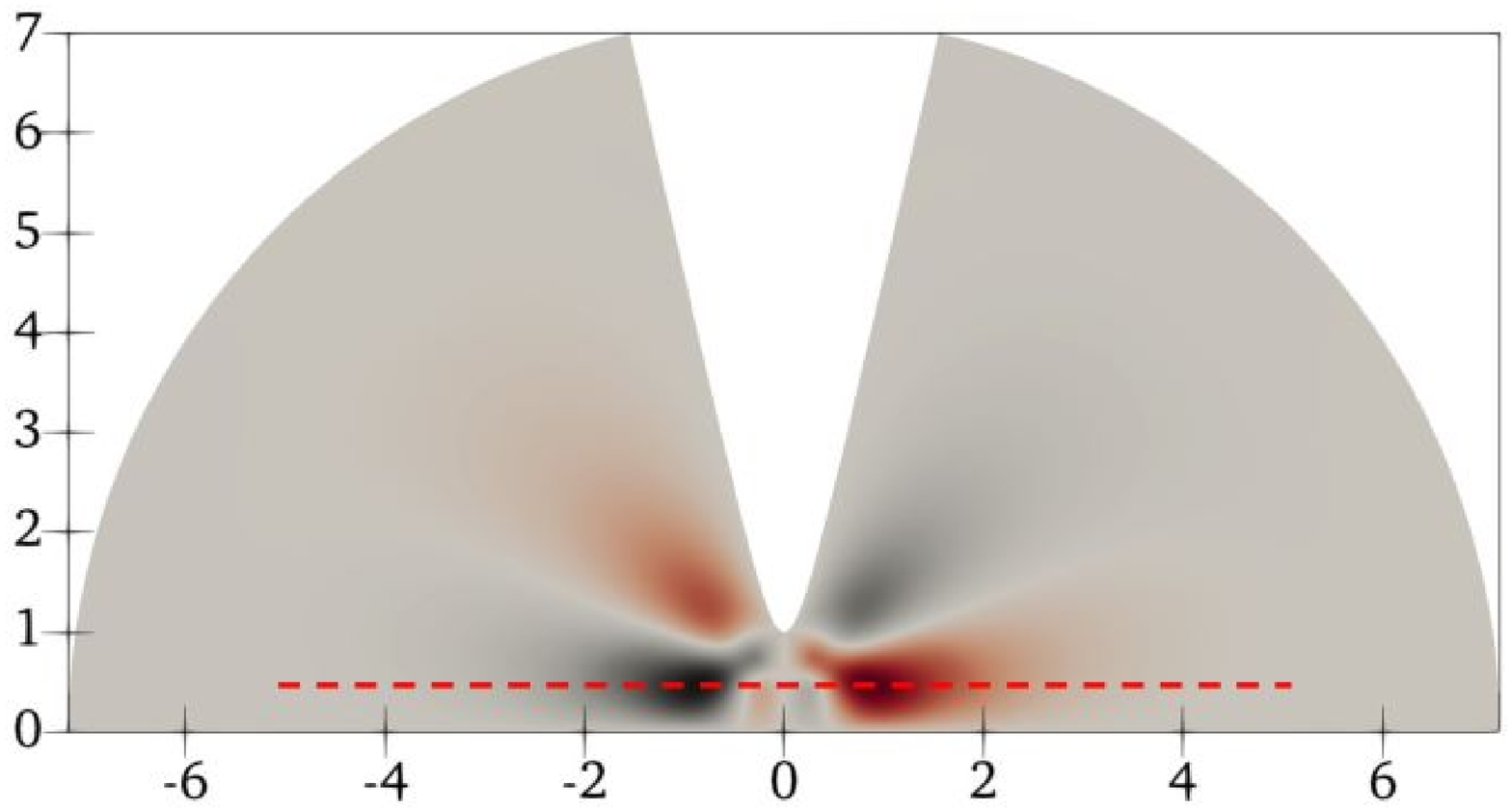}
			\put(-180,85){(a)}
			\put(-162,45){$ y $}
			\put(-77,-5){$ x $}
			\label{fig.LvxP}
	\end{minipage}}
	\hspace{20pt}
	\subfigure{
		\begin{minipage}[h]{0.4\linewidth}
			\centering
			\includegraphics[height=3.5cm]{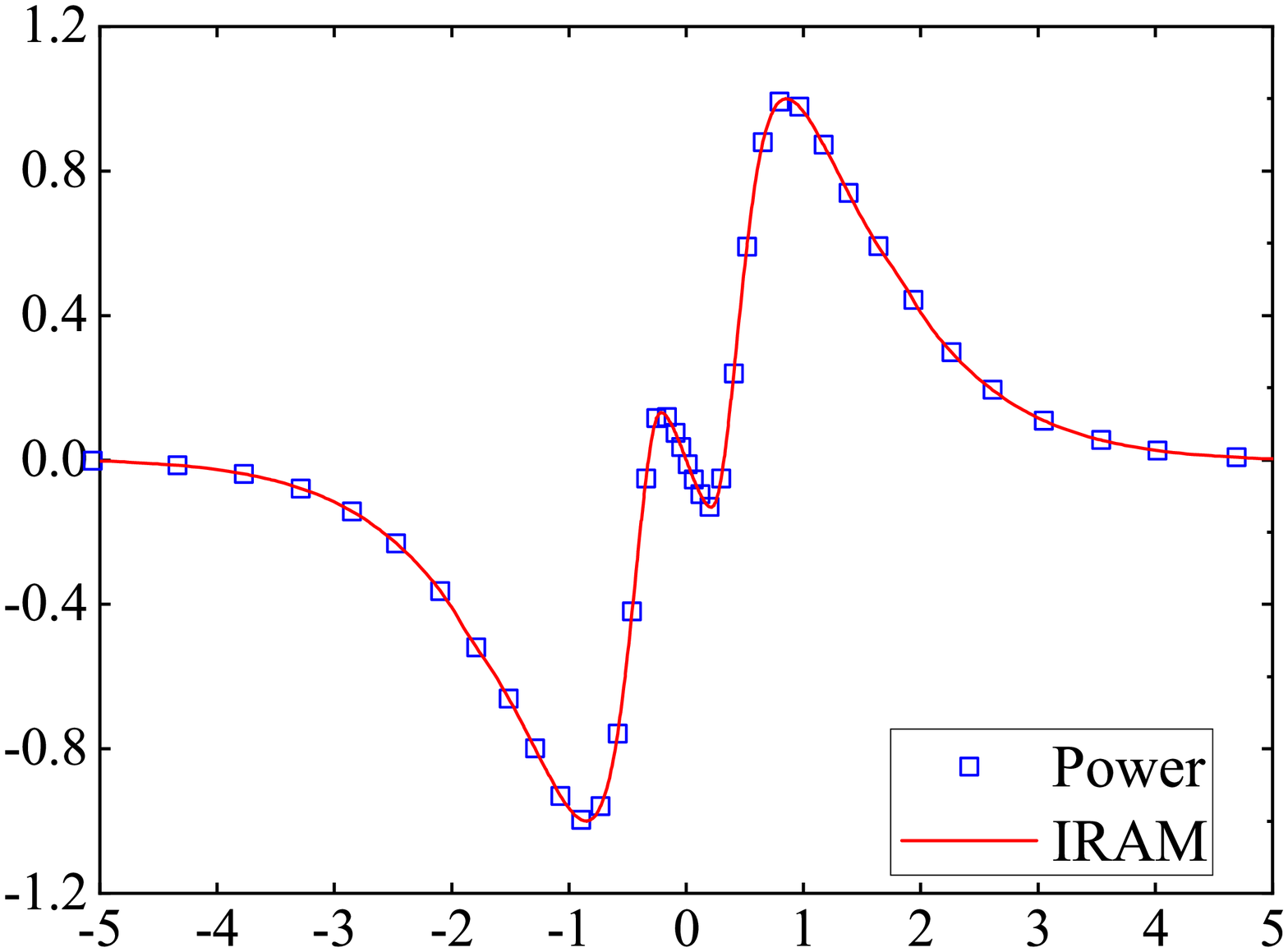}
			\put(-155,90){(b)}
			\put(-140,50){$ u $}
			\put(-65,-5){$ x $}			
			\label{fig.eigenvx}
	\end{minipage}}
	\hspace{20pt}
	\subfigure{
		\begin{minipage}[h]{0.4\linewidth}
			\centering
			\includegraphics[height=3cm]{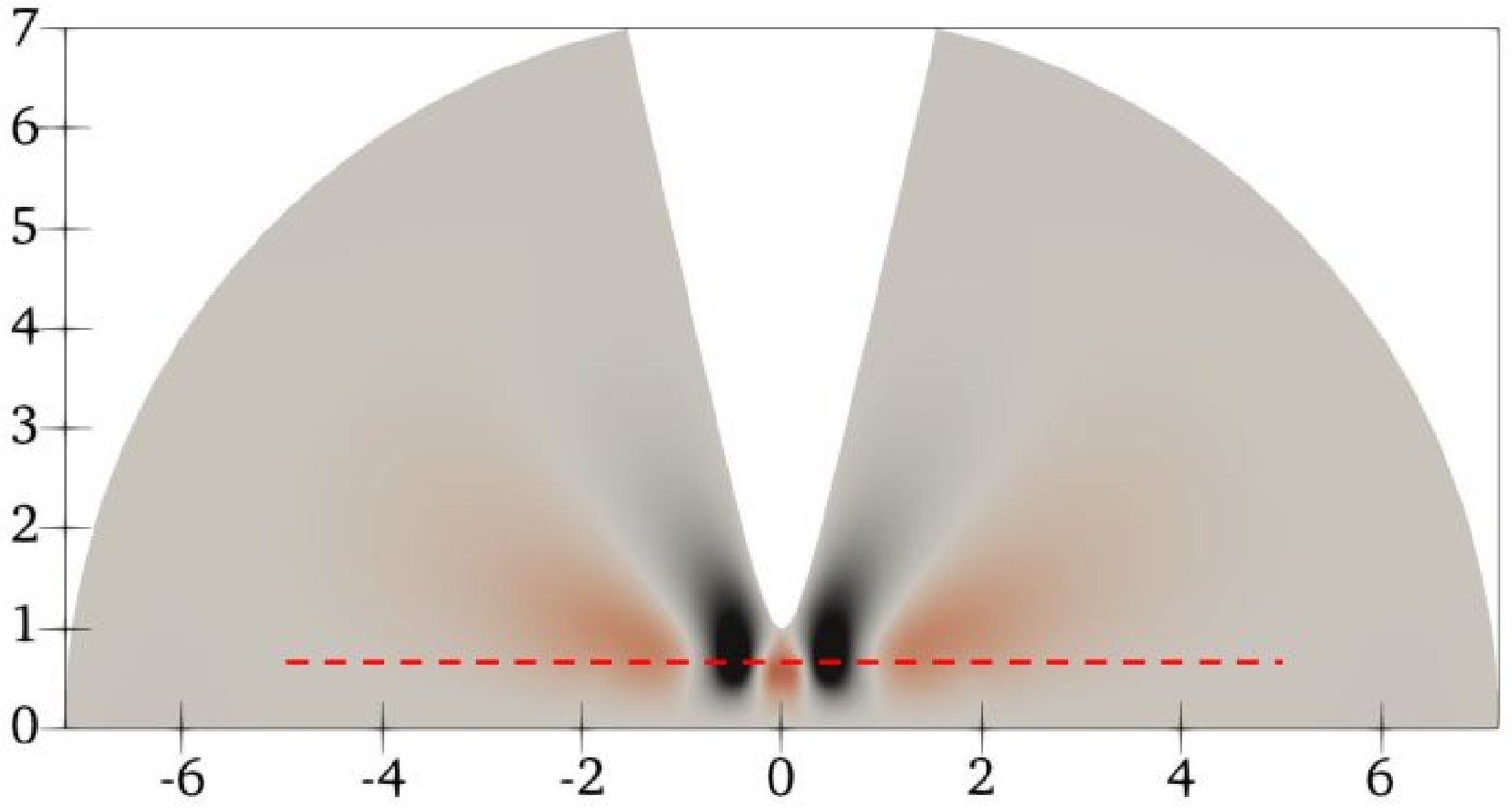}
			\put(-166,45){$ y $}
            \put(-77,-5){$ x $}
			\put(-183,85){(c)}
			\label{fig.LvyP}
	\end{minipage}}
	\hspace{20pt}
	\subfigure{
		\begin{minipage}[h]{0.4\linewidth}
			\centering
			\includegraphics[height=3.5cm]{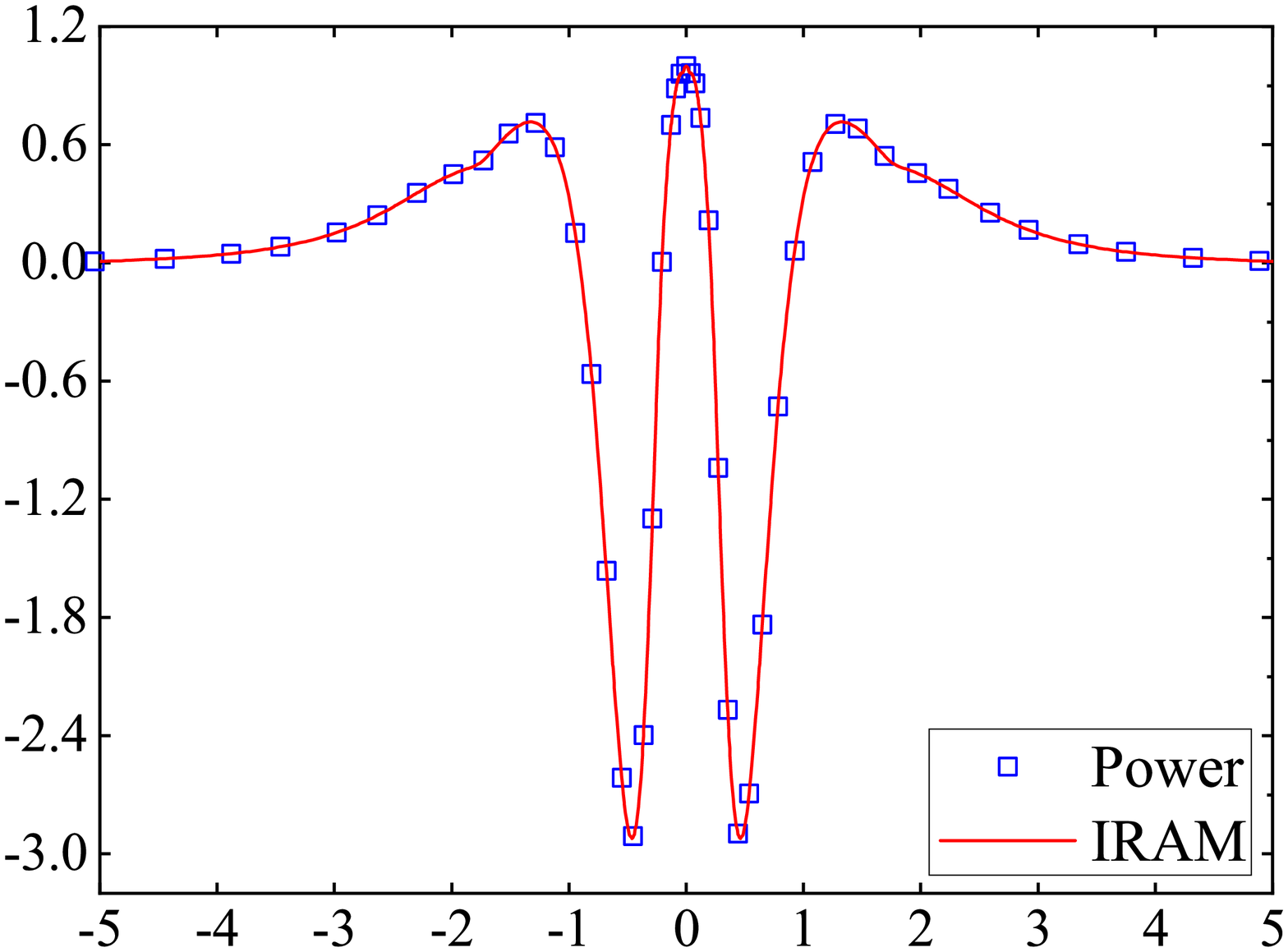}
			\put(-155,90){(d)}
			\put(-140,50){$ v $}
			\put(-65,-5){$ x $}	
			\label{fig.eigenvy}
	\end{minipage}}
	\hspace{20pt}
	\subfigure{
		\begin{minipage}[h]{0.4\linewidth}
			\centering
			\includegraphics[height=3cm,width=5.5cm]{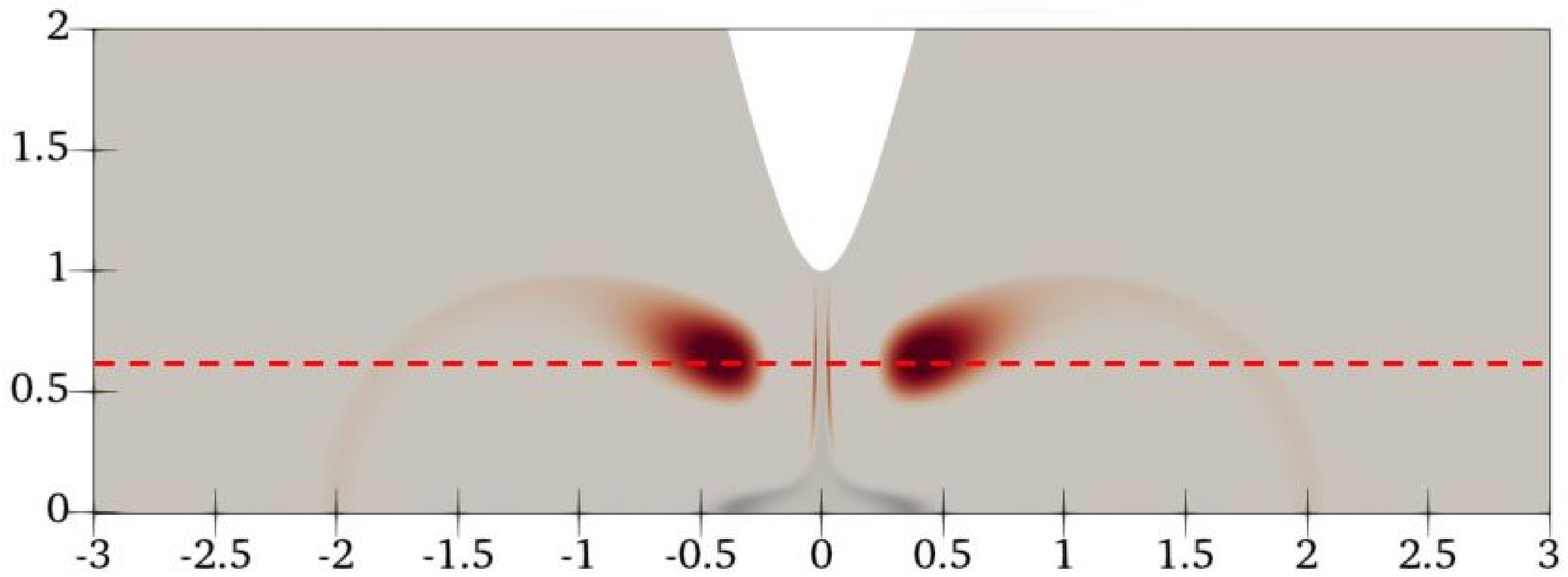}
			\put(-161,35){$ y $}
			\put(-77,-5){$ x $}
			\put(-180,85){(e)}
			\label{fig.LqP}
	\end{minipage}}
	\hspace{20pt}
	\subfigure{
		\begin{minipage}[h]{0.4\linewidth}
			\centering
			\includegraphics[height=3.5cm]{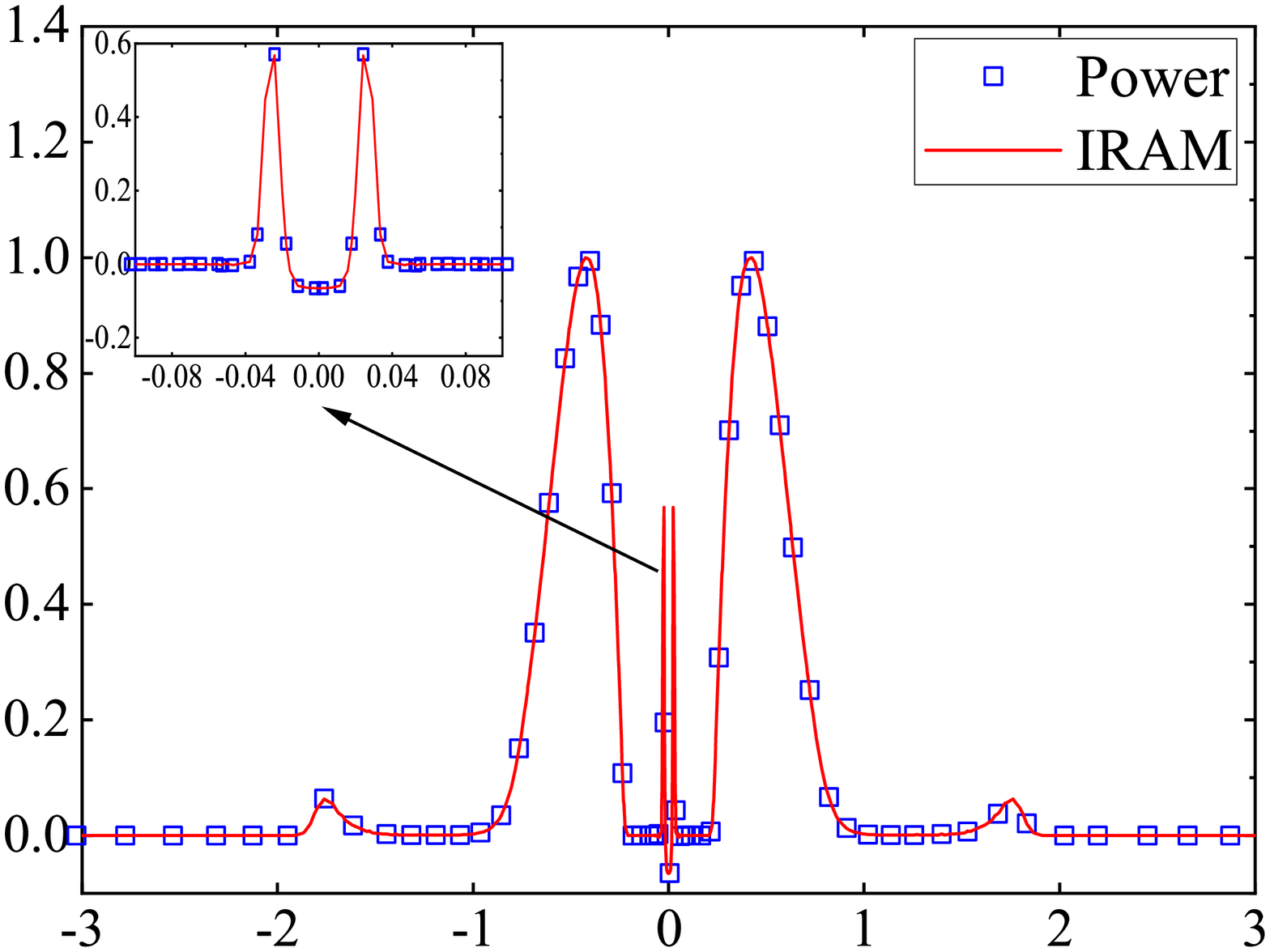}
			\put(-155,90){(f)}
			\put(-140,50){$ q $}
			\put(-65,-5){$ x $}	
			\label{fig.eigenq}
	\end{minipage}}
	\caption{Left column: Eigenvectors of most unstable mode for blade-plate EHD flow at $ T=500 $. Eigenvector of (a) horizontal velocity $ u $; (c) vertical velocity $ v $; (e) charge density $ q $. The eigenvectors can be arbitrarily scaled. Right column: Comparison between the results obtained by Power method (symbols) and IRAM (solid lines) of (b) $ u $ at $ y=0.5 $; (d) $ v $ at $ y=0.65 $; (f) $ q $ at $ y=0.65 $, and the maximum values of the eigenvectors are scaled to unity. The parameters are the same as figure \ref{fig.spectra}.}
	\label{fig.eigenF}
\end{figure}

\begin{figure}
	\centering
	\subfigure{
		\begin{minipage}[h]{0.4\linewidth}
			\centering
			\includegraphics[height=3.2cm]{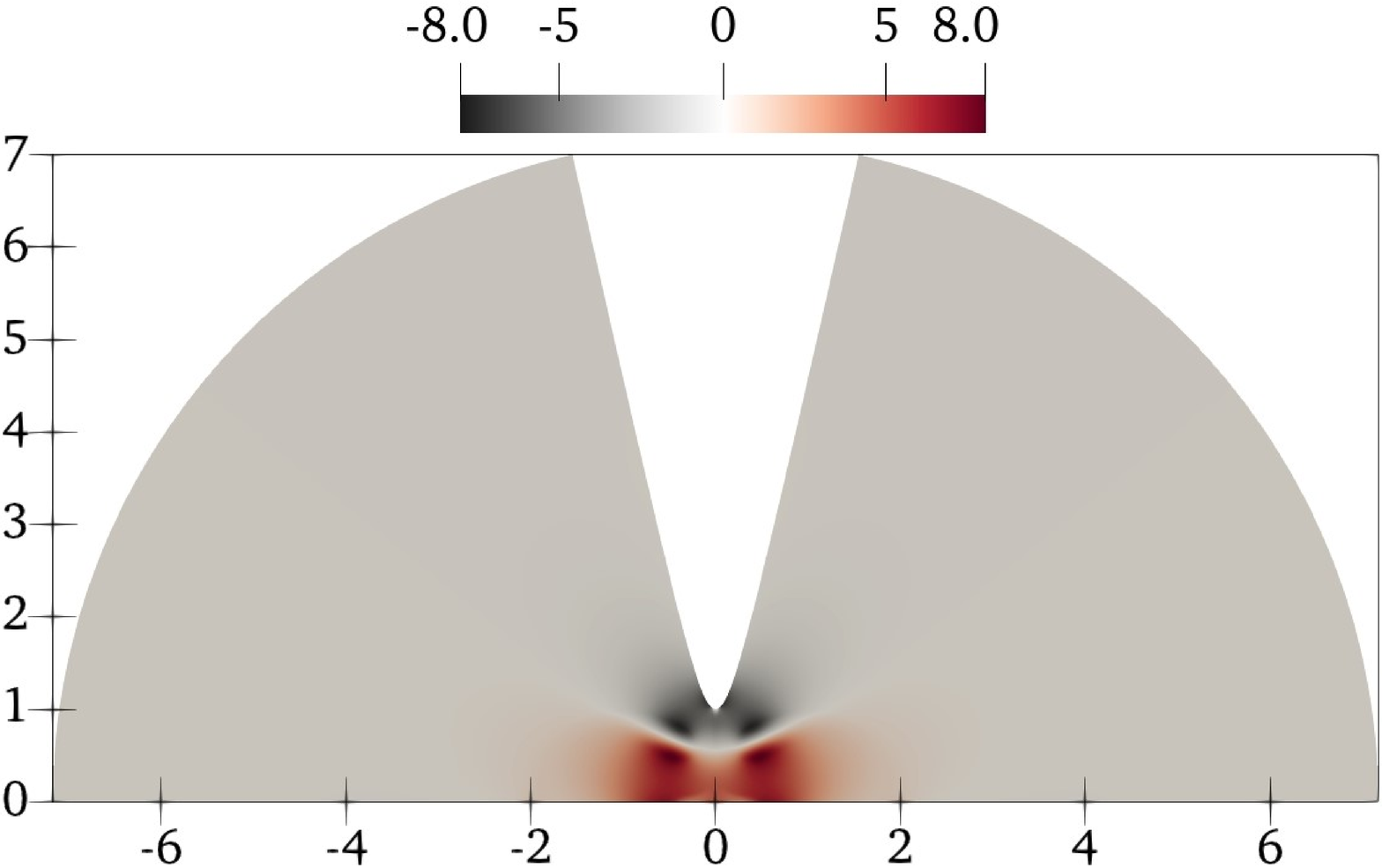}
			\put(-165,75){(a)}
			\put(-150,38){$ y $}
			\put(-71,-5){$ x $}
			\label{fig.LprP}
	\end{minipage}}
	\hspace{20pt}
	\subfigure{
		\begin{minipage}[h]{0.4\linewidth}
			\centering
			\includegraphics[height=3.2cm]{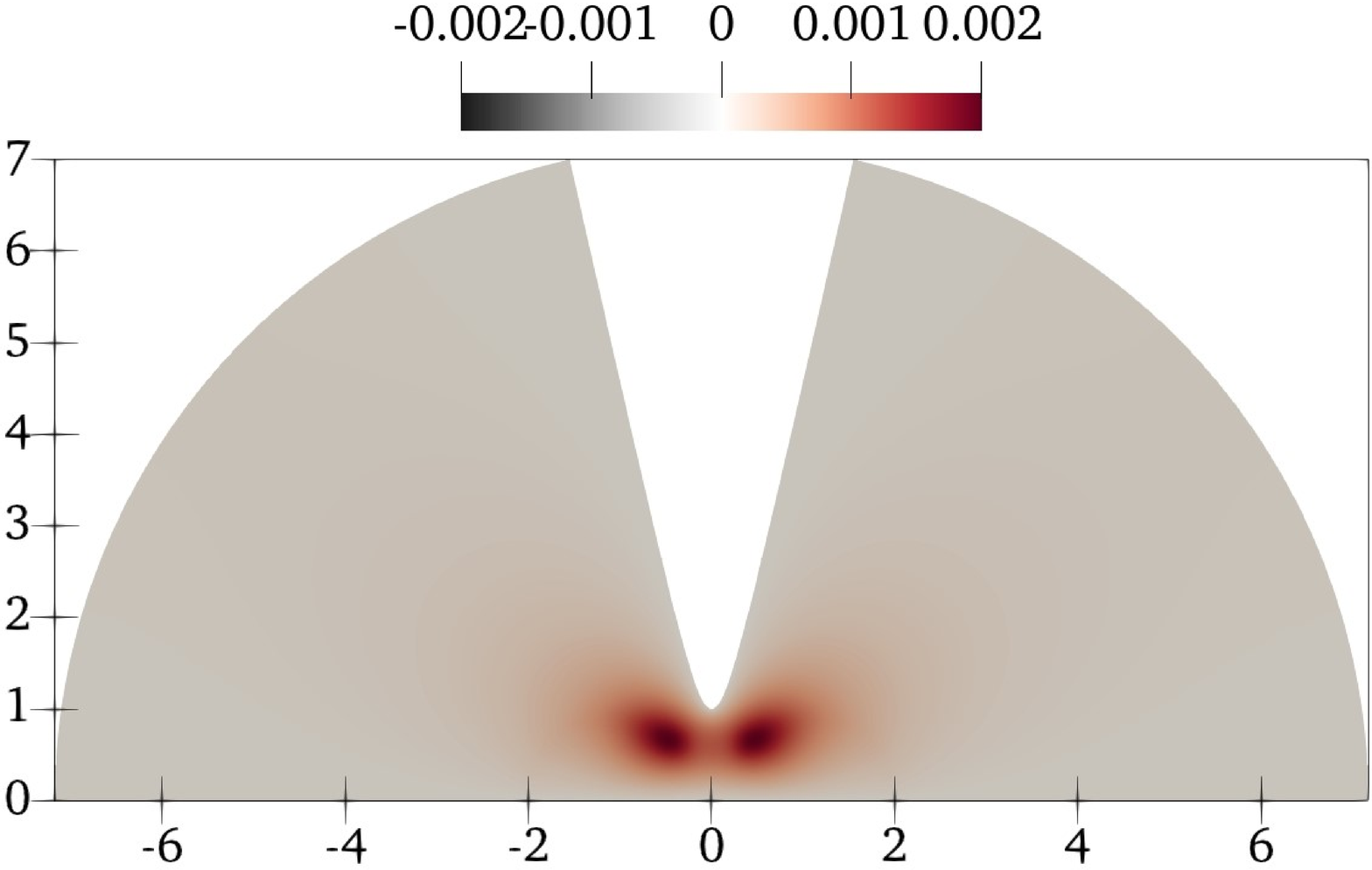}
			\put(-165,80){(b)}
			\put(-152,38){$ y $}
			\put(-73,-5){$ x $}
			\label{fig.LphiP}
	\end{minipage}}
	\caption{Eigenvectors of most unstable mode for blade-plate EHD flow at $ T=500 $. Eigenvector of (a) pressure perturbation $ p $; (b) electric potential perturbation $ \varphi $. The parameters are the same as figure \ref{fig.spectra}.}
	\label{fig.eigenFPr}
\end{figure}

The eigenvectors $ u $, $ v $ and $ q $ of the leading mode in the blade-plate EHD flow are presented in figure \ref{fig.eigenF}. In the right panel, we compare the leading eigenvectors at a certain line obtained by the power iteration method and the IRAM. The two agree with each other, which further validates our linear computations. We can see from the left panel that the perturbations are symmetric with respect to the central line $x=0$. This is a trivial observation because of the symmetry of the base flow in our case. In other flows, a self-excited axisymmetric mode was observed in the confined thermal plume by \cite{lesshafft2015linear}. In addition, it has been found that the leading mode of confined impinging jet is antisymmetric \citep{meliga2011global}. In figure \ref{fig.eigenF}(e), it is worth noting that in addition to the perturbation of charge density distribution in the central region resembling the nonlinear structure (figure \ref{fig.r03charge1}), there are also two tadpole-shaped structures on both sides. This special structure may stem from the strong convection nature of the charge density equation (Eq. \ref{eq.line3}). We find that the positions of the head of the 'tadpole' ($ [\pm 0.42,0.65] $) are approximately the same as the maximum value of the positive vertical velocity in the base flow, that is, the position where the fluids move up the fastest after impinging the plate, see figure \ref{fig.uyy05}. Additionally, in figure \ref{fig.LvyP} and \ref{fig.eigenvy}, we notice that the vertical velocity disturbance also has a local large absolute value. By examining the pressure perturbation, we find that this position is where the positive and negative pressure converts, as shown in figure \ref{fig.LprP}. Furthermore, the distribution of electric potential perturbation $\varphi$ is shown in figure \ref{fig.LphiP},  which is related to the perturbation $ q $. Note that in all the eigenvectors that we have studied here, the most important structures are located near the blade tip region, validating the consideration of using a smaller computational domain in the linear analysis.

\begin{figure}
	\centering
	\begin{minipage}[h]{0.4\linewidth}
		\centering
		\includegraphics[height=3cm,width=6cm]{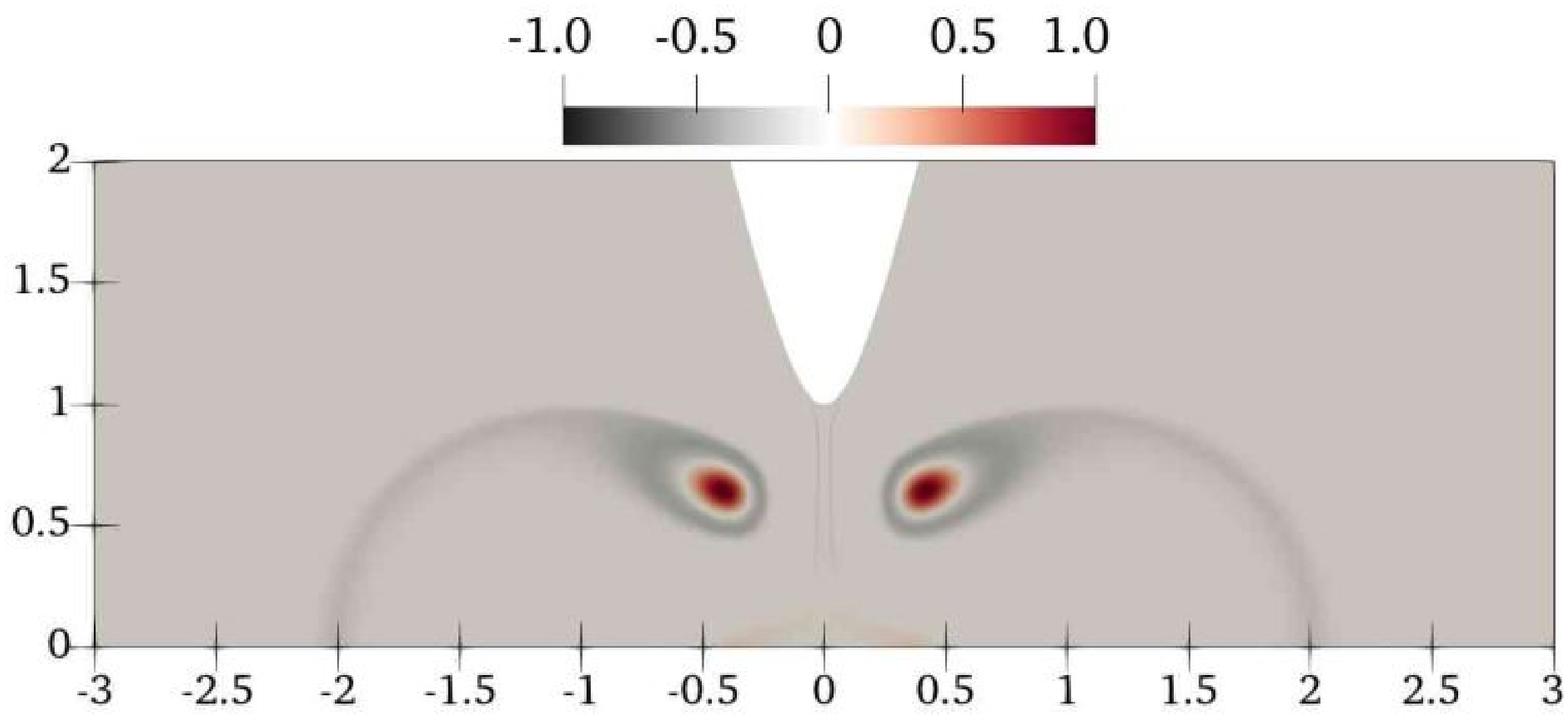}
		\put(-180,70){(a)}
		 \put(-177,35){$ y $}
        \put(-85,-5){$ x $}		
		\label{fig.mode2q}
	\end{minipage}
	\hspace{20pt}
	\begin{minipage}[h]{0.4\linewidth}
		\centering
		\includegraphics[height=3cm,width=6cm]{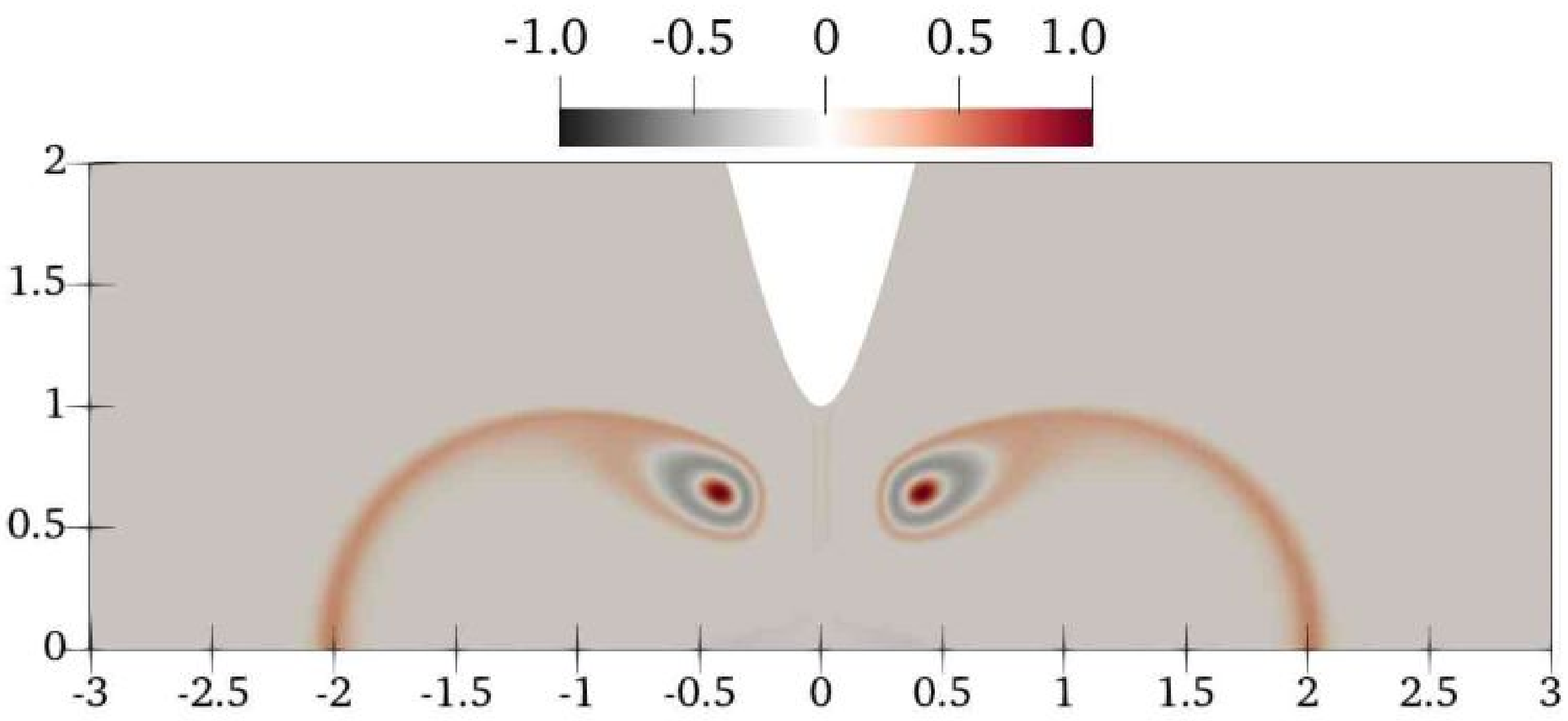}
		\put(-180,70){(b)}
        \put(-177,35){$ y $}
        \put(-85,-5){$ x $}	
		\label{fig.mode3q}
	\end{minipage}
	\hspace{20pt}
	\begin{minipage}[h]{0.4\linewidth}
		\centering
		\includegraphics[height=3cm,width=6cm]{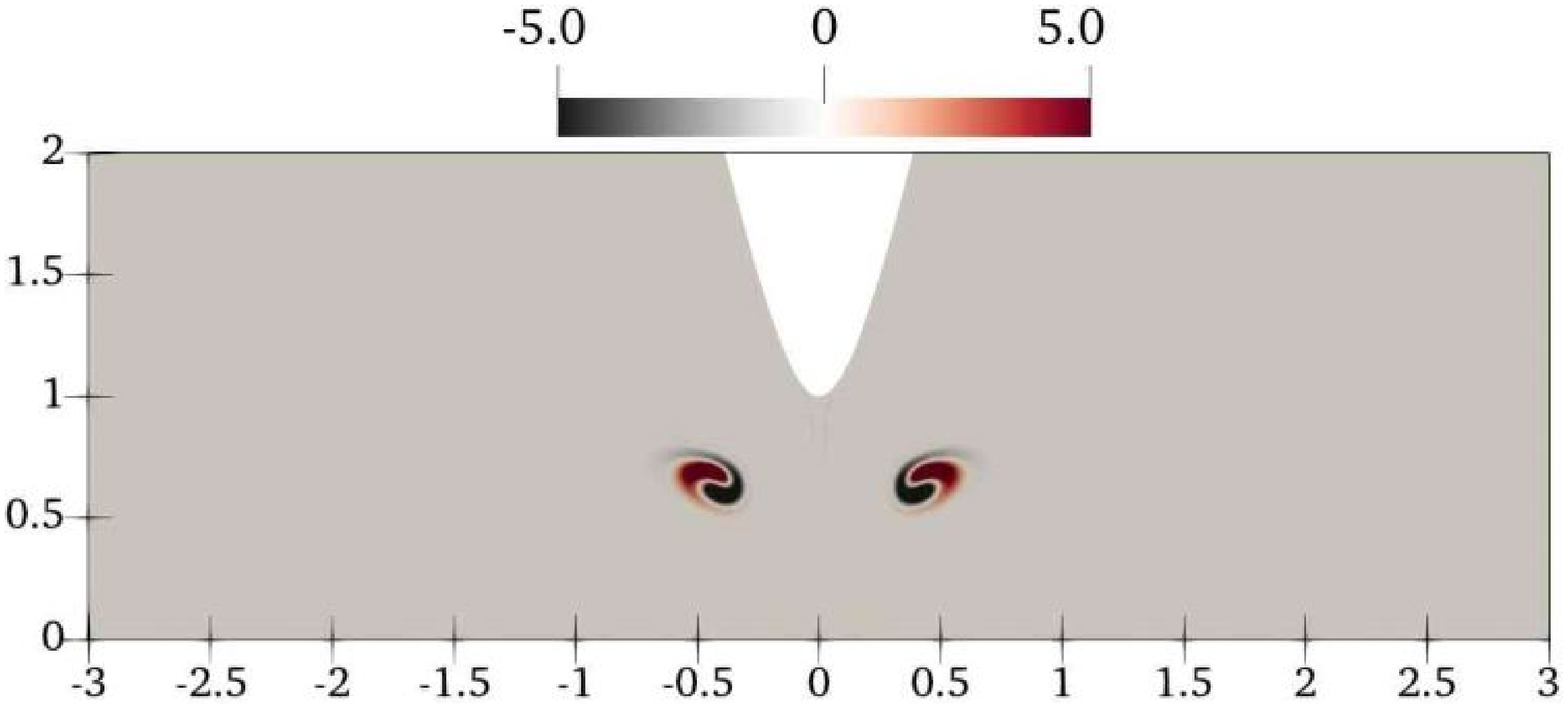}
		\put(-180,70){(c)}
        \put(-177,35){$ y $}
        \put(-85,-5){$ x $}	
		\label{fig.mode4pq}
	\end{minipage}
	\hspace{20pt}
	\begin{minipage}[h]{0.4\linewidth}
		\centering
		\includegraphics[height=3cm,width=6cm]{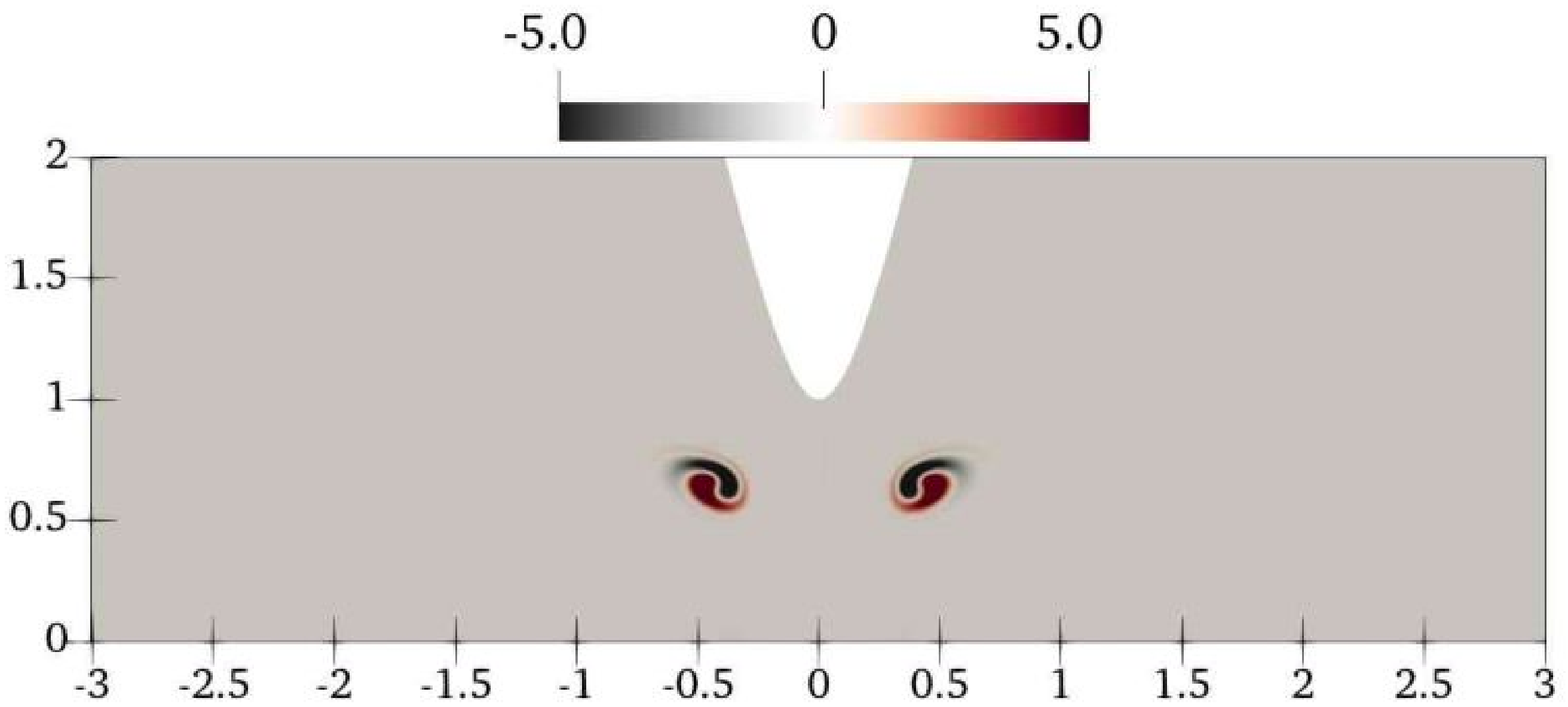}
		\put(-180,70){(d)}
        \put(-177,35){$ y $}
        \put(-85,-5){$ x $}	
		\label{fig.mode4nq}
	\end{minipage}
	\caption{Eigenvectors of charge density of mode 2-mode 5 for blade-plate EHD flow at $ T=500 $. The eigenvalues of (a)-(d) are (a) -0.0883+0i; (b) -0.245+0i; (c) -0.333+0.252i; (d) -0.333-0.252i. The parameters are the same as figure \ref{fig.spectra}.}
	\label{fig.mode}
\end{figure}

\begin{figure}
	\centering
	\begin{minipage}[h]{0.4\linewidth}
		\centering
		\includegraphics[height=4cm]{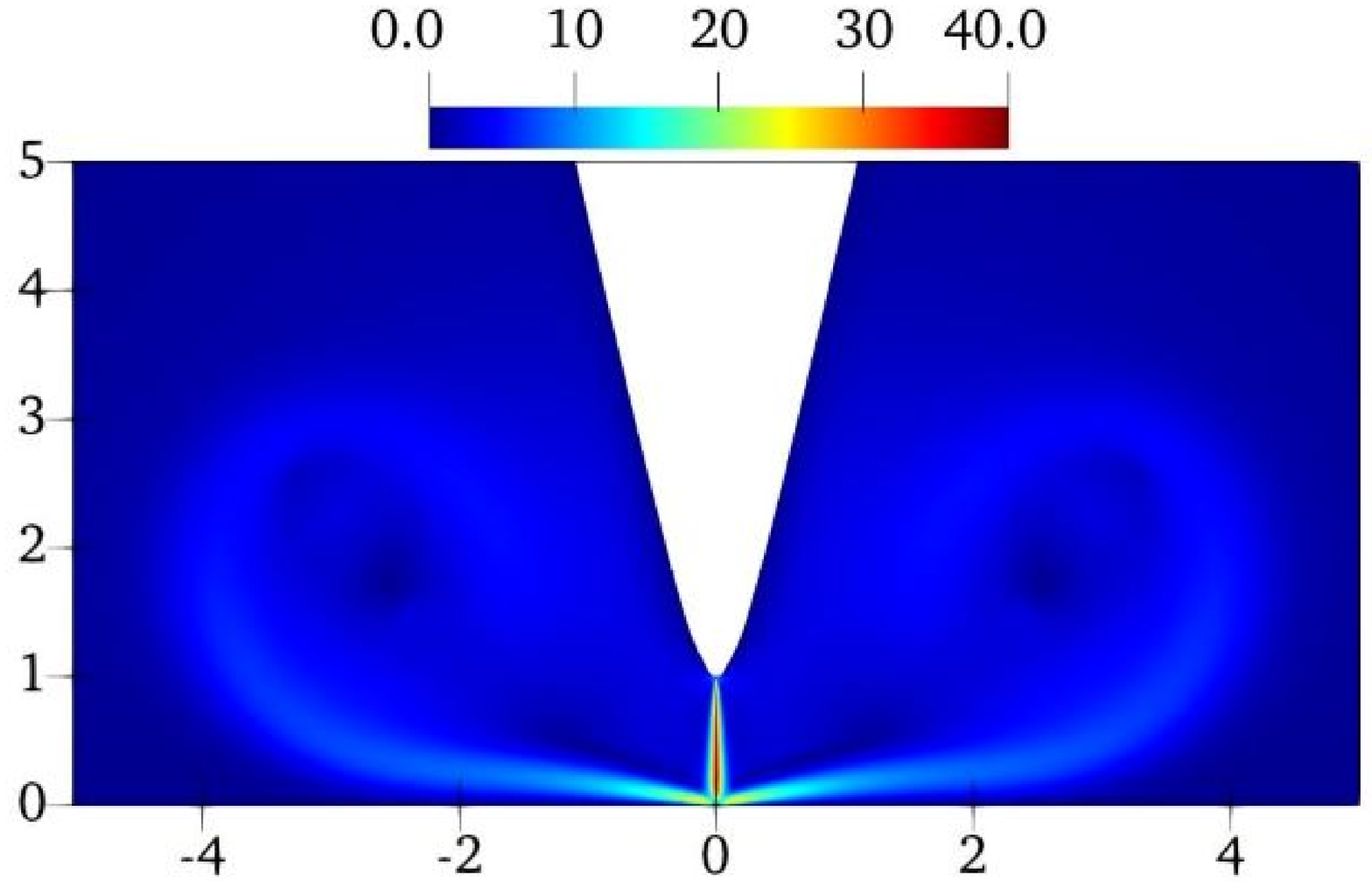}
		\put(-180,90){(a)}
		\put(-177,49){$ y $}
		\put(-85,-5){$ x $}		
		\label{fig.base}
	\end{minipage}
	\hspace{20pt}
	\begin{minipage}[h]{0.4\linewidth}
		\centering
		\includegraphics[height=4cm]{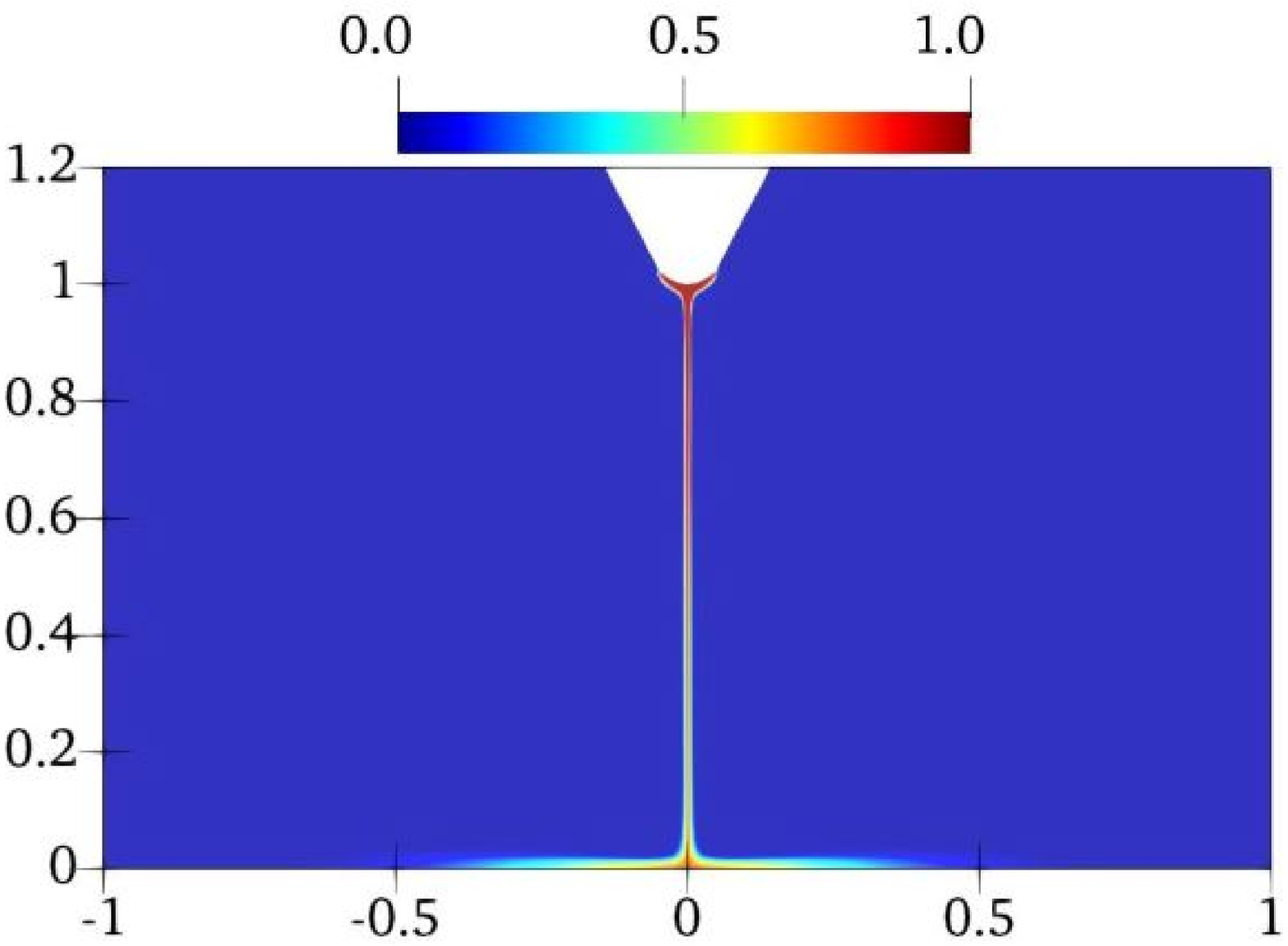}
		\put(-165,90){(b)}
		\put(-163,49){$ y $}
		\put(-75,-5){$ x $}	
		\label{fig.baseq}
	\end{minipage}
	\caption{The steady unstable base flow for blade-plate EHD flow at $ T=4\times10^4 $. (a) Velocity field; (b) charge density distribution. The other parameters are the same as those in figure \ref{fig.spectra}.}
	\label{fig.sfd}
\end{figure}

\begin{figure}
	\centering
	\begin{minipage}[h]{0.4\linewidth}
		\centering
		\includegraphics[height=3cm,width=6cm]{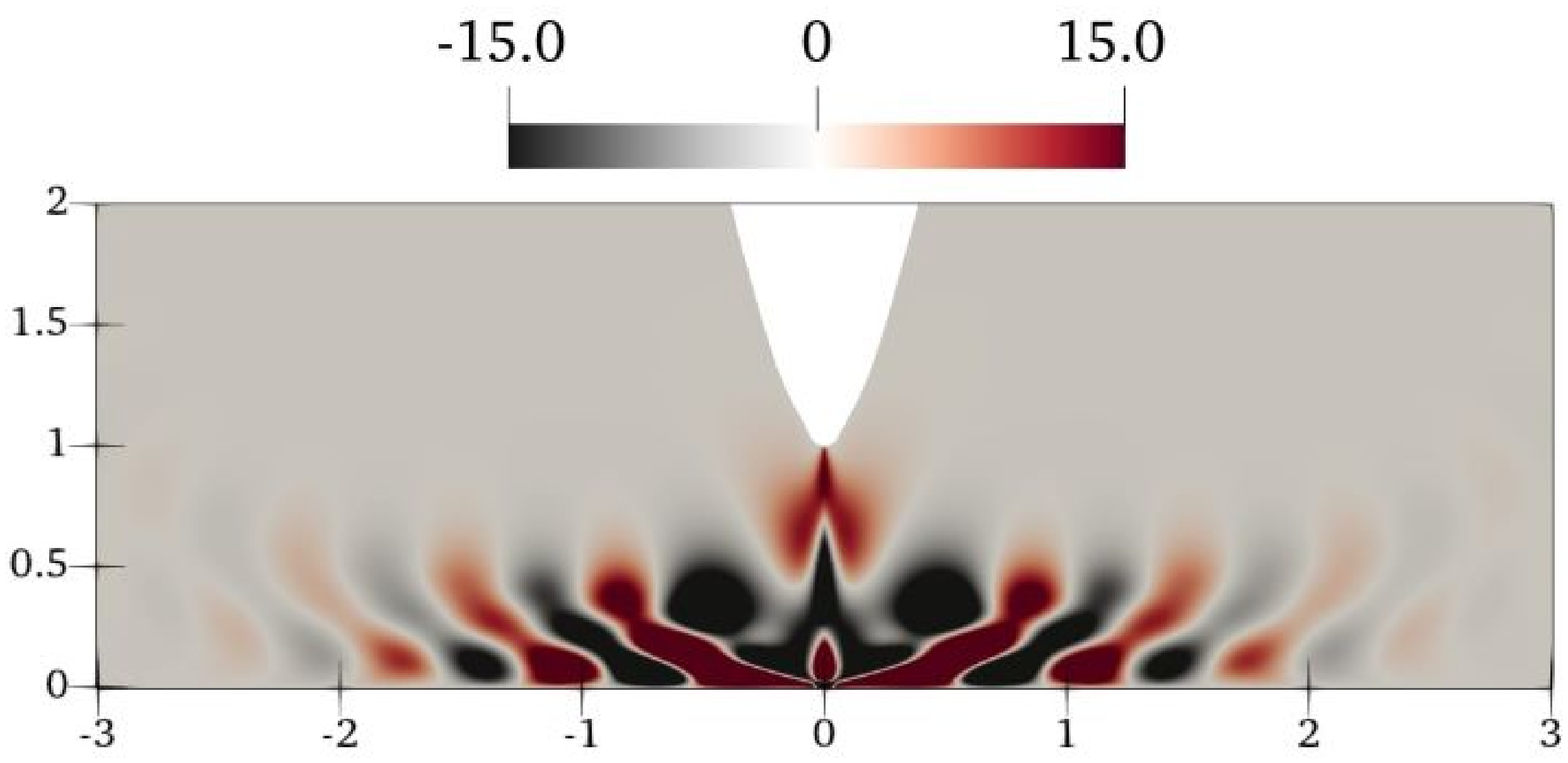}
		\put(-180,70){(a)}
        \put(-177,35){$ y $}
        \put(-85,-5){$ x $}		
		\label{fig.OAvx}
	\end{minipage}
	\hspace{20pt}
	\begin{minipage}[h]{0.4\linewidth}
		\centering
		\includegraphics[height=3cm,width=6cm]{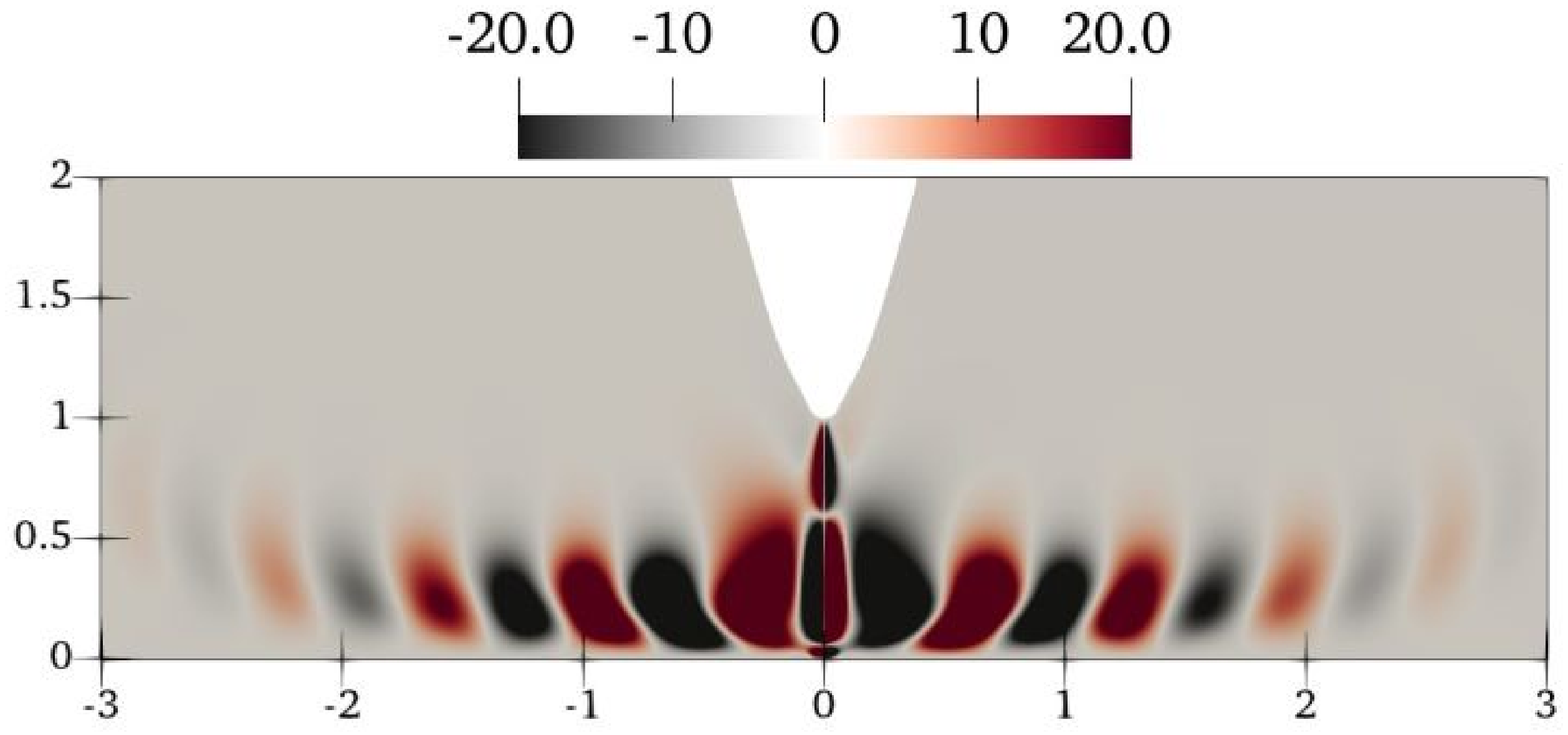}
		\put(-180,70){(b)}
        \put(-177,35){$ y $}
        \put(-85,-5){$ x $}		
		\label{fig.OAvy}
	\end{minipage}
	\hspace{20pt}
	\begin{minipage}[h]{0.4\linewidth}
		\centering
		\includegraphics[height=5cm]{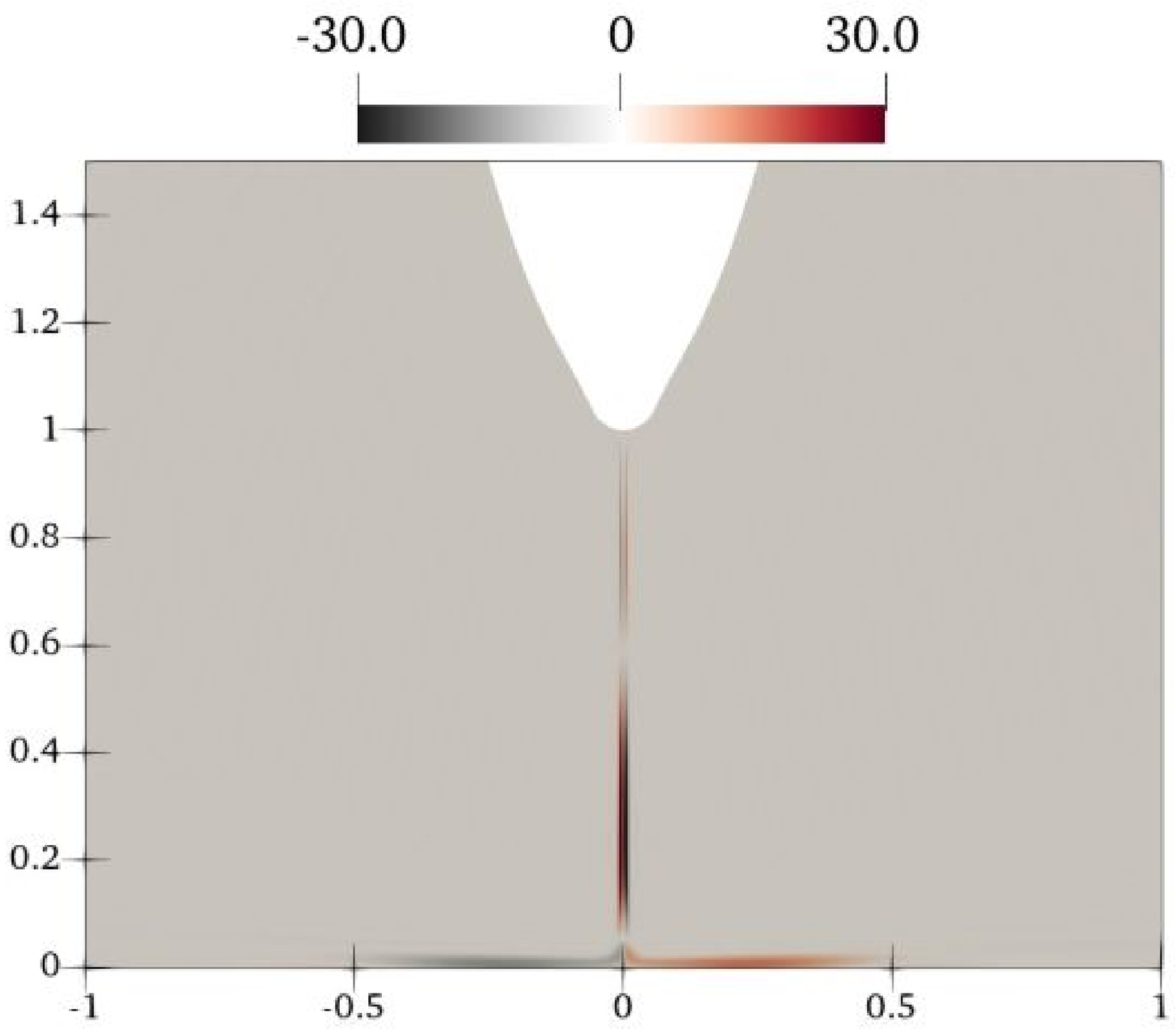}
		\put(-180,120){(c)}
		\put(-172,60){$ y $}
		\put(-80,-5){$ x $}	
		\label{fig.OAq}
	\end{minipage}
	\caption{Eigenvectors of most unstable mode (imaginary part of the eigenvalue is positive) for blade-plate EHD flow at $ T=4\times10^4 $. Eigenvector of (a) horizontal velocity $ u $; (b) vertical velocity $ v $; (c) charge density $ q $. The other parameters are the same as those in figure \ref{fig.spectra}.}
	\label{fig.Oscmode}
\end{figure}

Figure \ref{fig.mode} displays the charge density eigenvectors of mode 2 to mode 5 at $ T=500 $. It can be seen that the second and third eigenmodes are similar to the leading one (figure \ref{fig.LqP}), which are all stationary eigenmodes (see figure \ref{fig.spectra}), and the difference is only in the positive and negative regions around the heads of the tadpoles. Panels (c) and (d) show the eigenvectors of charge density for a pair of complex-conjugate eigenvalues with equal real parts and opposite imaginary parts. It is interesting to see that these oscillating eigenmodes (with non-zero imaginary part) possess two symmetrical Tai Chi shapes on both sides of the central axis.

Next, we also discuss an unstable flow at a higher $T=4\times10^4 $. This is the oscillating flow as we have analysed in figure \ref{fig.osc}. In order for its stability analysis, we solve for its steady unstable solution to the nonlinear governing equation using the selective frequency damping (SFD) method \citep{aakervik2006steady} (which damps the high-frequency content in the equation to stabilise the flow). The unsteady base flow and the base charge density are presented in figure \ref{fig.sfd}. We can see that now the charge density is a single steady jet impinging on the plate electrode, unlike the oscillating behaviour in figure \ref{fig.osc}. One can understand that this unstable flow will transition to another flow state, which can be analysed by a stability analysis. Such a methodology can also be applied to analyze the transient flow phenomenon in \cite{perri2020electrically}. We present its eigenvectors of the leading mode in figure \ref{fig.Oscmode}. It can be seen that the patterns of eigenvectors of the leading mode in this case are quite different from those in the low $T$ case. The eigenvalues for the leading modes at $T=4\times10^4 $ are calculated as $ 3.364 \pm 61.592i $, meaning that the linear growth rate is 3.364 and that the flow is indeed unstable. In addition, the eigenfrequency (which is the imaginary part of the eigenvalue) is 61.592. This value can also be approximately related to the DNS results in figure \ref{fig.T4e4}, which indicated that the dominant circle frequency of the flow oscillation is $9.89\approx61.592/(2\pi)$.

\begin{figure}
	\centering
	\begin{minipage}[h]{0.4\linewidth}
		\centering
		\includegraphics[height=4.5cm,width=5.5cm]{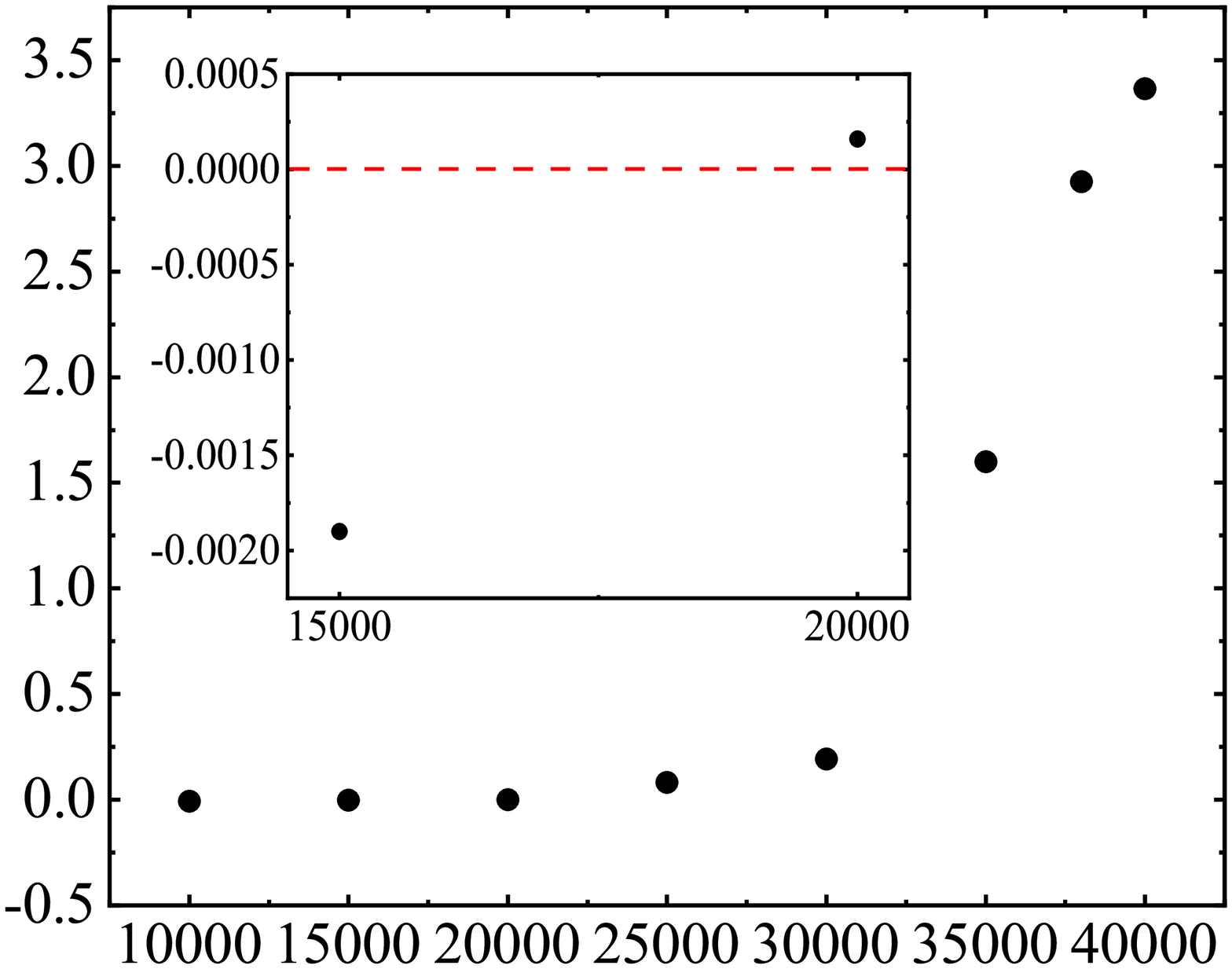}
		\put(-180,120){(a)}
		\put(-170,65){$ \omega_r $}
		\put(-75,-10){$ T $}		
		\label{fig.grLT}
	\end{minipage}
	\hspace{20pt}
	\begin{minipage}[h]{0.4\linewidth}
		\centering
		\includegraphics[height=4.5cm,width=5.5cm]{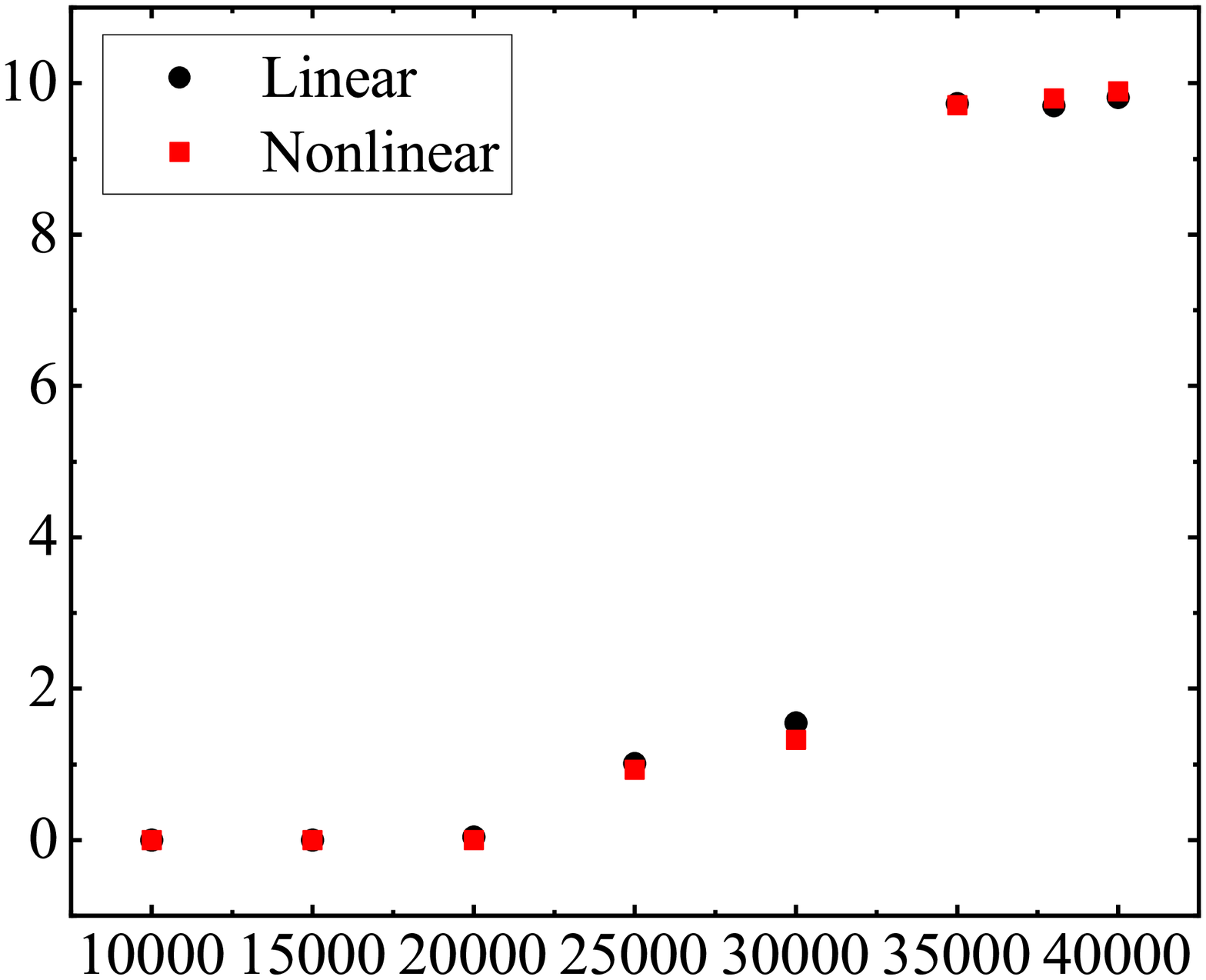}
		\put(-180,120){(b)}
		\put(-165,45){\rotatebox{90}{Frequency}}
		\put(-75,-10){$ T $}		
		\label{fig.frLT}
	\end{minipage}
	\caption{(a) Growth rate and (b) frequency of leading global mode in the blade-plate EHD flow of $ T $ in the interval $[10000,40000]$. The other parameters are $ C=5, M=50, R=0.05$ and $ Fe=5\times10^3 $.}
	\label{fig.LT}
\end{figure}

Finally, we present in figure \ref{fig.LT} the results of linear stability analysis at large $ T\in[10000,40000]$ to showcase the variation of the growth rate and the eigenfrequency in a large range of $T$ covering the linear critical condition. The growth rate versus $ T $ is displayed in panel (a), which shows that the critical $ T_{c}  $ for the onset of global linear instability is between 15000 to 20000 and very close to $T=20000$. When $T$ is larger than this critical condition, the time-periodic oscillation occurs (see also panel (b)). Panel (b) depicts the frequency of the leading global mode at different $ T $, obtained by both nonlinear simulations and linear stability analyses. Note that the eigenfrequency is  $ 2\pi $ times the actual frequency of oscillation. We can observe that the two methods generate consistent results. A peculiar phenomenon occurs of a sudden increase of the frequency when $T$ is varied from 30000 to 35000. To explain this, we resort to figure \ref{fig.osc} where we have plotted the time series of the flow oscillation at a point (0,0.5) and the representative flow snapshots. We can clearly observe that when $T=30000$, the oscillation is slow and almost monotonic with a single frequency from figure \ref{fig.T3e4}. The amplitude of the oscillation is also small. On the other hand, when $T$ is increased to 40000, from figure \ref{fig.T4e4}, we can see that the oscillation is featured by 2 dominant frequencies and is also more drastic. The charge beam strikes the flat plate forcefully and triggers nonlinear effects in the flow, as evidenced by the second spike in the FFT result in the inset of figure \ref{fig.T4e4}. From these observations, we conclude that the sudden increase of the frequency seems to more likely stem from the confinement effect in the center region. That is, at a larger $T$, the stronger charge jet will hit the flat plate more violently and get reflected, leading to the more drastic oscillation with an increased frequency. This phenomenon may deserve a further investigation in the future.

\section{Conclusions} \label{Conclusions}

In this work, we performed numerical simulations and conducted linear global stability analyses of the Moffatt-like eddies in 2-D blade-plate EHD flows, motivated by the recent experimental and numerical work of \cite{perri2020electrically,perri2021particle} in a different setting. Driven by a unipolar strong injection, an impinging flow motion occurs, issuing from the blade tip (subjected to a high voltage) to the grounded flat plate electrode due to the Coulomb force. The impingement of the EHD flow on the plate helps to form vortices in the space between the two electrodes. The vortices are studied and compared to the theoretical results of \cite{moffatt}, despite the different geometries and the additional Coulomb force in our EHD flow. The existence of the Moffatt-like eddies in the EHD flow has been proven and their characteristics have been investigated. 

We first presented the evolution of the EHD flow and its nonlinear behaviour. The results show that the first pair of vortices near the tip are formed due to the charge injection under the effect of the Coulomb force, and two larger pairs of vortices further away from the corner are formed because of the viscous force in the flow. We then analysed the sequence of eddies (in the current work, three vortices are resolved) using the case of the inter-electrode angle being $ 61.3^\circ $ as an example. Our quantitative results indicate that the ratios of size and intensity of the two successive eddies in the farfield (the second and third vortices) can be compared favourably to the theoretical results of \cite{moffatt}. Differences exist for the ratios of size and intensity of the two successive eddies in the nearfield, which is due to the unclosed corner in our geometry. 

We also studied the influence of the inter-electrode angle by changing the radius of the curvature of the hyperbolic blade to further validate the properties of EHD Moffatt-like eddies. We found that the above conclusion (that the ratios of size and intensity between the second and third vortices agree well with the results in \cite{moffatt}) is generally valid for all the inter-electrode angles investigated in this work. In addition, we also investigated the effect of the electric field intensity ($ T $) on the Moffatt-like eddies. Our results show that increasing $ T $ renders the centers of the vortices closer to the corner and increases the intensity of vortices. In addition, at a sufficiently large $ T $, a (periodic) flow oscillation occurs, meaning that the flow may transition to another type of state. This result is qualitatively consistent with the experimental observation of \cite{perri2020electrically} who also reported a transient flow experiencing bifurcation. Furthermore, we found that increasing the charge diffusion effect can increase the size of vortices in this EHD flow. It can be explained that the charge diffusion drives the lateral movement of the fluid towards farfield, thus extending the vortex. This effect was not studied in the numerical simulations of \cite{perri2020electrically}.

In the end, we investigated the linear global stability of the steady flows in this EHD flow by adopting the Implicitly Restarted Arnoldi Method. In order to improve the computational efficiency, a smaller computational domain was chosen, and a sponge layer was applied to damp the reflection of the outgoing waves. The eigenvectors of the leading mode at the steady flow ($ T=500 $) and the periodic oscillatory flow ($ T=40000 $) are presented, and they show evidently different patterns. Furthermore, it is interesting to find that with the increase of charge diffusion, the flow becomes more stable. The leading global eigenvector of the charge density field shows a tadpole-shaped structure whose amplitude is consistent with the base flow characteristics. Finally, the linear stability analyses on high-$ T $ EHD flows have been conducted to investigate the flow instability, and the critical $ T_c $ is found to be slightly smaller than 20000. When $T$ is sufficiently large, we also observe a strong flow oscillation with an increased frequency, which might be due to the confinement effect of the geometry in the center region.

Future work can consider numerically investigating the EHD Moffatt-like eddies in the needle-plate configuration (in a cylindrical coordinate), which is the geometry adopted in the experiments of \cite{perri2020electrically,perri2021particle}. This will enable a more consistent and quantitative comparison to their results. Another future work can consider determining more accurately the linear instability criterion and bifurcation diagram in the transition process of this EHD flow. The stability analysis of the oscillating flow can be further refined by analyzing the time-averaged flow of the oscillating jet at a large $T$, which may be compared favourably to the experimental results (following the previous research on the cylindrical wake flow, e.g. \cite{Barkley2006}). 

To sum up, our work takes a step further to quantitatively characterise in detail the Moffatt-like eddies in blade-plate EHD flow with favourable comparisons with Moffatt's original result, supplementing the recent work of \cite{perri2020electrically,perri2021particle} to study this new phenomenon in EHD. It numerically demonstrates the relevance of the Moffatt's theoretical results in multi-physical flows. We hope that the current work can facilitate the observation of Moffatt-like eddies in other multi-physical flows in the future.

\begin{acknowledgments}
The financial support from the Ministry of Education, Singapore is acknowledged (the WBS No. R-265-000-689-114). X.H. is supported by a doctoral research scholarship from the National University of Singapore and a scholarship from the China Scholarship Council. The computational resources of the National Supercomputing Centre, Singapore are acknowledged.
\end{acknowledgments}
Declaration of Interests. The authors report no conflict of interest.

\section*{Appendix A: Verification of the domain size and the grid resolution in the simulations of Moffatt-like eddies}
\label{MoffattVali}
This appendix determines the size of the computational domain in our numerical simulations of the Moffatt-like eddies. We take $R=0.05$ (corresponding to $2\alpha=77.4^\circ) $ and $ T=500, Fe= 5\times10^3, C=5, M=50 $ to verify that the computational domain is large enough to obtain accurate results. We investigate three vortices, requiring a large computational domain that needs to be resolved. In addition, the inter-electrode angle is also a factor influencing the computation domain. We consider three sizes of the computational domain: they are G70, G80 and G90, corresponding to $H_w=70,80,90$ respectively, see figure \ref{fig.fig2a} for $H_w$.
 The ratio $ r_3/r_2 $ is evaluated in three different geometry sizes, as shown in table \ref{table:Mgeo2}. We can see that the results calculated in G70, G80 and G90 are close to the theoretical solution and the errors relative to the theoretical solution are all lower than $ 1\% $. Considering the calculation efficiency and the accuracy of the results, we choose G80 to perform the numerical simulations in this part.

\begin{figure}
	\centering
	\subfigure{
		\begin{minipage}[h]{0.4\textwidth}
			\centering
			\includegraphics[height=3.5cm]{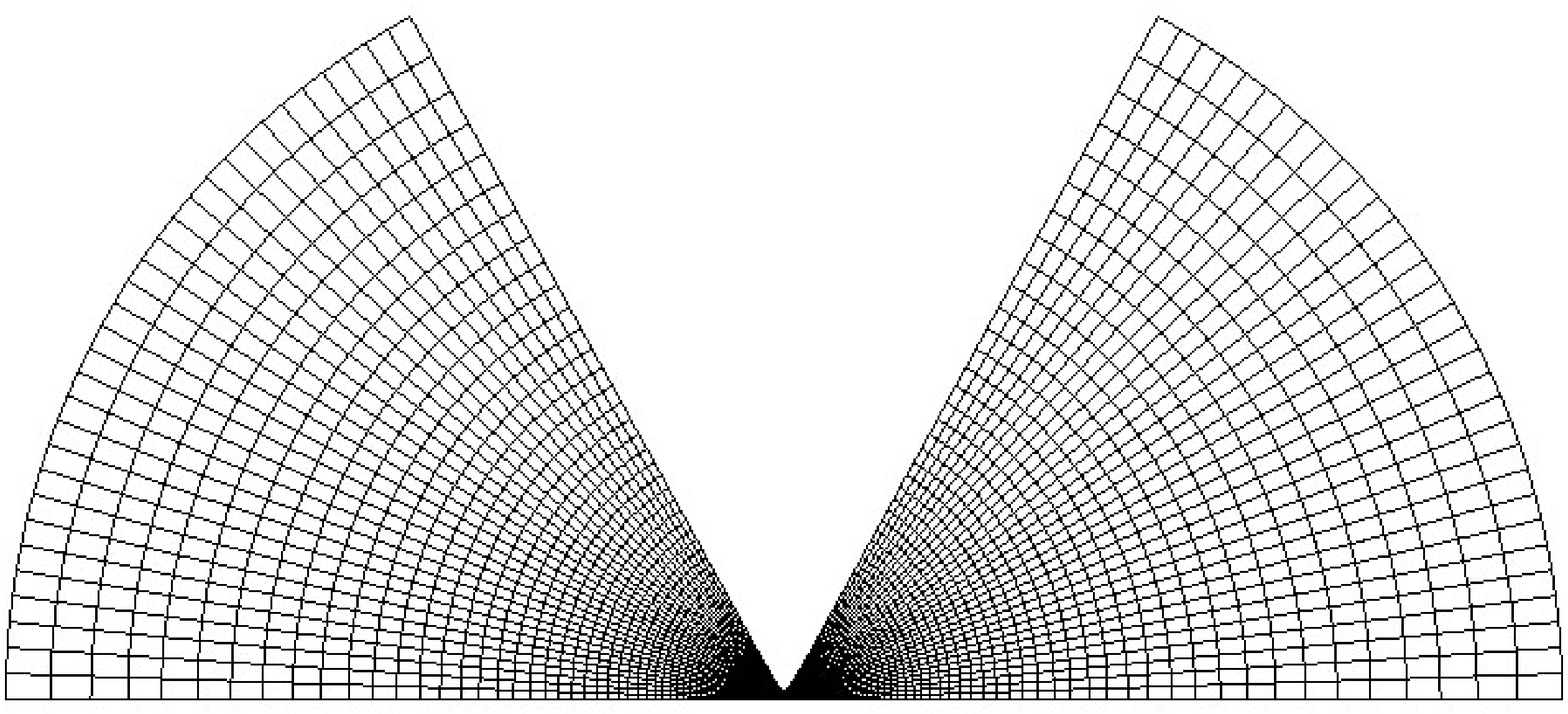}
			\put(-220,90){(a)}
			\label{fig.Mmesh}
		\end{minipage}
	}
	\hspace{40pt}
	\subfigure{
		\begin{minipage}[h]{0.4\textwidth}
			\centering
			\includegraphics[height=3.5cm]{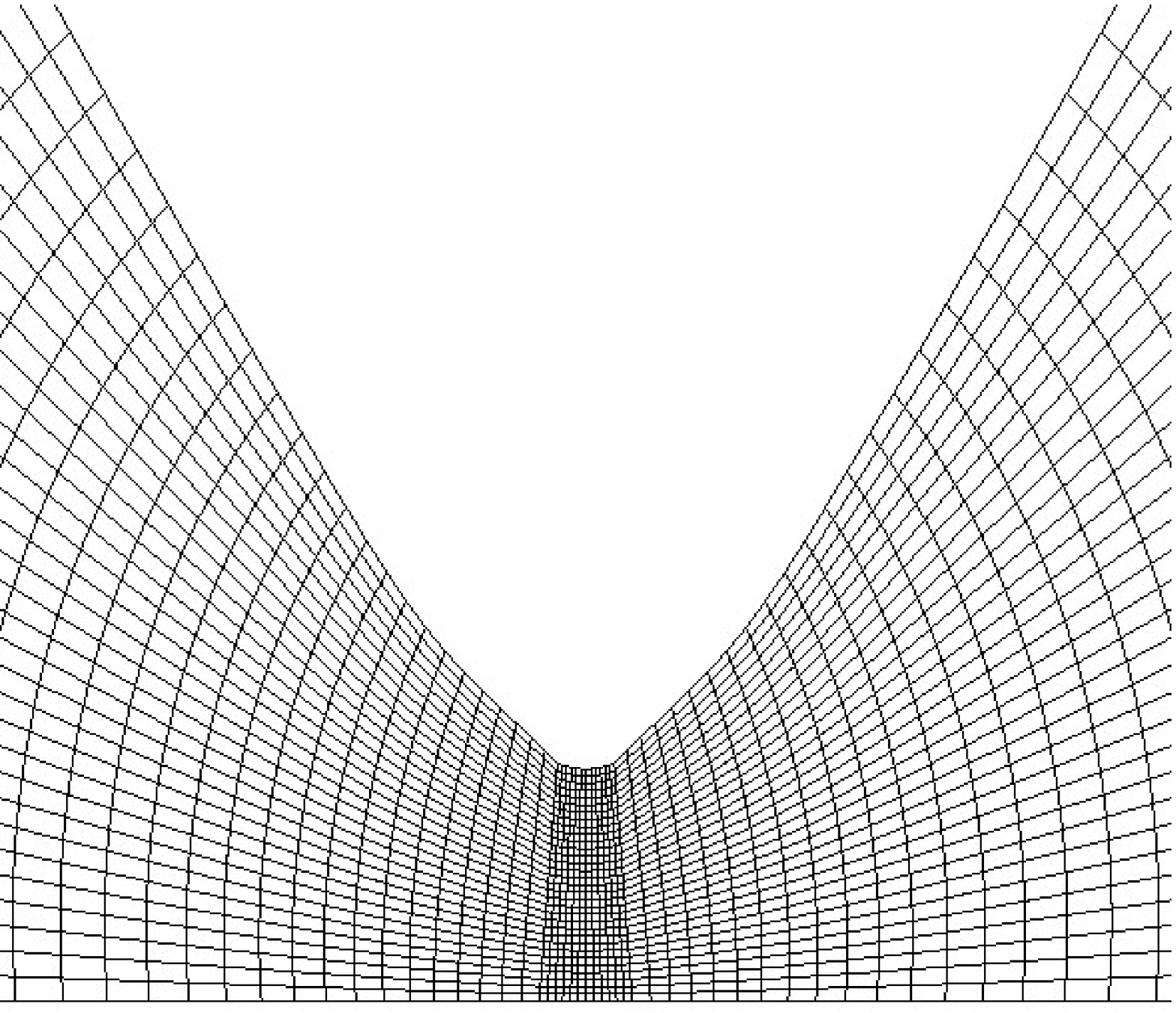}
			\put(-140,90){(b)}
			\label{fig.Mmeshs}
		\end{minipage}
	}
	\caption{(a) Distribution of the Legendre spectral elements for the numerical simulation of Moffatt eddies in the blade-plate EHD flow. (b) Zoom in display of the mesh refinement in the central region.}
	\label{fig.Mmeshl}
\end{figure}

We then perform the mesh independence test for the simulations. It is noted that the spectral elements mesh depends on two factors, one being the number of spectral elements $ N_e $ and the other the polynomial order $ N $. We test six sets of grid sizes; as shown in table \ref{table:Mmesh}. The ratio $ r_3/r_2 $ at $ R=0.3, T=500 $ and $ Fe= 5\times10^3 $ has been calculated for different meshes.  We can see that relative errors of MM2-MM6 to the theoretical solution are all less than $ 1\% $, and MM4 is adopted in this section. In addition, the distribution of the Legendre spectral elements is shown in figure \ref{fig.Mmeshl}.

\begin{table}
	\begin{center}
		\def~{\hphantom{0}}
		\begin{tabular}{l c c c}
			Geometry size & ~G70~ & ~G80~ & ~G90~\\
			$ H_w $ & ~70~ & ~80~ & ~90~\\
			Numerical $ r_3/r_2 $(present results) & ~9.30~ & ~9.31~ &~9.31~\\
			Theoretical $ r_3/r_2 $\citep{moffatt}& ~9.37~ & ~9.37~ &~9.37~\\
			Error relative to the theoretical results & ~0.75\%~ & 0.64\% & ~0.64\%~ \\
		\end{tabular}
		\caption{Independence of computational domain size validation at $ Fe= 5\times10^3, T=500 $, $ R= 0.05 $ ($ 2\alpha=77.4^\circ $), $ C=5 $ and $ M=50 $.}
		\label{table:Mgeo2}
	\end{center}
\end{table}

\begin{table}
	\begin{center}
		\def~{\hphantom{0}}
		\begin{tabular}{l c c c c c c}
			Mesh & MM1 & MM2 & MM3 & MM4 & MM5 & MM6\\
			$ N $ & 3 & 5 & 7& 7 & 7 & 9\\
			$ N_e $ & 5056 & 5056&3432 & 5056& 6660 & 5056\\
			Numerical $ r_3/r_2 $(present results) & ~5.31~ & ~5.26~ &~5.25~&~5.25~&~5.25~ &~5.25~\\
            Theoretical $ r_3/r_2 $\citep{moffatt}& ~5.22~ & ~5.22~& ~5.22~ &~5.22&~5.22~~&~5.22~\\
			Error relative to the theoretical solution & 1.72\% & 0.77\% & 0.57\%& 0.57\%& 0.57\%&0.57\%\\
		\end{tabular}
		\caption{ Grid independence validation at $ Fe= 5\times10^3, T=500 $, $ R= 0.3 $ ($ 2\alpha=61.3^\circ $), $ C=5 $ and $ M=50 $.}
		\label{table:Mmesh}
	\end{center}
\end{table}

Since we use a similar electrode configuration (blade-plate) with the experiment \citep{perri2020electrically} (needle-plate), we perform a comparison with experimentally reported values on peak vertical velocity here. The liquid used in the experiment \citep{perri2020electrically} was food-grade canola oil, and its physical properties were listed in their table 1: kinematic viscosity $ \nu^*=9\times 10^{-5}\mathrm{m^2/s} $, mass density $ \rho^*=900\mathrm{Kg/m^3} $, relative dielectric permittivity $ \epsilon_r^*=3.14 $, conductivity $ \sigma^*= 4\times10^{-10} \mathrm{S/m}$. The diffusion coefficient for the charges is $ D_\nu^*=1\times10^{-11} \mathrm{m^2/s} $. The distance between the needle tip and the plate in the experiment was $ H^*=1.8\mathrm{mm} $. In addition, the experiments were performed at $ 300\mathrm{K} $ and the inter-electrode angle is $ 76^\circ $. Therefore, the corresponding values of the non-dimensional parameters in our scaling method for $ \Phi_0^*=12 $kV (that is a typical value in the experiment) can be calculated as $ C_0=10, M=500, T=10000, R=0.06$ and $ Fe=5\times10^5 $. The maximum vertical velocity in our numerical results is 48.43, corresponding to the dimensional value $ 125\mathrm{mm/s} $. The maximum vertical velocity is about $ 45\mathrm{mm/s} $ indicated from figure 5 in \cite{perri2020electrically}. The discrepancy may mainly be due to the different geometry of the two works as well as the 3-D effect in the experiments (whereas we conducted 2D simulations).

\section*{Appendix B: Verification of the domain size and the grid resolution in the global stability analysis}
\label{velidationGlo}

When conducting linear global stability analysis for the blade-plate EHD flow, we adopt a smaller computational domain, and a sponge layer is used to prevent the reflections of the outgoing disturbances from the farfield boundary. Here we perform the verification of the domain size and the grid independence to ensure the credibility of the results. In this section, the value of $T$ is 2900, which is the largest value studied in our linear stability analysis. The other parameters are $ C=5, M=50, R=0.05$ and $ Fe=5\times10^3 $. Note that in the global stability analysis, two steps are involved: one should first obtain the nonlinear steady base flow and then conduct its stability analysis. We present below the verification of both steps.

We first consider the computational domain size, which should be sufficiently large. This is confirmed by examining the saturated maximum velocity magnitude $ |\Ub|_{max}^s $ appearing in the flow for five different geometry sizes as shown in table \ref{table:geo}. The notation of $H_p$ (controlling the size of the physical domain) can be found in figure \ref{fig.fig2a}. In addition, the size of the sponge region remains the same, namely $ \Delta H_s=2 $ (also see figure \ref{fig.fig2a}). From table \ref{table:geo}, we can see that the relative error between G3 and G5 is less than $ 0.5\% $. Thus, we choose G3 as our computational domain, that is $  H_p=5 $.

\begin{table}
	\begin{center}
		\def~{\hphantom{0}}
		\begin{tabular}{l c c c c c}
			Geometry size & G1 & G2 & G3 & G4 & G5\\
			$ H_p $ & 3 & 4 & 5 & 6 & 7\\
			$ |\Ub|_{max}^s $ & 17.8054 & 18.0383 &18.0596 & 18.0738 &18.1163\\
			Relative error & 1.72\% & 0.43\% & 0.31\%&  0.23\%& \\
		\end{tabular}
		\caption{Independence of computational domain size validation at $ T=2900 $. The other parameters are $ C=5, M=50, R=0.05$ and $ Fe=5\times10^3 $.}
		\label{table:geo}
	\end{center}
\end{table}

\begin{figure}
	\centering
	\includegraphics[height=5.5cm]{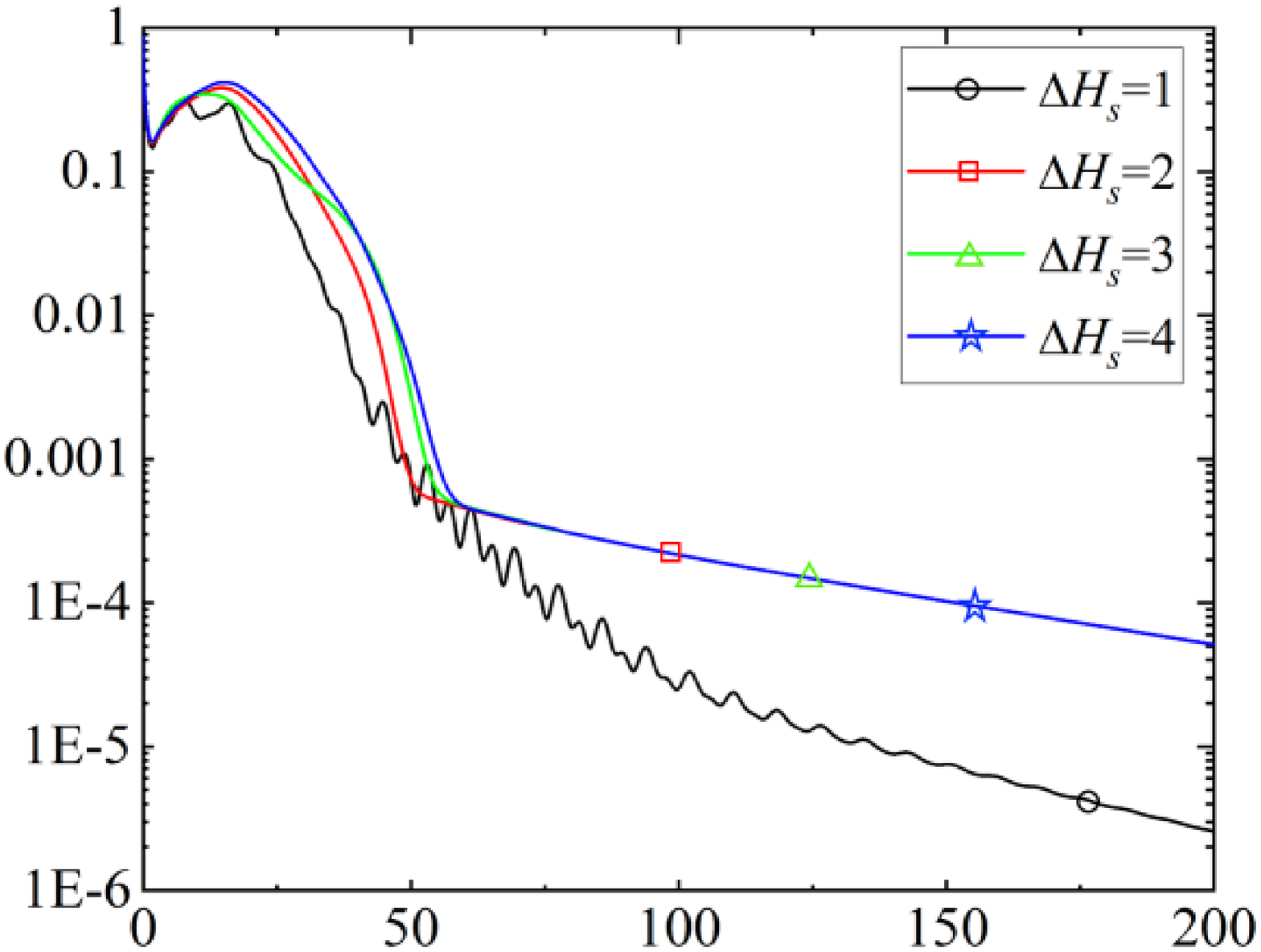}
	\put(-220,50){\rotatebox{90}{$ Energy $}}
	\put(-95,-10){$ t $}
	\caption{Energy evolution of linear blade-plate EHD flow with different sponge zones at $ T=2900 $. The parameters are the same as table \ref{table:geo}.}	
	\label{fig.sponge}
\end{figure}

We then determine the size of the sponge region. The strategy is to keep the size and the grid resolution of the physical area unchanged ($  H_p=5 $) and test four values of sponge region with different sizes, i.e. different $\Delta H_s $, as shown in table \ref{table:sponge}, and the number of elements in the sponge area changes in proportion to the size. The saturated maximum velocity magnitude $ |\Ub|_{max}^s $ of nonlinear EHD flow with different sponge layers are shown in table \ref{table:sponge}, we can see that the size of the sponge layer almost has no effect on the nonlinear results. Since the sponge region mainly acts on the linear wave, we compare the energy evolution of linear EHD plume with different sponge regions, as shown in figure \ref{fig.sponge}. According to the results presented in table \ref{table:sponge} and figure \ref{fig.sponge}, S2 is adopted. To summarise, we have decided a proper computational domain with $  H_p=5 $ and $  \Delta H_s=2 $.

\begin{table}
	\begin{center}
		\def~{\hphantom{0}}
		\begin{tabular}{l c c c c}
			Geometry size & S1 & S2 & S3 & S4\\
			$ \Delta H_s $ & 1 & 2 & 3 & 4 \\
			$ |\Ub|_{max}^s $ & 18.0489 & 18.0596 &18.0586 & 18.0585 \\
			Relative error & 0.053\% & 0.0061\% & 0.00055\%& \\
		\end{tabular}
		\caption{Independence of sponge region size validation at $ T=2900 $. The parameters are the same as table \ref{table:geo}.}
		\label{table:sponge}
	\end{center}
\end{table}

We next verify the independence of numerical results with respect to the grid resolution by using the same method in the appendix. We test seven sets of grid sizes, as shown in table \ref{table:mesh}. The saturated maximum velocity magnitude $ |\Ub|_{max}^s $ inside the flow domain has been calculated for different meshs.  It can be seen that relative errors of M2-M7 compared to M7 are all less than $ 1\% $, indicating that results converge at these mesh sizes. Thus, considering both the computational accuracy and efficiency, we  generate results with the mesh M4 in most cases in the result section. We note that when calculating the results for larger $ T $, a larger computational domain is adopted ($  H_p=20 $ and $  \Delta H_s=10 $), and a refined mesh ($ N_e=2950, N=9 $) is used.

\begin{table}
	\begin{center}
		\def~{\hphantom{0}}
		\begin{tabular}{l c c c c c c c}
			Mesh & M1 & M2 & M3 & M4 & M5 & M6& M7\\
			$ N $ & 3 &  7 & 5 & 7 & 9 & 7& 11 \\
			$ N_e $&2816&1944& 2816&2816&2816&3710&2816\\
			$ |\Ub|_{max}^s $  & 19.0105& 18.1302 & 18.0628&18.0596 & 18.0027&17.9966 &17.9700\\
			Relative error & 5.79\%  & 0.89\%& 0.52\% & 0.50\%&  0.18\% &0.15\% &\\
		\end{tabular}
		\caption{ Grid independence validation at $ T=2900 $. The parameters are the same as table \ref{table:geo}.}
		\label{table:mesh}
	\end{center}
\end{table}

\section*{Appendix C: Nomenclature}\label{Nomenclature}
This work deals with a multi-physical flow with many different symbols to denote flow parameters and the geometry. They are summarized in table \ref{table:symbol}.

\begin{table}
	\begin{center}
		\def~{\hphantom{0}}
		\begin{tabular}{l l }
			Symbol & Definition \\
			\hline
			$ C $ & Dimensionless charge injection intensity\\ 
			$ Fe $ & Inverse of the charge diffusion coefficient \\ 
			$ M $ & Dimensionless charge mobility  \\ 	
			$ T $ & Electric Rayleigh number \\ 			
			\\		
			$ H^* \ [m]$ & Distance from the blade tip to the plate electrode, used for nondimensionalisation\\
			$ H_p $ & Height of the physics domain  \\ 
			$ H_w $ & Height of the whole domain \\
			$ \Delta H_s $ & Height of the sponge layer  \\ 
			$ O $ & Original point of the Cartesian coordniate\\
			$ O_1,O_2,O_3 $ & Center of vortex 1, 2, 3 \\ 
			$ r_1,r_2,r_3 $ & Distance from the center of vortex 1, 2, 3 to the corner \\
			$ R $ & Radius of curvature of the blade tip \\ 
			$ v_\theta $ &  Azimuthal velocity\\
			$ |v_{\theta}|_{n+1/2} $  &  Absolute value  of the local maximum azimuthal velocity of $n$-th vortex\\		
			$ 2\alpha $ &Inter-electrode angle\\	
			$ \rho $ & Size ratio of adjacent vortices\\ 		
			$ \omega $ & Eigenvalue  \\ $  \Omega $ & Intensity ratio of adjacent vortices \\
		\end{tabular}
		\caption{Symbols used in this paper for the parameters and geometry.}
		\label{table:symbol}
	\end{center}
\end{table}

\bibliographystyle{jfm}
\bibliography{Reference}

\end{document}